\documentclass[prl,amsmath,amssymb,twocolumn, showpacs, superscriptaddress,10pt]{revtex4-1}

\usepackage{amsmath}
\usepackage{hyperref}
\usepackage{graphicx}
\usepackage{amsfonts}
\usepackage{amsthm}
\usepackage{cases}
\usepackage{bm}

\usepackage[normalem]{ulem}

\usepackage{color}
\definecolor{Blue}{rgb}{0.00, 0.00, 1.00}
\definecolor{Red}{rgb}{1.00, 0.00, 0.00}
\definecolor{Green}{rgb}{0.00, 0.70, 0.00}

\hypersetup{
    colorlinks=true,       
    linkcolor=red,          
    citecolor=blue,        
    filecolor=magenta,      
    urlcolor=cyan           
}

\newcommand{\nn}{\nonumber}
\newcommand{\be}{\begin{equation}}
\newcommand{\ee}{\end{equation}}
\newcommand{\bea}{\begin{eqnarray}}
\newcommand{\eea}{\end{eqnarray}}

\newcommand{\beq}{\begin{equation}}
\newcommand{\eeq}{\end{equation}}
\newcommand{\beqn}{\begin{eqnarray}}
\newcommand{\eeqn}{\end{eqnarray}}

\DeclareMathOperator{\Det}{Det}

\newcommand{\dd}{\ensuremath{\mathrm d}}

\def\Xint#1{\mathchoice
   {\XXint\displaystyle\textstyle{#1}}%
   {\XXint\textstyle\scriptstyle{#1}}%
   {\XXint\scriptstyle\scriptscriptstyle{#1}}%
   {\XXint\scriptscriptstyle\scriptscriptstyle{#1}}%
   \!\int}
\def\XXint#1#2#3{{\setbox0=\hbox{$#1{#2#3}{\int}$}
     \vcenter{\hbox{$#2#3$}}\kern-.5\wd0}}

\def\dashint{\Xint-}

\begin{document}

\title{
A transition in the hole probability at finite temperature for free fermions in $d$ dimensions}

\author{Giuseppe Del Vecchio Del Vecchio}
\affiliation{Laboratoire de Physique de l'Ecole Normale Sup\'erieure, CNRS, ENS and PSL Universit\'e, Sorbonne Universit\'e, Universit\'e Paris Cit\'e,
24 rue Lhomond, 75005 Paris, France}

\author{Pierre Le Doussal}
\affiliation{Laboratoire de Physique de l'Ecole Normale Sup\'erieure, CNRS, ENS and PSL Universit\'e, Sorbonne Universit\'e, Universit\'e Paris Cit\'e,
24 rue Lhomond, 75005 Paris, France}

\author{Gr\'egory \surname{Schehr}}
\affiliation{Sorbonne Universit\'e, Laboratoire de Physique Th\'eorique et Hautes Energies, CNRS UMR 7589, 4 Place Jussieu, 75252 Paris Cedex 05, France}
\date{\today}

\begin{abstract}
In a free Fermi gas at temperature $T$ much higher than the Fermi temperature one expects that the fluctuations of the number of particles in a given region has Poissonian/classical statistics. On the other hand at low temperature the Pauli exclusion principle leads to non trivial counting statistics. It is of great interest
from a theoretical and experimental point of view to characterize the crossover between these two limits. {Here we focus on the hole probability $P(R,T)$, i.e. the probability that a region of size $R$ is devoid of particles, in dimension $d$, and on the case of a spherical region of large radius $R$. We show that at low temperature it takes the scaling form
$P(R,T)\sim \exp\big[-(k_F R)^{d+1}\Phi_d(u=2R\,T/k_F)\big],$
where $k_F$ is the Fermi momentum.
By mapping the problem to an effective Coulomb gas, we compute exactly the scaling function $\Phi_d(u)$ in any dimension. Remarkably, it exhibits a transition of order $\tfrac{3}{2}(d+1)$ at the universal critical value $u_c=2/\pi$, signaling a sharp change in the mechanism of rare fluctuations, associated with the emergence of a macroscopic gap in the optimal density of the associated Coulomb gas. Our analytical predictions are supported by precise numerical evaluations of the corresponding Fredholm determinants.}

\end{abstract}



\maketitle

Free fermions, i.e., non-interacting fermions in the absence of an external potential, are among the most fundamental systems in nature~\cite{Landau1980,Huang1987,Mahan,Castin2007}. Despite their textbook status in quantum statistical physics and the wealth of existing results, they can still hold some surprises, as we demonstrate here. In fact, even without interactions, they exhibit rich collective behavior induced by the Pauli exclusion principle \cite{Pauli}. In particular, they display nontrivial correlations, entanglement properties, and full counting statistics, which have been extensively studied 
theoretically \cite{Levitov1996,Klich,Klich2006,Peschel2009,KlichLevitov2009,CalabreseMinchev,LeHur2011,AbanovIvanovQian2011,
CalabreseMinchev2012,DeanPLDReview}, notably using tools from random matrix theory and determinantal point processes~\cite{MehtaBook,Forrester,Macchi,Joh_det,Boro_det,Eisler1,MMSV14,CalabreseLeDoussalMajumdar2015,DeanReview2019,SmithLMS2021}. As an emblematic exactly solvable class of quantum many-body systems, with connections to a broad range of problems in statistical physics, they are currently attracting a renewed interest at the interface between physics and mathematics~\cite{PrahoferSpohn2002,tiling_fermion2019,Bardenet2022,Borodin_freefermion2023}.

With the advent of Fermi quantum microscopes, it is now possible to directly image the positions of the fermions~\cite{BDZ08,GrossBloch2017,Fermicro1,Fermicro2,Fermicro3,Omran2015,Greif2016,deJongh2025,flattrap,Pauli}. In a free Fermi gas, these positions form a determinantal point process known as the ``Fermi sphere'', which exhibits statistical properties markedly different from those of an independent Poisson point process~\cite{Macchi,Torquato1,Scardicchio2009}. 
Its defining property (i.e., Wick's theorem) is that all the spatial correlation functions can be expressed as determinants of an elementary kernel function, the one body correlation function $g_1(x,y) = \langle \Psi^+(x) \Psi(y) \rangle$~\cite{Mahan,Macchi,Boro_det,Joh_det,DeanPLDReview,GaudinWick}. 
The homogeneous free Fermi gas with density $\rho$ and a single spin component
is characterized by three length scales:
(i) the inter-particle distance, which at zero temperature 
is the inverse Fermi wavevector $k_F^{-1} \propto \rho^{-1/d}$; 
(ii) the thermal de Broglie wave length $\lambda_T= \sqrt{2 \pi \hbar^2/(m k_B T)}$ and 
(iii) the thermal correlation length $\xi_T= \frac{\hbar^2 k_F}{\pi m k_B T}$~\cite{Castin2007,Giam_book,Mahan,Goedeker1998}. The length $\xi_T$ controls the decay of the one body correlation at finite temperature, $g_1(r)\sim k_F^d \frac{T}{T_F} 
(r k_F)^{- (d-1)/2} \cos(k_F r + \phi_d) e^{-r/\xi_T}$, where $T_F=\hbar^2k_F^2/(2m\,k_B)$ is the Fermi temperature~\cite{footnote_xiT}. In the following we set $m=\hbar=k_B=1$.
\\
\begin{figure}[t]
\includegraphics[width=\linewidth]{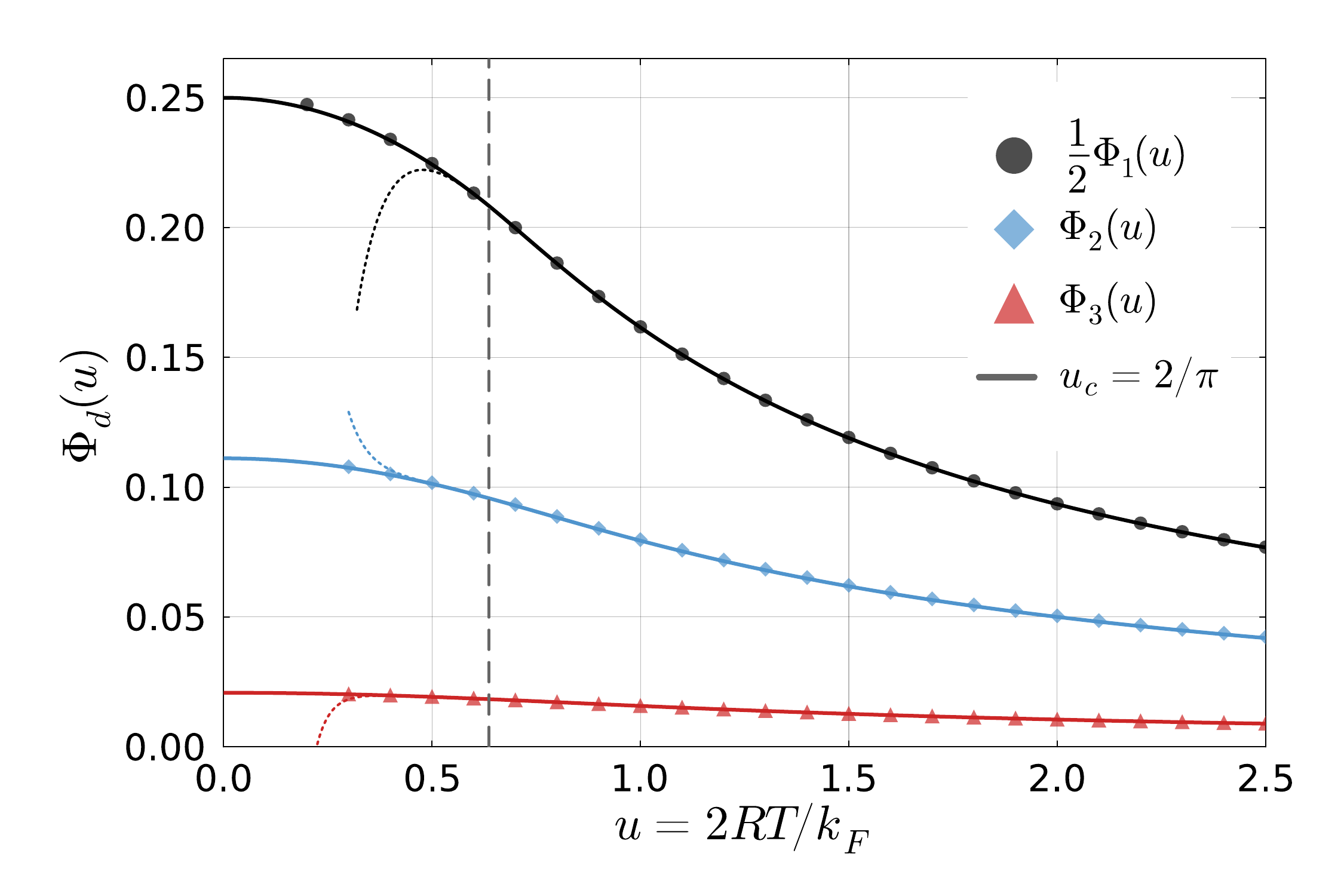}
\caption{Scaling function $\Phi_d(u)$ for the hole probability $P(R,T)$ of a spherical domain of radius $R$ at low temperature 
defined in \eqref{lowT} for $d=1,2,3$. The solid line is the theoretical prediction
\eqref{phid}-\eqref{phi_dm2}. The dots are obtained from a numerical evaluation of the Fredholm determinant in \eqref{prod_FD}-\eqref{Pell}, followed by an extrapolation from finite $z$ data, see Fig. \ref{Fig_conv} in End Matter.
The location of the predicted transition of order 
$\frac{3}{2}(d+1)$ at $u=u_c = 2/\pi$, see \eqref{deltaphi_txt}, is shown by the vertical dotted line. The analytic continuation of the $\Phi_{d,+}(u)$ for $u<u_c$ is indicated with a dash-dotted line. }\label{Fig}
\end{figure}


A useful way to characterize such point processes is to study the full counting statistics (FCS), e.g. the fluctuations of the number of fermions in a given domain. For free fermions, this is of particular interest since, e.g., in the ground state, the second cumulant gives the leading behavior of the bipartite entanglement entropy \cite{Klich2006,CalabreseMinchev,LeHur2011,CalabreseLeDoussalMajumdar2015}. Another fundamental observable, which probes higher-order correlation functions, is the so-called hole probability, i.e. the probability that the domain is empty of fermions.
For a Poisson point process, it decays exponentially with the volume, and for a spherical domain of radius $R$ in dimension $d$ it reads
\be \label{Poisson}
P^{\rm Poi}(R)= e^{- \rho V_d R^d} \;,
\ee
where $V_d=S_d/d= \pi^{d/2}/\Gamma(1+d/2)$, is the volume of the unit sphere and $\Gamma(z)$ is the gamma function.

For free fermions at zero temperature, the Pauli principle makes the system more rigid, and
the hole probability $P(R,T=0)$ decays faster, as was predicted in \cite{Torquato1} and 
shown only recently in Ref.~\cite{Gabriel_gap} by an exact calculation 
leading to the asymptotic behavior for $k_F R \gg 1$
\be \label{holeT0}
P(R,T=0) \sim e^{- \kappa_d (k_F R)^{d+1}} ~,~ \kappa_d=\frac{2}{(d+1)^2 \Gamma(d+1)} \;.
\ee 
This prediction has recently been probed experimentally in a two-dimensional ultracold Fermi gas using a continuum quantum gas microscope \cite{DixmeriasHoleProba}. Temperatures as low as $T/T_F \approx 0.15$ were achieved, allowing one to explore the crossover, in $d=2$ and for $k_F R = O(1)$, from the degenerate low-temperature regime \eqref{holeT0} to the classical high-temperature limit described by the Poisson law in~(\ref{Poisson}). Excellent agreement with theoretical predictions was found \cite{DixmeriasHoleProba}, based on determinantal point process techniques (i.e., Fredholm determinants).




In this Letter we show that the finite temperature hole probability
exhibits a crossover on the scale $R \sim \xi_T$ 
which, remarkably, sharpens into a genuine transition 
in the simultaneous large-radius/low temperature limit. 
Let us present our main results. From the exact expression for the finite temperature hole probability $P(R,T)$ as a product 
of Fredholm determinants valid in any dimension [see Eqs. (\ref{prod_FD})-(\ref{hat_BesselKernel_txt})], we show that 
at large $R$ two distinct regimes emerge:  

(i) {\it a high temperature regime} with fixed $T \sim T_F$, where
the characteristic length is $\lambda_T$ and 
for $R/\lambda_T \gg 1$
\be \label{highT}
\log P(R,T) \simeq - (R \sqrt{T})^d A_d\left(\frac{T}{T_F}\right) + (R \sqrt{T})^{d-1} B_d\left(\frac{T}{T_F}\right)  
\ee 
The leading and subleading scaling functions are found explicitly as \cite{thermo}
\bea  \label{Adt} 
 A_d(t) &=& - \frac{1}{2^{d/2} \Gamma(1+d/2)} 
{\rm Li}_{\frac{d}{2}+1} (- e^{\beta \tilde \mu})  \\
 B_d(t) &=& \frac{2^\frac{d-3}{2}}{\pi \Gamma(d)} 
\int_0^{+\infty} dv \, v^{\frac{d-1}{2}} 
[ {\rm Li}_{1/2}(- e^{\beta {\tilde \mu}- v } ) ]^2 \;, \label{Bdt}
\eea 
where ${\rm Li}_{s}(z)$ denotes the
polylogarithm function of index $s$. Here $\beta = 1/T$ and $\tilde \mu$ is the chemical potential fixed by the relation
$e^{\beta \tilde \mu}  = \Upsilon_d^{-1}(1/t^{d/2})$
with $\Upsilon_d(z)$ defined in \eqref{muvsT}.
At very high temperature $t=T/T_F \gg 1$ (equivalently $\beta \tilde \mu \to -\infty$)
$A_d(t) \simeq \frac{1}{2^{d/2} \Gamma(1+d/2)^2} t^{-d/2} $
and one recovers the Poisson result \eqref{Poisson} \cite{footnotePoisson};

(ii) {\it a low temperature regime} $T \ll T_F$, where 
the characteristic length is the interparticle distance 
$1/k_F$, and for $k_F R \gg 1$ 
\bea \label{lowT}
\log P(R,T) \simeq - (k_F R)^{d+1} \Phi_d(u) \,, \, u = \frac{2 T R}{k_F} = \frac{2}{\pi}\,\frac{R}{\xi_T} \;.
\eea 
This regime corresponds to $R \sim \xi_T$, where $\xi_T$ is the thermal correlation length defined above, i.e. to a fixed scaling variable $u$. 
Remarkably, the scaling function $\Phi_d(u)$ exhibits
a phase transition {\it in any dimension} $d \geq 1$ at the
universal critical value $u=u_c=2/\pi$ (which corresponds to $R=\xi_T$)
\be \label{phid}
\Phi_d(u) = \begin{cases} \Phi_{d,+}(u) \quad , \quad u>2/\pi  \\
\Phi_{d,-}(u) \quad , \quad u<2/\pi \;.
\end{cases} 
\ee 
{As we show below, the formation of a large empty region is controlled by an effective Coulomb gas in momentum space, whose momentum density undergoes a transition at $u_c$. For $u>u_c$, the support of this density extends down to zero momentum, while for $u<u_c$ it is shifted away from the origin, corresponding to a gap in (scaled) momenta -- see Fig. \ref{plot_rtilde} in the End Matter. This macroscopic gap opening in momentum space underlies the nonanalytic behavior of $\Phi_d(u)$.}

We find that for $u>u_c$
the scaling function
is a polynomial of $1/u$ of degree $d+1$, explicitly
\bea \label{phi_dp}
\Phi_{d,+}(u) 
=
\begin{cases} 
\hspace*{-0.5cm}& \frac{4}{3 \pi u} - \frac{1}{\pi^2 u^2} \,, \, \hspace*{0.cm}d= 1\;, \\ 
\hspace*{-0.5cm}& \frac{1}{8 u}-\frac{8}{15 \pi ^2
   u^2}+\frac{1}{12 \pi ^2
   u^3} \,, \, \hspace*{0.cm}d= 2\;, \\ 
\hspace*{-0.5cm}& \frac{4}{45 \pi  u} -\frac{1}{6 \pi ^2
   u^2}+\frac{16}{105 \pi ^3
   u^3}-\frac{1}{18 \pi ^4
   u^4}\,, \, d= 3\;.
\end{cases} 
\eea 
In the phase $u<u_c$ the scaling function for
odd $d$ is a polynomial of $u^2$ with highest order
term $u^{d+1}$, explicitly
\bea \label{phi_dm}
\Phi_{d,-}(u) 
=
\begin{cases} 
\hspace*{-0.5cm}& \frac{1}{2} - \frac{\pi^2 u^2}{48} \;, \; d= 1 \\
\hspace*{-0.5cm}& 
\frac{1}{48} -\frac{\pi ^2 u^2}{1440} 
+ \frac{\pi^4 u^4}{53760} \;, \; d= 3 \;.
\end{cases} 
\eea  
Its expression is more involved in even $d$,
e.g. in $d=2$
\bea \label{phi_dm2}
&& \Phi_{2,-}(u) = \sqrt{4-\pi ^2 u^2} \left(-\frac{\pi ^2
   u^2}{1440}+\frac{9}{40 \pi ^2
   u^2}+\frac{7}{360}\right) \nonumber \\
   && -\frac{8}{15 \pi ^2
   u^2}+\left(\frac{1}{6 \pi ^3 u^3}+\frac{1}{4 \pi 
   u}\right) {\rm sin}^{-1}\left(\frac{\pi  u}{2}\right) \;,
\eea 
with, in particular, $\Phi_{2,-}(u) \simeq \frac{1}{9}-\frac{\pi ^2 u^2}{240}$
at small $u$. The above formulae reproduce for any $d$ the zero temperature
result \eqref{holeT0}, 
with $\Phi_{d,-}(0)=\kappa_d$. Next, close to the transition at $u=u_c=2/\pi$
one finds that for $u<u_c$
\be \label{deltaphi_txt}
\Phi_{d,+}(u) - \Phi_{d,-}(u) \propto  (u_c-u)^{\frac{3}{2}(d+1)} \;,
\ee 
where here $\Phi_{d,+}(u)$ is given by \eqref{phi_dp} continued
to $u<u_c$, and the prefactor in (\ref{deltaphi_txt}) is given in \cite{SM}. The transition is thus of ``order'' $\frac{3}{2}(d+1)$.
Around the critical point setting $y = (2 k_F R)^{2/3}(u-u_c)/u_c$
there is a critical scaling region $y=O(1)$ where the hole probability 
takes the form
\bea \label{hole_TW_txt}
P(R,T) \simeq   G_d(y) e^{- (k_F R)^{d+1} \Phi_{d,+}(u) } \;.
\eea 
The scaling function behaves as $G_d(y) \to 1$ for $y \to +\infty$ 
and $\log G_d(y) \propto - |y|^{(3/2)(d+1)}$
for $y \to -\infty$ to ensure matching with the 
singular behavior \eqref{deltaphi_txt}. The critical
region can be probed by varying temperature around
$T_c= k_F/(\pi R)$, with a width $\delta T \sim T_c (k_F R)^{-2/3}
\sim T_F (k_F R)^{-5/3}$, in any $d$.

Finally, one can match the high-$T$ regime (i) in Eq.~(\ref{highT}) with the low-$T$ regime (ii) in Eq. (\ref{lowT}).
Indeed for $t \ll 1$, one has~\cite{betad} 
\be 
\log P(R,T) \simeq - \beta_d \frac{(k_F R)^{d}}{t}  +
\gamma_d \frac{(k_F R)^{d-1}}{t^2} \;,
\ee 
using $t = T/T_F=u/(k_F R)$ one finds that it exactly reproduces
the $1/u$ and $1/u^2$ terms in $\Phi_{d,+}(u)$ in Eq.~(\ref{phi_dp}).
i.e. one has $\Phi_{d,+}(u)=\frac{\beta_d}{u}- \frac{\gamma_d}{u^2}
+ O(1/u^2)$. 
We have tested our predictions in the low $T$ regime (ii) against a numerical evaluation of the 
Fredholm determinant \eqref{Pell} using the Bornemann method, as in \cite{DixmeriasHoleProba}. 
Taking into account the finite $z$ corrections, see End Matter and \cite{SM}, 
we obtain an excellent agreement, as shown in Fig. \ref{Fig}.

\vspace*{0.5cm}

We now sketch the main steps of the derivation, starting
by recalling the model. We consider free fermions in $d$ dimensions described
by the $N$-body Hamiltonian $\hat {\cal H}_N = \sum_{i=1}^N \frac{{\bf p}_i^2}{2}$, where the ${\bf p}_i$'s are the fermion's momenta.
We describe the system in the grand canonical ensemble at temperature $T=1/\beta$ and
chemical potential $\tilde \mu$. The quantum averages are thus defined as
$\langle {\cal O}  \rangle= \frac{1}{Z} {\rm Tr} \, {\cal O} e^{- \beta( {\cal H}_N - \tilde \mu N)}$ with $Z={\rm Tr} e^{- \beta( {\cal H}_N - \tilde \mu N)}$.
We define ${\cal N}_R$ the number of fermions in a spherical
domain of radius $R$. We compute the hole probability $P(R,T)$,
i.e. the probability that ${\cal N}_R=0$, 
\be 
P(R,T) = \lim_{s \to + \infty} \langle e^{- s {\cal N}_R} \rangle \;.
\ee 
%
%
%
%
We work at fixed mean density $\rho$, 
where 
$T_F= \frac{k_F^2}{2} = \frac{1}{2} (\rho/c_d)^{2/d}$ with
$c_d=1/((4 \pi)^{d/2} \Gamma(1+ d/2))$. Then the chemical potential
$\tilde \mu\equiv \tilde \mu(T)$ is determined by the relation
\be \label{muvsT}
1 = \left(\frac{T}{T_F}\right)^{\frac{d}{2}} \Upsilon_d(e^{\beta \tilde \mu}) \;, \; \Upsilon_d(z)=- \Gamma\left(1+\frac{d}{2}\right) {\rm Li}_{\frac{d}{2}}(-z)\;.
\ee
In the limit $T/T_F \to 0$, since $\Upsilon_d(z) \simeq (\log z)^{d/2}$ for 
$z \gg 1$, one has $\tilde \mu \to \mu= k_F^2/2=T_F$
where $\mu$ is the Fermi energy.

We use two different methods to analyse the high $T$ (i) and
low $T$ (ii) regimes. To treat the high $T$ regime it is convenient to use a formula
due to Widom~\cite{Widom_d_dim}. In the grand canonical ensemble, the positions of
the free fermions form a determinantal point process (DPP) with kernel 
\be \label{Fermi_factor}
K({\bf x},{\bf x}') = \int \frac{d^d {\bf k} }{(2 \pi)^d} 
e^{i {\bf k} \cdot ({\bf x}-{\bf x}') } \sigma(k) \;, \; \sigma(k)= \frac{1}{e^{\beta (\frac{k^2}{2} - \tilde \mu) }+1 }\\
\ee 
where $\sigma(k)$ is the Fermi factor and $k = |{\bf k}|$. 
Using standard formula for DPP's we can write the hole probability as
a Fredholm determinant (FD) in $\mathbb{R}^d$ \cite{Joh_det,Boro_det}
\be \label{FD_start}
P(R,T)= {\rm Det}( I - \Pi_R K)  \;,
\ee 
where $\Pi_R$ denotes the projector on the $d$-dimensional sphere of radius $R$ and centered at the origin. The large $R$ asymptotics of a broad class of such FD 
was obtained by Widom \cite{Widom_d_dim}. Specifying to the spherical
geometry and performing the $d$-dimensional integrals we obtain 
the results \eqref{highT}-\eqref{Bdt}, see \cite{SM} for details. 

It turns out, however, that Widom's formula {\it does not allow} to describe the
low temperature regime (ii) since it is valid only for sufficiently
smooth function $\sigma(k)$. In particular it is well known that 
it fails to describe the $T=0$ limit, where {the Fermi factor} $\sigma$ becomes singular. 
To study the regime (ii), we develop in this paper a 1D Coulomb gas method
specially tailored to compute this finite temperature hole probability.
To this aim we first reduce the $d$-dimensional problem to a collection of 1D systems. 

The first step is to use the angular decomposition of the fermion coordinates (see e.g. \cite{Farthest}).
The single particle eigenstates 
are then obtained by standard methods from those of a collection
of 1D radial Hamiltonians
$\hat H_\ell = - \frac{1}{2} \partial_r^2  + \frac{\nu^2- \frac{1}{4} }{2 r^2}$
with $\nu=\ell+ \frac{d}{2}-1$ with $r \geq 0$, each with degeneracy
$g_d(\ell) = \frac{(2 \ell + d-2) \Gamma(\ell+d-2)}{\Gamma(\ell+1) \Gamma(d-1)}$, 
$\ell \geq 1$, and $g_d(0)= 1$. Let us denote ${\cal N}^{(\ell)}_{[0,R]}$
the number of noninteracting fermions in the interval $[0,R]$ in each sector.
Let us define $P^{(\ell)}(R,T)$ the probability that ${\cal N}^{(\ell)}_{[0,R]}=0$
in the grand canonical ensemble associated to $\hat H_\ell$
at chemical potential $\tilde \mu$ and temperature $T$. Each angular sector being
independent, the total hole probability $P(R,T)$ factorizes over the different $\ell$-sectors and one has 
\be \label{prod_FD}
P(R,T) = \prod_{\ell \geq 0} [P^{(\ell)}(R,T)]^{g_d(\ell)} \;.
\ee
Using the determinantal structure of noninteracting fermions
each of the hole probabilities $P^{(\ell)}(R,T)$
can be expressed as a Fredholm determinant (FD)
\be \label{Pell}
P^{(\ell)}(R,T)  = {\rm Det}( I - \Pi_{[0,R]} \hat K_\nu)  
\quad , \quad \nu = \ell + \frac{d}{2}-1 
\ee 
involving the finite $T$ Bessel kernel 
\bea  \label{hat_BesselKernel_txt}
\hat K_\nu(r,r')  &=& \sqrt{rr'}  \int_0^{+\infty} dk \, k J_\nu(k r) J_\nu(k r') \sigma(k) 
\eea 
where $r,r'>0$, and $\sigma(k)$ is the Fermi factor (\ref{Fermi_factor}). Formula \eqref{prod_FD}-\eqref{hat_BesselKernel_txt} were used for $d=2$ in \cite{DixmeriasHoleProba}.
The FD associated to the finite temperature Bessel kernel
\eqref{Pell} has been studied in a few mathematics papers \cite{BE2003,Ruzza2025,XuBessel},
but not in the regime of interest here.

A crucial step \cite{SM} is to exhibit a duality relation between real and momentum space
which allows us to rewrite the real space FD in \eqref{Pell} as a 
linear statistics problem in momentum space \cite{footnotepi}
\be \label{linstatbesselmain} 
P^{(\ell)}(R,T) =  \Big\langle \exp\left(  \sum_i \log( 1 + e^{- \beta (\frac{k_i^2}{2} - \tilde \mu) })  
\right)  \Big\rangle_{K_{\rm Be,\nu}} \;,
\ee 
with $k_i=\sqrt{p_i}/R^2$, where the expectation value is taken over the set of 
$\{ p_i \}_{i \in \mathbb{N}}$ which form a
DPP on the positive real
axis with the {\it zero
temperature} Bessel kernel $K_{\rm Be,\nu}$ (i.e.
obtained from 
\eqref{hat_BesselKernel_txt} 
by replacing $\sigma(k) \to \theta(\sqrt{2 \mu}-|k|)$). 
The advantage of this representation (\ref{linstatbesselmain}) is that $K_{\rm Be,\nu}$
describes the statistics of the eigenvalues of the
complex Wishart-Laguerre ensemble of random matrices near the
so-called hard edge \cite{Forrester}, whose probability distribution
is explicitly known and related to Coulomb gases (CG). 

To make use of this connection to CG,
we generalize an approach pioneered by Dyson in
\cite{Dyson_2015} for the case of the sine-kernel. Although less known than the CG method based on the Gaussian unitary ensemble (GUE) at finite $N$ \cite{Dyson_GUE}, it has the advantage to access directly the properties of the infinite volume free Fermi gas, described by the sine kernel. 
In \cite{Dyson_2015} the asymptotics of the 
hole probability in $d=1$ at large $R$ and $T=0$
was obtained by that method. 
In the present work we extend this approach
to (a) the Bessel kernel $K_{\rm Be,\nu}$ 
on the positive half-axis and 
(b) finite temperature using~\eqref{linstatbesselmain} and (c) {arbitrary spatial dimension $d$}.

Introducing
the potential function
$V(p)= \log( 1 + e^{- \beta (\frac{p}{2 R^2} - \tilde \mu)})$,
the right hand side (RHS) in
\eqref{linstatbesselmain} can be seen
an expectation value of the exponential of the total potential 
energy $e^{\sum_i V(p_i)}$. Rewriting the 
distribution of the $p_i$'s of the $T=0$ Bessel DPP (BDPP)
as the Gibbs measure of a CG, see \cite{SM} for details, 
and defining the empirical density 
$\rho(p)= \sum_i \delta(p-p_i)$, 
we arrive 
at the following representation of the hole probability in the sector $\ell$, which we
expect to be valid for large $R$
\be  \label{P0_Eminbessel0}
P^{(\ell)}(R,T) \sim e^{-2 E_{0,\nu}} \quad , \quad E_{0,\nu} = 
\min_{\rho \geq 0} E_\nu \;,
\ee 
where 
\bea 
2 E_\nu &=& - \int_0^{+\infty} dp \int_0^{+\infty} dp' \log|p-p'| \hat \rho(p) \hat  \rho(p') \nonumber \\
&+& \int_0^{+\infty} dp  \, \rho(p)
\log\left ( 1 + e^{- \beta  (\frac{p}{2 R^2} - \tilde \mu) } \right )\;, \label{E_deltarho_besselmain}
\eea
with $\rho(p) = \rho_0(p)  + \hat \rho(p)$.
Here, defining $(x)_+=\max(0,x)$, 
$\rho_0(p) = \frac{\sqrt{(p-\nu^2)_+}}{2 \pi \,p}$
is the equilibrium density
in the absence of potential $V(p) \to 0$
(such that the second term in \eqref{E_deltarho_besselmain} vanishes). 
The energy minimization in \eqref{P0_Eminbessel0} is subject to the constraint
$\rho(p) \geq 0$, i.e. $\hat \rho(p) \geq - \rho_0(p)$.
Note that $\rho_0(p)$ has support on $[\nu^2,+\infty)$, 
but since $V(p)$ is repulsive, we find below that 
the optimal density $\rho^*(p)$ has instead support on $[a,+\infty)$ with $a \geq \nu^2$. 

We now study the minimization problem in the low temperature regime (ii)
where $z= k_F R \gg 1$, $\beta \tilde \mu \simeq \beta \mu \gg 1$ with $u=z/(\beta \mu)=O(1)$
fixed. We first consider a given sector $\ell = O(1)$ in Eq. (\ref{prod_FD}) which corresponds to the hole probability of the finite $T$ BDPP. 
One finds that there is a {dramatic} change of behavior in the support
of the optimal density, such that for 
$a =\nu^2 (\frac{\pi u}{\pi u - 2})^2 = O(1)$ for 
$u > \frac{2}{\pi}$, 
while for 
$u < \frac{2}{\pi}$, $a$ grows with $R$, i.e. $a \simeq \tilde a z^2$ 
and one finds 
\be \label{atilde}
\tilde a = \tilde a_c(u) := (1 - \frac{\pi^2 u^2}{4})_+ \;.
\ee 
In other words, for $u<2/\pi$, a {\it macroscopic} hole appears, i.e
a gap $k_{\rm gap}=k_F \sqrt{\tilde a_c(u)}$ 
in the original (physical) momenta variables $k_i = \sqrt{p_i}/R^2$ (see Fig. \ref{plot_rtilde} in End Matter).
The optimal density takes the form
$\rho^*(p) \sim \frac{1}{z} r\left( \frac{p}{z^2}\right)$
with
\be \label{rho_hat_explicitmain}
 r(\tilde p) = \frac{(1-\frac{2}{\pi u})_+}{2\pi \sqrt{\tilde p}}  + \frac{{\rm tanh}^{-1}\left( \min\left( \sqrt{\frac{\tilde p-\tilde a}{1-\tilde a}}, \sqrt{\frac{1-\tilde a}{\tilde p-\tilde a}}\right) \right)}{\pi^2 u} \;,
\ee 
where $\tilde a= \tilde a_c(u)$. It 
exhibits a log divergence as $\tilde p \to 1$,
while $r(\tilde p)=0$ for $\tilde p<\tilde a$.
Inserting $\rho^*(\tilde p)$ in \eqref{E_deltarho_besselmain}, after some calculations, one finds the 
low temperature scaling 
form for the hole probability of the BDPP for $\ell = O(1)$
\bea \label{finiteell}
\log P^{(\ell)}(R,T) \simeq - z^2 
\begin{cases}
    \frac{2}{ 3 \pi u} - \frac{1}{2 \pi^2 u^2} \quad , \quad u > \frac{2}{\pi} \\
    \frac{1}{4} - \frac{\pi^2 u^2}{96} \quad , \quad \quad \; u < \frac{2}{\pi} \;,
\end{cases}
\eea 
which is independent of $\ell$. This result is of
interest for the BDPP (which describes e.g. fermions in
a $1/r^2$ potential \cite{BLACT2018}) but is not sufficient for
our purpose. Note that it already exhibits a phase transition
at $u=u_c=2/\pi$, and turns out to be very similar to $d=1$ 
(it differs only by a factor of 2) -- 
see Eqs. \eqref{phi_dp} and \eqref{phi_dm} and below.

In fact, to compute the hole probability in dimension $d$, one 
must sum over the different $\ell$-sectors [see \eqref{prod_FD}], 
\be \label{sumsectors}
\log P(R,T) \simeq - 2 \sum_{\ell \geq 0} g_d(\ell) E_{0,\nu} \;.
\ee 
It turns out that for large $R$ this sum is dominated by large 
values of $\ell \sim \nu \sim R$. This was already the case
at $T=0$, which was studied by a different method in \cite{Gabriel_gap}.

Hence we must now study the regime $\ell \simeq \nu =\lambda z \gg 1$
with $0 \leq \lambda \leq 1$ fixed \cite{lambda}. 
The optimal density now depends both on $u$
and $\lambda$. Its support is such that $a \simeq \tilde a z^2$
where $\tilde a = \tilde a(u,\lambda)$ is the unique solution of 
\be 
\lambda = f_u(\tilde a) = \sqrt{\tilde a} - \frac{2 \sqrt{\tilde a(1-\tilde a)}}{\pi u} \label{lambdaamain} \;,
\ee 
for $\lambda \in [0,1]$. One finds that $\tilde a(u,\lambda)$ is an increasing function of $\lambda$ at fixed $u$,
with $\tilde a \in [\tilde a_c(u),1]$ where $\tilde a_c(u)$ was defined in 
\eqref{atilde}.
The optimal density $\rho^*(\tilde p)$ now takes the same scaling form as above with, for $\tilde p = p/z^2 > \tilde a$
\be \label{rho_hat_explicitmain2}
 r(\tilde p) = \frac{\lambda \sqrt{\tilde p - \tilde a}}{2\pi \tilde p \sqrt{\tilde a}}  
 + \frac{{\rm tanh}^{-1}\left( \min\left( \sqrt{\frac{\tilde p-\tilde a}{1-\tilde a}}, \sqrt{\frac{1-\tilde a}{\tilde p-\tilde a}}\right) \right)}{\pi^2 u} \;,
\ee 
and $r(\tilde p)=0$ for $\tilde p < \tilde a$, with $\tilde a=\tilde a(u,\lambda)$.
Inserting $\rho^*(\tilde p)$ in \eqref{E_deltarho_besselmain}
after some nontrivial calculations one finds the 
low temperature scaling 
form for the hole probability of the BDPP for $\ell = O(k_F R)$
\bea \label{def_phimain}
&& \log P^{(\ell\sim z \lambda)}(R,T) \simeq - z^2 \phi(\lambda, \tilde a(u,\lambda))  \\ 
&& \phi(\lambda, \tilde a) = \frac{1}{24} \Bigg[\frac{3 (3 \tilde a-1) \lambda^2}{\tilde a}-\frac{2 (5 \tilde a+1)
   \lambda}{\sqrt{\tilde a}}+\tilde a+5 \nonumber \\
&&   +6 \lambda \Bigg(\lambda \log \left(\frac{\tilde a}{\lambda^2}\right)+\frac{2
   \left(\lambda-\sqrt{\tilde a}\right) {\cos}^{-1}\left(\sqrt{\tilde a}\right)}{\sqrt{((1-\tilde a) \tilde a)}}\Bigg)\Bigg] \;,\nonumber 
\eea 
where we recall that $\tilde a=\tilde a(u,\lambda)$ is the root of Eq. \eqref{lambdaamain}. In that case $k_{\rm gap}=k_F \sqrt{\tilde a(u,\lambda)}$ 
for all $u$. In the limit $\lambda \to 0$, using $\lambda/\sqrt{\tilde a} \simeq
(1- \frac{2}{\pi u})_+$ from \eqref{lambdaamain}, one
recovers the finite $\ell$ result \eqref{finiteell} and 
$k_{\rm gap}=k_F \sqrt{\tilde a_c(u)}$ [see Eq. (\ref{atilde})]. Furthermore, in the zero temperature limit, i.e., $u\to 0$, one sees from \eqref{lambdaamain} that $\tilde a =\tilde a(0,\lambda) \to 1$ 
and one finds 
$\log P^{(\ell= \lambda z)}(R,T=0) \sim e^{- z^2 \phi(\lambda,1) }$
with 
\bea \label{GWW}
\phi(\lambda,1)  
   =   - \frac{\lambda^2}{2} \log \lambda + \frac{3}{4} \lambda^2 - \lambda + \frac{1}{4} \;,
\eea
for $\lambda \in [0,1]$, which agrees with Eq. (16) in \cite{Gabriel_gap}.
It was originally derived in the context of lattice QCD in \cite{GW1980,Wadia},
not as a hole probability, but as a partition function, by a completely
different CG technique. It was also proved later in the study of the longest increasing subsequence of random permutations \cite{Joh1998} (see also \cite{KC2010}).
It is quite remarkable that it is recovered here by a different method. 

We are now ready to perform the sum over the angular
sectors in \eqref{sumsectors}. With the scaling $\ell=\lambda z$
in the large $z=k_F R$ limit the sum becomes an integral.
Using the asymptotics $g_d(\ell) \simeq 2\,\ell^{d-2}/\Gamma(d-1)$ for $\ell \gg 1$, one finds
\be 
- \log P(R,T) \simeq  \frac{2}{\Gamma(d-1)} z^{d+1} \int_0^1 d \lambda \lambda^{d-2} 
\phi(\lambda, \tilde a(u,\lambda)) 
\ee 
One can obtain an alternative formula by performing
the change of variable from $\lambda$ to $\tilde a$, which
gives the scaling function defined in 
\eqref{lowT} for $d>1$ as
\be \label{eq:Pd_txt}
\Phi_d(u)=  \frac{2}{\Gamma(d-1)}  \int_{\tilde a_c(u)}^1 d \tilde a f_u'(\tilde a) [f_u(\tilde a)]^{d-2} 
\phi(f_u(\tilde a), \tilde a) 
\ee 
where $f_u$ is defined in \eqref{lambdaamain} and $\phi$ in \eqref{def_phimain}. This 
function exhibits a transition at $u=u_c=2/\pi$ {originating}
from the lower bound $\tilde a_c(u)$ defined in \eqref{atilde}.
Interestingly
the integral in \eqref{eq:Pd_txt} can be performed explicitly in 
integer dimension $d \geq 2$, and yields the results
for $d=2,3$ given in \eqref{phi_dp}-\eqref{phi_dm2}. 

In $d=1$ the same method based on the CG can be used.
The key formula \eqref{linstatbesselmain} also holds, with the substitution
$P^{(\ell)}(R,T) \to P(R,T)$ and $K_{\rm Be,\nu}$
replaced by the sine-kernel (where we recall that the size of the interval is $2 R$ in $d=1$). The same steps lead to a finite temperature version of the CG considered by Dyson \cite{Dyson_2015}.
Upon computing the optimal density and the corresponding energy, we
obtain $\Phi_1(u)$ displayed in Eqs. \eqref{phid}-\eqref{phi_dm}.  
Strikingly, the evaluation of the FD (\ref{FD_start}) in $d=1$ 
was also considered very recently in mathematics using quite different Riemann-Hilbert techniques \cite{Xu2025}. Our prediction for $\Phi_1(u)$, including the existence of a transition in $d=1$, agrees with the rigorous result of \cite{Xu2025}. 
It is remarkable that this result can also be obtained using a physics approach, which provides a nontrivial test for the CG method introduced in this paper. 
An additional outcome of \cite{Xu2025} is to show that, remarkably, the exact scaling function in \eqref{hole_TW_txt} is the GUE-Tracy-Widom distribution $G_1(y)=F_2(y)$ \cite{TW1994}. This is in line with the third order nature of the transition in $d=1$, which ubiquitously leads to the Tracy-Widom statistics \cite{SatyaReview}.
In general dimension $d$, the fact that the location of the transition
$u=u_c=2/\pi$ is independent of $\ell$ (see e.g. Eq. \eqref{finiteell}) leads us to conjecture that the form \eqref{hole_TW_txt} holds for any $d$, see \cite{SM} for details.

In this paper, we showed that the hole probability of free fermions displays two distinct regimes in the large radius limit $z= k_F R \gg 1$: (i) a high temperature regime for $T \sim T_F=k_F^2/2$ and (ii) a low temperature regime for $T \sim k_F/R$. Quite remarkably, we found that the second regime exhibits a transition
at the critical value $u=2 T R/k_F=2/\pi$.
One can ask whether that property extends to other counting statistics observables.
The simplest one is the variance $\Delta(R,T):={\rm Var} {\cal N}_R= \langle {\cal N}_R^2 \rangle^c$. Using the standard formula $\Delta(R,T):= {\rm Tr}[ \Pi_R K- (\Pi_R K)^2]$ we show 
\cite{SM} that it also takes a low temperature scaling form in any $d$, 
at large $z$ and fixed $u$, namely
$\Delta(R,T)-\Delta(R,T=0) \simeq z^{d-1} {\cal V}_d(u)$, where
$\Delta(R,T=0) \simeq \frac{z^{d-1}}{\pi^2 \Gamma(d)}(\log z + b_d)$ is the variance
at zero temperature and $b_d$ is given in \cite{Castin2007,SmithLMS2021}. However, the scaling function ${\cal V}_d(u)$
{\it does not show any transition}, {see Eq. (\ref{eq:varsd1}) End Matter}. It only exhibits 
a smooth crossover from ${\cal V}_d(u) \sim u^2$ for $u \ll 1$,
to ${\cal V}_d(u) \sim u$ for $u \gg 1$, where it matches 
the high $T$ regime, $T =O(T_F)$, for which all cumulants are 
easily obtained from Widom's formula, {see End Matter}. Presumably all cumulants (which depend only on finite order correlations) behave smoothly, while 
our preliminary work \cite{UsToBePublished} shows that the FCS observable, $P({\cal N}_R= n z^d,T)$ (which requires correlations to arbitrary order),
does exhibit a similar transition. It is likely that the
origin of these transitions can be traced {back} to a {\it limit shape
phenomenon} \cite{Abanov2006,Stephan2016,Gangardt2022} in the space-time word-line (polymer) picture, as was recently uncovered in a discrete tight-binding model in $d=1$ in Ref. \cite{AbaLimitShape25} at zero temperature. 
In that work a macroscopic hole in the fermion density creates an astroid-shaped hole in the space time optimal density. One can conjecture that a finite temperature introduces
an additional length scale $v_F \beta \hbar$, where $v_F$ is the Fermi velocity, which allows to control
the transition. It would be of great interest to understand this
connection in any space dimension.

Recent experiments~\cite{DixmeriasHoleProba,PaperTail} show that the regime $R\sim \xi_T$ should be within reach of the quantum gas microscope experiments, in combination with flat-bottom traps.
From a theoretical point of view we expect our results to extend to more general confining potentials, see \cite{Gabriel_gap}
for the $T=0$ case. Furthermore, although we can check that in $d=1$ the function ${\cal V}_1(u)$ agrees with predictions from conformal field theory \cite{CalabreseCardy2004},
the question of universality of the low temperature scaling 
functions ${\cal V}_d(u)$ and $\Phi_d(u)$ computed here, 
in higher $d$ remains to be explored. 
In particular, it would be interesting to extend the present results
for domains of various shapes,
or in the presence of interactions \cite{Kitanine}.




{\it Acknowledgments:} We thank A. Abanov and D. Gangart for stimulating discussions on related topics. We thank S. X. Xu for useful exchanges related to the present work. We thank T. Yefsah and his group for fruitful collaboration on cold atom experiments \cite{DixmeriasHoleProba}, and for helpful comments on the manuscript.
This research was supported by ANR Grant No. ANR-23-CE30-0020-01 EDIPS.

\vspace*{0.5cm}

\newpage
\begin{widetext}
\begin{center}
{\bf End Matter}
\end{center}

\subsection{Numerical determination of the scaling function $\Phi_d(u)$}

We show here in Fig. \ref{Fig_conv} the scaling functions $\Phi_d(u)$ in $d=1$ and $d=2$ determined
by numerical evaluation of the Fredholm determinant
(see \eqref{prod_FD} for $d=2$). 
The convergence as a function of $z$ is shown, as well as
the extrapolated value based on the presence 
of $O(1/z^2)$ corrections.
For the details of the numerical method see \cite{SM}. 

\begin{figure}[h]
\includegraphics[width = 0.45\linewidth]{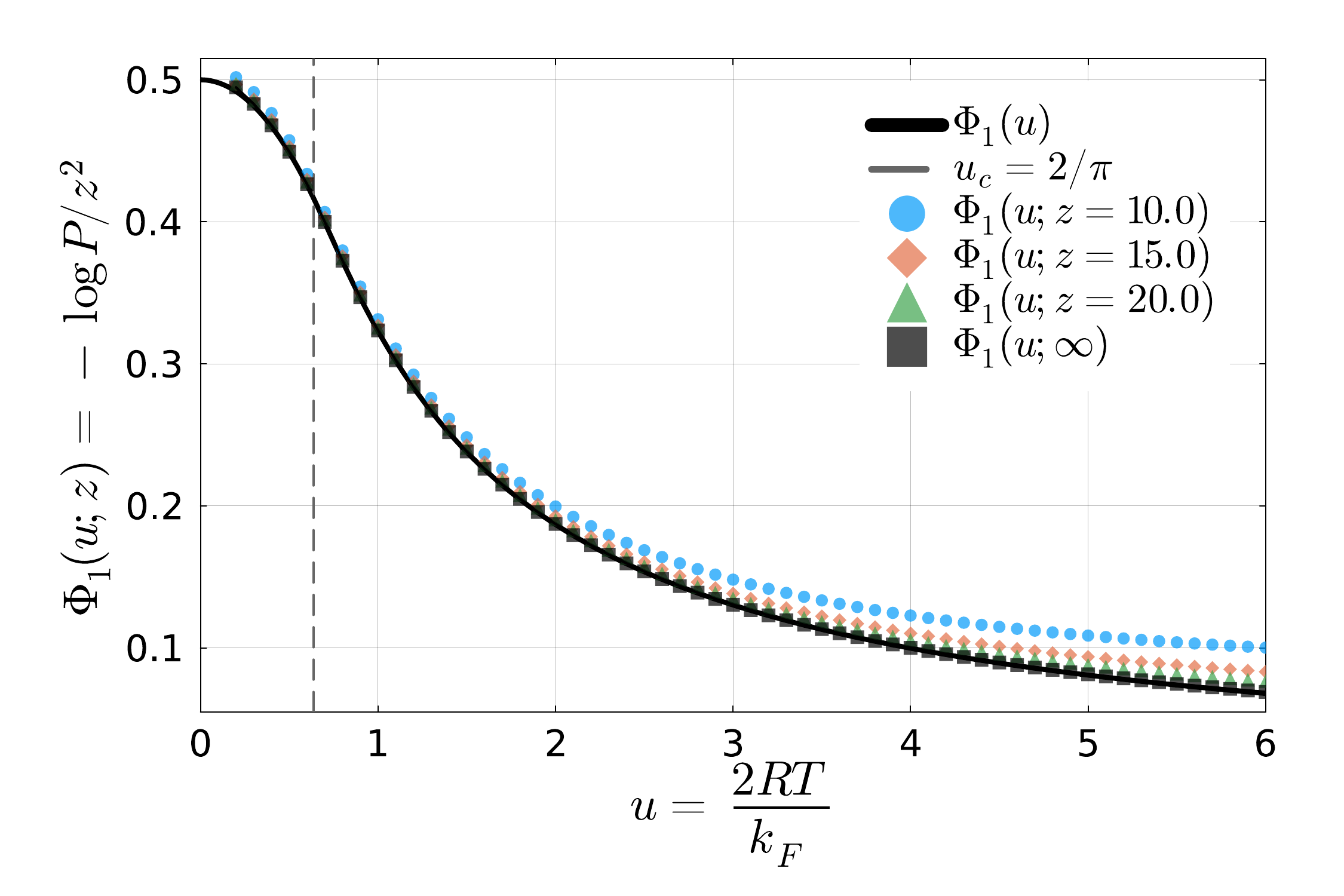}
\includegraphics[width = 0.45\linewidth]{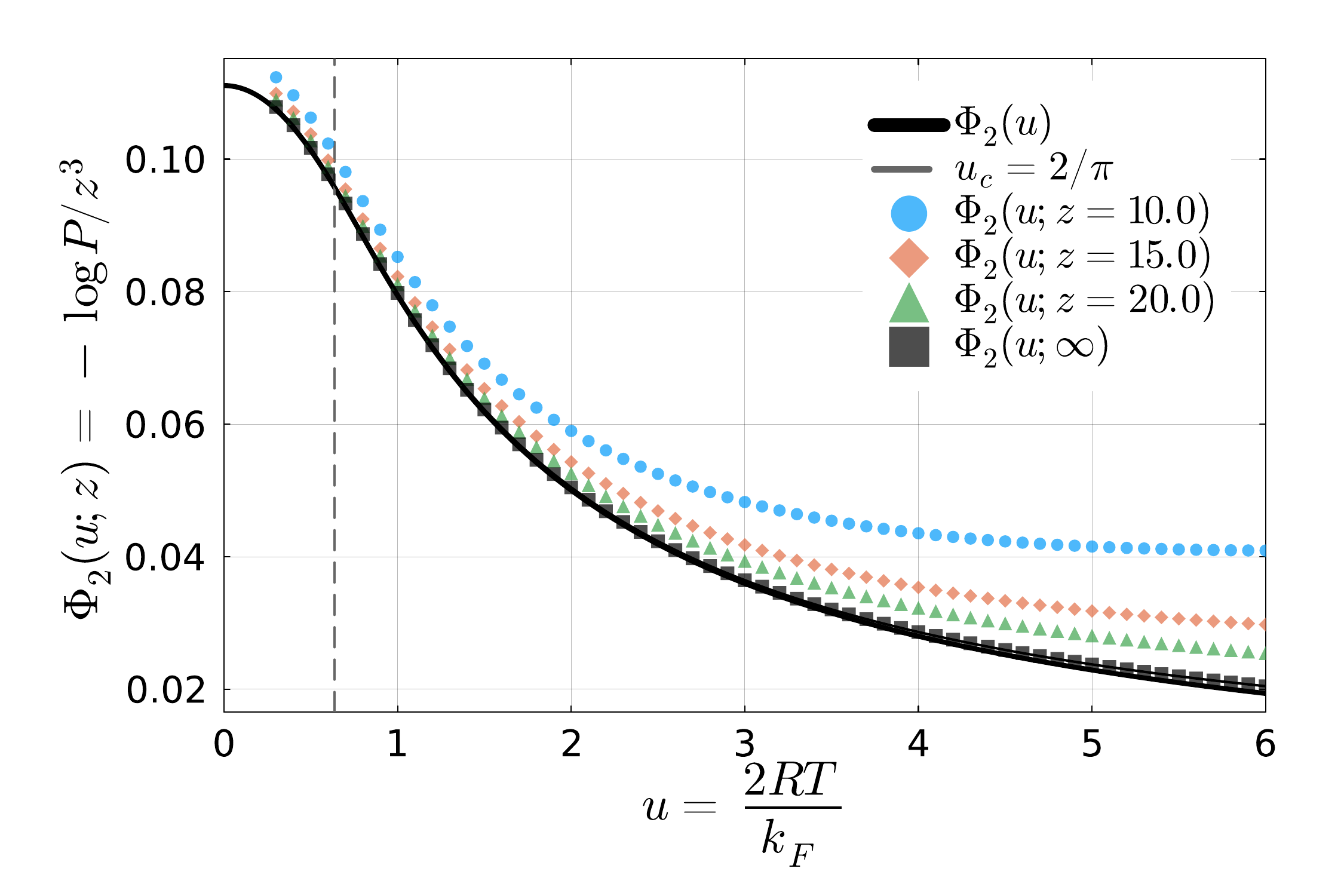}
\caption{Plots of $- \log P(R,T)/z^{d+1}$ in $d=1$ (left) and $d=2$ (right) as a function of $u$ for different values of $z=10,15,20$ from top to bottom. We recall that in $d=1$ one has $P(R,T) = {\rm Det}( I - \Pi_{[-R,R]} \hat K)$ where $\hat K$ is the finite $T$ sine-kernel (\ref{def_sineK}) while in $d\ge 2$ we use the angular decomposition Eq.~\eqref{prod_FD} in the main text. In both cases, $d=1,2$, the markers correspond to the numerical evaluation of the Fredholm determinant as explained in \cite{SM}. The solid black line corresponds to the analytic predictions in Eqs.~\eqref{phi_dp}-\eqref{phi_dm2} specialized to $d=1$ (left) and $d=2$ (right). In both plots the dashed vertical line indicates the critical point $u_c = 2/\pi$. The extrapolated value $\Phi_{1,2}(u;\infty)$ (squares) is obtained taking into account the finite $z$ corrections as explained in \cite{SM}.}\label{Fig_conv}
\end{figure}

\subsection{Optimal density profile of the associated Coulomb gas}

A plot of the scaled density profile $\tilde r(\tilde p)$ as given in Eq. (\ref{rho_hat_explicitmain}), is shown Fig. \ref{plot_rtilde}.
\begin{figure}[h]
\includegraphics[width=0.7\linewidth]{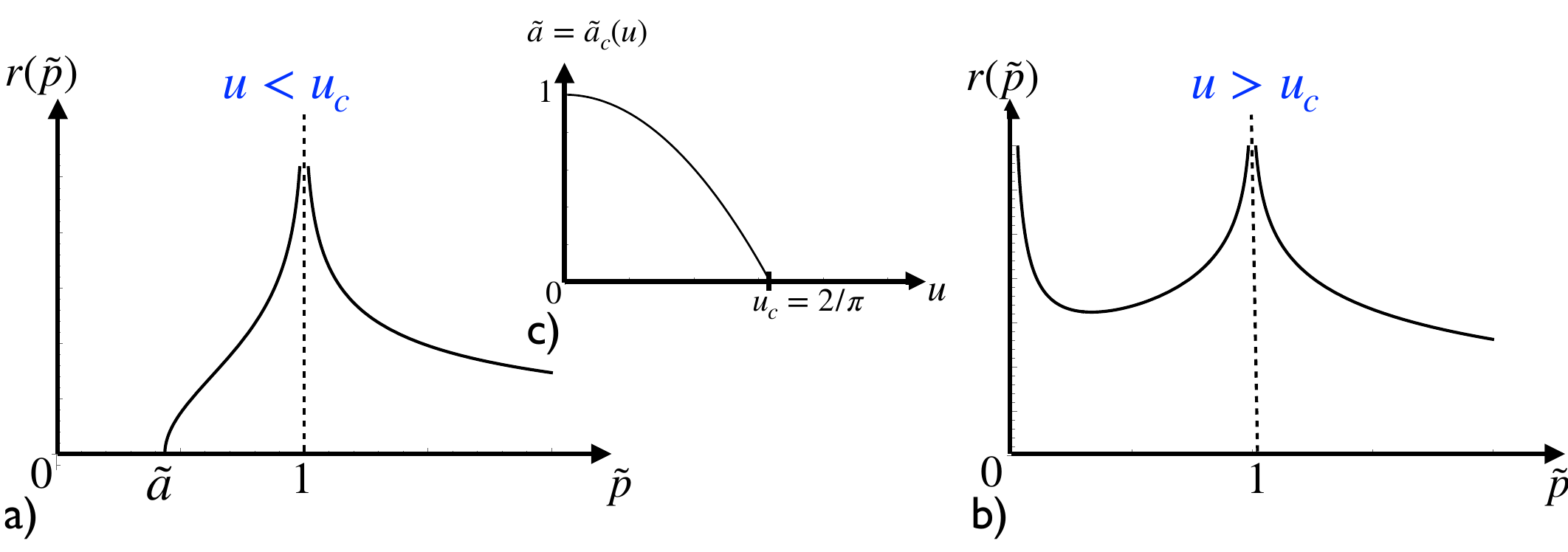}
\caption{Plot of the scaled density $\tilde r(\tilde p)$ of the underlying CG in the low temperature regime, as given in Eq. (\ref{rho_hat_explicitmain}), vs $\tilde p$ for $u<u_c$ (left panel a) and $u>u_c$ (right panel b). The middle panel c) shows a plot of the gap $\tilde a \equiv \tilde a_c(u)$ vs $u$, as given in Eq.~(\ref{atilde}). {Expressed in the original momentum variables, the gap is $k_{\rm gap}=\sqrt{\tilde a_c(u)} k_F$.} }\label{plot_rtilde}
\end{figure}

\subsection{High $T$: cumulants and entanglement entropy}

In the high $T$ regime (i), $T=O(T_F)$, the Widom formula allows to obtain 
the FCS generating function $\chi(s)=\log\langle e^{-s {\cal N}_R} \rangle$ \cite{SM},
and from it the cumulants, which read, to leading order at large $R$ 
\bea  
\langle {\cal N}_R^n \rangle^c 
\simeq  -  \frac{(R \sqrt{T})^d}{2^{d/2} \Gamma(1+d/2)} 
{\rm Li}_{\frac{d}{2}+1-n} 
(-e^{\beta \tilde \mu})
\eea  
This allows to obtain the Von Neumann bipartite entanglement entropy $S_1$ of a sphere of radius $R$. Using the standard relation \cite{Gamayun2020}
between $\chi(s)$ and $S_1$ {for non-interacting fermions (Gaussian state)}
we obtain \cite{SM}, to leading order in $R$ in the high $T$ regime $T=O(T_F)$
\be 
S_1 \simeq  \frac{(R \sqrt{T})^d}{2^{1 + \frac{d}{2}} \Gamma(1 + \frac{d}{2}) }
\left(2 \beta \tilde \mu
   \text{Li}_{\frac{d}{2}}\left(-e^{\beta \tilde \mu}\right)-(d+2)
   \text{Li}_{\frac{d}{2}+1}\left(-e^{\beta \tilde \mu}\right)\right) \;.
\ee 
One can check \cite{SM} that to leading order at large $R$, the bipartite entanglement entropy is given by $S_1 = s(T,\mu) V_d R^d + O(R^{d-1})$, where $s(T,\mu)$ is the thermodynamic entropy density \cite{entropy_spitzer}. 

\subsection{Low $T$: scaling form of the variance}

In the low $T$ regime (ii), with $z \gg 1$ and fixed $u$, an explicit
calculation \cite{SM} gives the scaling function ${\cal V}_d(u)$ for the variance in $d=1,2,3$ as  
\begin{equation}\label{eq:varsd1}
    {\cal V}_1(u) = \frac{1}{\pi^2} \log \left(\frac{\sinh(\pi u)}{\pi u} \right) \;, \; {\cal V}_2(u)= \frac{2}{\pi} \int_0^{+\infty} dk J_1(k)^2 \int_0^{\frac{\pi}{2}} 
d\theta \frac{\cos \theta }{e^{2 k \cos \theta/u}-1}  \;, \; {\cal V}_3(u)= \frac{1}{\pi^2 u^2}
\int_0^u ds\, s\,
\ln\!\left(\frac{\sinh(\pi s)}{\pi s}\right) \;.
\end{equation}
%
These predictions are tested numerically 
in Fig. \ref{fig:varsd}. 
The result in $d=1$ is consistent with the conformal field theory prediction of Calabrese and Cardy \cite{CalabreseCardy2004} for $S_1$, setting the central charge $c=1$ for free fermions, and using the relation $S_1 \simeq \frac{\pi^2}{3}
\langle {\cal N}_R^{2} \rangle^c$.

\begin{figure}
    \centering
\includegraphics[width=0.5\linewidth]{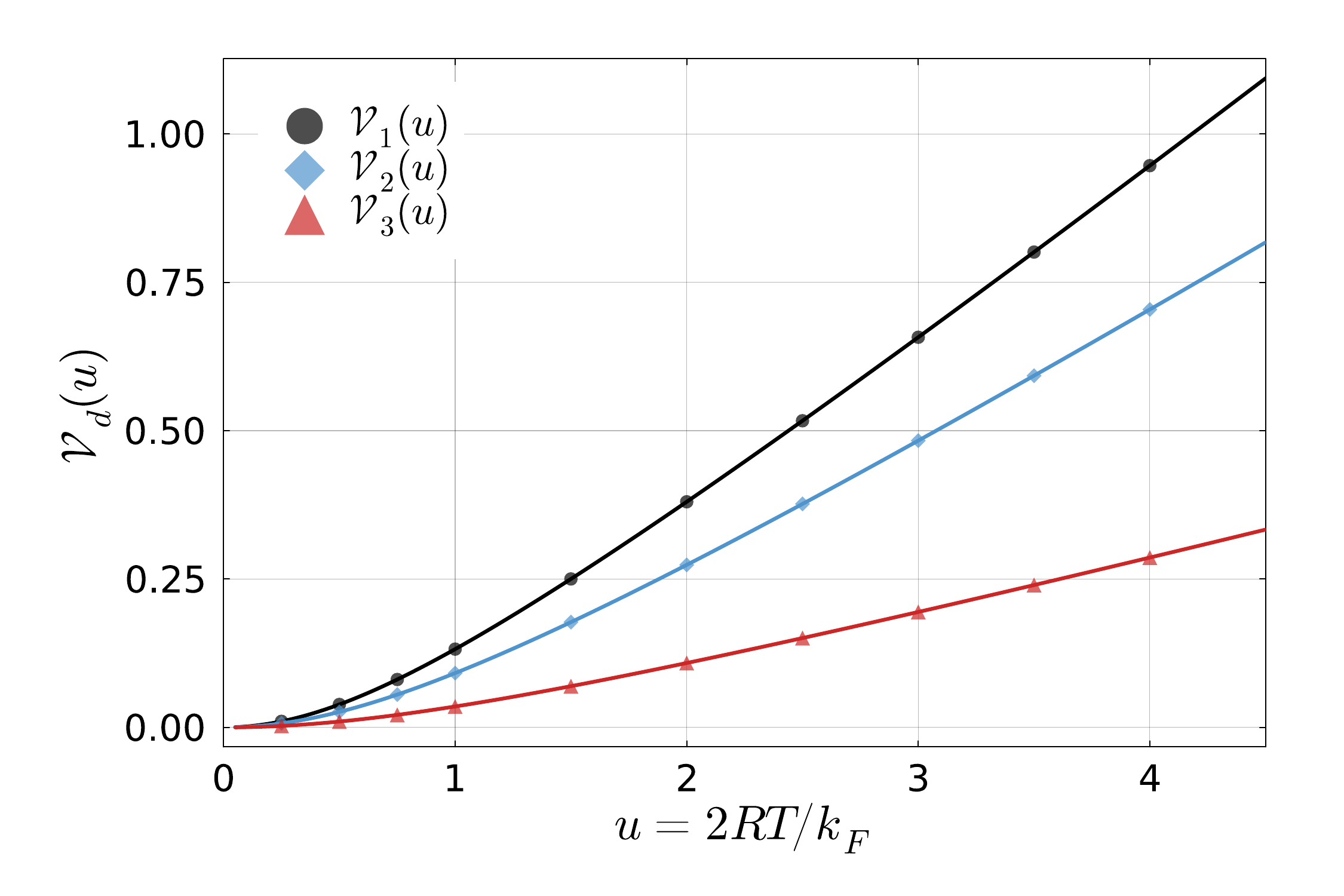}
    \caption{Scaling functions ${\cal V}_d(u)$ of the variance of the number of particles obtained via numerical quadrature of Eq.~\eqref{var1} and subtracting the zero temperature variance $\langle {\cal N}_R^2 \rangle^c|_{T=0}$ in Eq.~\eqref{varT0} (markers) compared to our analytic prediction (full lines) given in Eq.~\eqref{eq:varsd1}. See see \cite{SM} for details.}
    \label{fig:varsd}
\end{figure}

\newpage 

\setcounter{secnumdepth}{2}

\begin{large}
\begin{center}

Supplementary Material for\\  {\it Hole probability for noninteracting fermions in a $d$-dimensional trap}

\end{center}
\end{large}

\bigskip

We give the principal details of the calculations described in the main text of the Letter. We also present some related numerical results. 

\bigskip

\tableofcontents

\section{Relation between mean density and chemical potential}

In the grand canonical ensemble 
the mean density $\rho$ is a function of $T$ and $\tilde \mu$ given by
\be
 \rho = \int \frac{d^d k}{(2 \pi)^d} \frac{1}{1 + e^{\beta (\frac{k^2}{2} - \tilde \mu)}}
= \frac{-1}{(2 \pi \beta)^{d/2} }  {\rm Li}_{d/2}(- e^{\beta \tilde \mu}) 
\underset{d=2}{=} \frac{T}{2 \pi} \log(1+ e^{\beta \tilde \mu}) \quad , \quad d=2 
\ee
which simplifies in $d=2$. In this paper we use as variables the mean density $\rho$ and
the temperature $T$. For all $(\rho,T)$, we define both the Fermi temperature $T_F$ and the
Fermi momentum $k_F$ as
\be 
T_F : = \frac{k_F^2}{2} : = \frac{1}{2} (\rho/c_d)^{2/d} \quad , \quad c_d =  \frac{1}{(4 \pi)^{d/2} \Gamma(1+ d/2)} \;,
\ee 
so that $\rho= c_d k_F^d$. 
Then the chemical potential $\tilde \mu$ is determined by the relation
\be \label{muversusrho}
1 = \left(\frac{T}{T_F}\right)^{d/2} \Upsilon_d(e^{\beta \tilde \mu}) \quad , \quad 
\Upsilon_d(z)=- \Gamma(1+d/2) {\rm Li}_{d/2}(-z) 
\quad , \quad \Upsilon_2(z)= \log(1 + z) \;,
\ee 
where ${\rm Li}_s(z)$ is a polylogarithm function of index $s$, i.e., ${\rm Li}_s(z) = \sum_{k \geq 1}z^k/k^s$. We will use the following asymptotics for $z \to +\infty$
\bea \label{Liasympt}
{\rm Li}_s(- z) \simeq  \sum_{k=0}^\infty (-1)^k (1 - 2^{1-2 k}) (2 \pi)^{2 k} \frac{B_{2k}}{(2 k)!} \frac{ [\log(z)]^{s-2 k} }{\Gamma(s+1-2 k)} + O(1/z) 
\eea 
where $B_{2k}$'s are the Bernoulli numbers. Note that for integer $s$
the series terminates and the correcting terms are exponential in $\log z$, i.e.
as $1/z$. For $s+1$ equal to minus a positive integer the expansion is regular in $1/z$. One then obtains 
the following asymptotics for the function $\Upsilon_d(z)$
\be \label{Upsilonasympt}
\Upsilon_d(z) \simeq \begin{cases} (\log z)^{d/2} \left( 1 + \frac{\pi^2}{6 (\log z)^2} \frac{d}{2} (\frac{d}{2}-1) 
+ O((\log z)^{-4}) \right) \quad , \quad z \to + \infty  \\
\Gamma(1+d/2) \left(z-\frac{z^2}{2^{d/2}} + O(z^3)\right) \quad , \quad \hspace*{2.63cm}z \to 0 \;.\end{cases} 
\ee 
In the limit $T/T_F \to 0$, using $\Upsilon_d(z) \simeq (\log z)^{d/2}$ as $z \to + \infty$ to leading order, one checks 
that 
\be 
\tilde \mu \to \mu = T_F = \frac{k_F^2}{2} \;,
\ee 
where, by definition, $\mu = \tilde \mu|_{T=0}$ is the
Fermi energy. 

In the opposite
limit, at fixed $\rho$ and for $T \to +\infty$ one obtains (for simplicity
in $d=2$) 
\be 
\tilde \mu \sim T \log(2 \pi \rho/T) \quad , \quad d=2 
\ee 
hence $\beta \tilde \mu \to - \infty$. This can be understood
from the fact that at fixed $\tilde \mu$, the density
diverges as $T \to \infty$ as $\rho \simeq T/(2 \pi)$.
Hence to keep the mean density constant one needs to consider
$\tilde \mu<0$.

\section{High temperature regime using Widom's formula}\label{sec:widom}

In this section we compute the hole probability for large $R$ in the high temperature regime
$T = O(T_F)$ using Widom's formula. In the notation of \cite{Widom_d_dim} one has
\bea
F({\bf y}) = \frac{1}{e^{\beta(\frac{{\bf y}^2}{2}-\tilde \mu)} + 1} \quad, \quad \rho({\bf x})  = \int_{E_n} \frac{d^n{\bf k}}{(2\pi)^n} e^{-i {\bf k}. {\bf x}} F({\bf k})  \;,
\eea
where $E_n$ is $\mathbb{R}^d$ with $n=d$. Given a 
a convex domain ${\sf K}$ of $\mathbb{R}^d$ and denoting 
$a {\sf K}$ the domain scaled by the positive parameter $a=R$
the main theorem of \cite{Widom_d_dim} gives an expansion
of the Fredholm determinant ${\rm Det}(I - \lambda \Pi_{a} \rho({\bf x}- {\bf y} ) )$
where $\Pi_{a}$ denotes the projector on the domain $a {\sf K}$. In our main
application ${\sf K}$ is the unit ball, but the formula is valid for a more general convex domain
or polyhedron. This formula allows to compute the FCS generating function, in our notations
\bea \label{chisFD}
\chi(s) = \log \langle e^{- s {\cal N}_R } \rangle = \log {\rm Det}(1 - (1-e^{-s})  \Pi_R K ) \;,
\eea 
where ${\cal N}_R$ is the number of fermions in the domain $R {\sf K}$. 
It gives the asymptotics
\bea 
&& \chi(s) \simeq  R^d V({\sf K}) 
 \int_{{\mathbb R}^d} \frac{d^d {\bf k}}{(2 \pi)^d}
\log\left(1 - \frac{1- e^{-s}}{e^{{ \beta} 
(\frac{{\bf k}^2}{2}-\tilde \mu)}+1} \right)  \label{eq:Widom1}\\
&& + \frac{1}{2} R^{d-1} 
\int_{\partial K} d^{d-1} \sigma 
\int_{{\bf x} \in \mathbb{R}^d, {\bf x}  \cdot {\bf n}(\sigma)>0}   \, ({\bf x}  \cdot {\bf n}(\sigma)) \,\left( \int_{{\mathbb R}^d}
\frac{d^d {\bf k}}{(2 \pi)^d} 
\log\left(1 -  \frac{1- e^{-s}}{e^{{ \beta} 
(\frac{{\bf k}^2}{2}-\tilde \mu)}+1} \right) \, e^{i {\bf k}\cdot {\bf x}}
\right)^2 \, d^d{\bf x} \;,\label{eq:Widom2}
\eea 
where $V({\sf K})$ is the volume of the domain ${\sf K}$
and ${\bf n}(\sigma)$ is the unit outer normal at the point
$\sigma$ on the boundary $\partial {\sf K}$ of the domain ${\sf K}$.

We can now make explicit the temperature dependence by rescaling 
${\bf k} \to {\bf k}/\sqrt{\beta}$ and ${\bf x} \to {\bf x} \sqrt{\beta}$.
We also specialize to $s=+\infty$ to compute the hole probability 
$\chi(+\infty)=P(R,T)$. This leads to 
\bea \label{expansionlogP1}
\log P(R,T) \simeq - (R \sqrt{T})^d A_d\left(\frac{T}{T_F}\right) + (R \sqrt{T})^{d-1} B_d\left(\frac{T}{T_F}\right) \;,
\eea 
at large $R$, where we recall the definition $T_F= \frac{1}{2} (\rho/c_d)^{2/d}$ where 
$c_d=1/((4 \pi)^{d/2} \Gamma(1+ d/2))$ and that 
$\tilde \mu$ is determined by \eqref{muversusrho}, i.e.
one has 
\be \label{mut} 
e^{\beta \tilde \mu}  = \Upsilon_d^{-1}(1/t^{d/2}) \underset{d=2}{=} e^{1/t} - 1 \;.
\ee 
It is useful to give the asymptotics at small and large $t$ respectively
\be \label{Ups} 
\Upsilon_d^{-1}(1/t^{d/2}) = 
\begin{cases}  e^{\frac{1}{t} - \frac{\pi^2}{6} (\frac{d}{2}-1) t + O(t^3)} \\
\frac{t^{-d/2}}{\Gamma(1+ \frac{d}{2})}  \left( 1+ \frac{t^{-d/2}}{2^{d/2} \Gamma(1+ \frac{d}{2}) }  + O(t^{-d} \right) \;.
\end{cases} 
\ee 
The two functions $A_d(t)$ and $B_d(t)$ are obtained as discussed below. 

\subsection{The function $A_d(t)$}
One has
\be \label{leading_Widom}
A_d(t) = V({\sf K}) 
 \int_{{\mathbb R}^d} \frac{d^d {\bf k}}{(2 \pi)^d}
\log\left(1 + e^{ \beta \tilde \mu-\frac{{\bf k}^2}{2}} \right) 
= - \frac{ V({\sf K}) }{(2 \pi)^{d/2}} {\rm Li}_{\frac{d}{2}+1} (-e^{\beta \tilde \mu}) \;,
\ee 
where $e^{\beta \tilde \mu}$ is expressed in function of $t$ from 
\eqref{muversusrho}. In the case where ${\sf K}$ is the unit sphere, 
$V({\sf K}) = V_d=S_d/d= \pi^{d/2}/\Gamma(1+d/2)$, in which case
\be 
A_d(t) = - \frac{1}{2^{d/2} \Gamma(1+d/2)} 
{\rm Li}_{\frac{d}{2}+1} (- \Upsilon_d^{-1}(1/t^{d/2}))  
\quad , \quad 
\Upsilon_d(z)=- \Gamma(1+d/2) {\rm Li}_{d/2}(-z)
\ee \;.
It simplifies in $d=2$
\be 
A_2(t)= - \frac{1}{2} {\rm Li}_2(1-e^{1/t}) \simeq \begin{cases} 
 \frac{1}{4 t^2} + \frac{\pi^2}{12} +
+ O(e^{-1/t}) \quad , \quad t \to 0 \\
\frac{1}{2 t} + \frac{1}{8 t^2} + O(t^{-3} ) \quad , \quad \quad \;\;\; t \to +\infty \;,
\end{cases} 
\ee 
where we used that $-{\rm Li}_2(-e^x)=\frac{\pi^2}{6}+ \frac{x^2}{2} + O(e^{-x})$
at large $x$.

In the high temperature limit $T/T_F \gg 1$ one has
\be 
A_d(t) = \frac{1}{2^{d/2} \Gamma(1+d/2)^2} t^{-d/2} 
\left( 1 + \frac{ 2^{-d/2}}{d \Gamma
   \left(\frac{d}{2}\right)}  t^{-d/2} 
   +\frac{\left(2^{-d}-2 \times 3^{-\frac{d}{2}-1}\right)}{\Gamma
   \left(\frac{d}{2}+1\right)^2} t^{-d} + O(t^{-3 d/2}) \right)
\quad , \quad t \to +\infty
\ee 
which matches the Poisson distribution 
\be 
P^{\rm Poi}(R)= e^{- \rho V_d R^d} \;.
\ee 
More precisely, one has for $T/T_F \gg 1$ 
\be 
\log P(R,T) = - \rho V_d R^d \left(  1 + \frac{ 2^{-d/2}}{d \Gamma
   \left(\frac{d}{2}\right)} \left(\frac{T_F}{T}\right)^{d/2} 
    +\frac{\left(2^{-d}-2 \times 3^{-\frac{d}{2}-1}\right)}{\Gamma
   \left(\frac{d}{2}+1\right)^2} \left(\frac{T_F}{T}\right)^{d}  + O(\left(\frac{T_F}{T}\right)^{3 d/2})
   \right) {+ O(R^{d-1}) } \;.
\ee 
{Equivalently, in terms of the density, we obtain the finite temperature, $O(R^d)$ i.e. volume corrections to the Poisson formula as}
\be \label{Poissoncorrection1}
\log P(R,T) = - \rho V_d R^d
\left(  1 + \frac{1}{2} \pi^{d/2} \rho  T^{-d/2}  + \frac{1}{3} \left(3-2^{d+1}
   3^{-d/2}\right) \pi ^d \rho^2 T^{-d} + O(\rho^3 T^{-3 d/2}) \right) 
    + O(R^{d-1}) \;,
\ee 
{which is the leading order in $R$. The subleading order $O(R^{d-1})$, i.e. the surface correction, is given below in \eqref{Poissoncorrection2}.}

In the low temperature limit $t = T/T_F \ll 1$ we find 
\be   \label{AlowT}
A_d(t) = \frac{1}{2^{d/2} \Gamma(1+d/2) \Gamma(2+d/2) } t^{-1-d/2}  \left( 1 + \frac{\pi^2}{12} (d+2) t^2 + O(t^4) \right)  \;.
\ee 
{This implies, to leading order in $R$}
\bea 
\log P(R,T) \simeq - R^d T_F^{d/2} \frac{T_F}{T} 
\frac{1}{2^{d/2} \Gamma(1+d/2) \Gamma(2+d/2) } 
\left( 1 + \frac{\pi^2}{12} (d+2) (\frac{T}{T_F})^2 + O((\frac{T}{T_F})^4) \right)  \;.
\eea  
Note that the function $A_d(t)$ can be obtained purely from the thermodynamics of the free Fermi gas, see the remark around Eq. \eqref{thermowidom}.

\subsection{The function $B_d(t)$}
Next one has 
\bea 
B_d(t) = 
\frac{1}{2} 
\int_{\partial {\sf K}} d^{d-1} \sigma 
\int_{{\bf x} \in \mathbb{R}^d, {\bf x}  \cdot {\bf n}(\sigma)>0}  d^d{\bf x}
({\bf x}  \cdot {\bf n}(\sigma))
\left( \int_{{\mathbb R}^d}
\frac{d^d {\bf k}}{(2 \pi)^d} 
\log\left(1 + e^{\beta {\tilde \mu}- \frac{{\bf k}^2}{2} }  \right) e^{i {\bf k}\, \cdot {\bf x}}
\right)^2 \;.
\eea 
We now specify to ${\sf K}$ being the unit ball.
Then, noting that $\frac{S_d}{\int_0^{\pi} d\theta (\sin \theta)^{d-2} }=S_{d-1}$
where we recall $S_d=2 \pi^{d/2}/\Gamma(d/2)$, we have for $d>1$, denoting $x^2=r^2$
\bea 
&& B_d(t) = 
\frac{S_d S_{d-1} }{2} 
\int_0^{\pi/2} d\theta (\sin \theta)^{d-2} \cos \theta 
\int_0^{+\infty}  dr r^{d}  
\left( \int_{{\mathbb R}^d}
\frac{d^d {\bf k}}{(2 \pi)^d} 
\log\left(1 + e^{\beta {\tilde \mu}- \frac{{\bf k}^2}{2} }  \right) e^{i {\bf k}\, \cdot {\bf x}}
\right)^2 \;.
\eea 
Let us rewrite as a series
\bea \label{id_log}
\int_{{\mathbb R}^d}
\frac{d^d {\bf k}}{(2 \pi)^d} 
\log\left(1 + e^{\beta {\tilde \mu}- \frac{{\bf k}^2}{2} }  \right) e^{i {\bf k}\, \cdot {\bf x}}
= \sum_{m=1}^\infty \frac{(-1)^m e^{m \beta {\tilde \mu}}}{m}  
\int_{{\mathbb R}^d}
\frac{d^d {\bf k}}{(2 \pi)^d} e^{- m \frac{{\bf k}^2}{2}
+ i {\bf k}\, \cdot {\bf x}} = \frac{1}{(2 \pi)^{d/2}} 
\sum_{m=1}^\infty \frac{(-1)^m e^{m \beta {\tilde \mu}}}{m^{1+\frac{d}{2}}} 
e^{- \frac{r^2}{2 m}}
\eea 
This leads to 
\bea \label{expl_Bd}
&& B_d(t) =  \frac{S_d S_{d-1} }{2 (d-1)} \frac{1}{(2 \pi)^{d}} \sum_{m_1,m_2 \geq 1 }
\frac{(-1)^{m_1+m_2} e^{(m_1+m_2) \beta {\tilde \mu}}}{m_1^{1+\frac{d}{2}}
m_2^{1+\frac{d}{2}}}
\int_0^{+\infty}  dr r^{d}   
e^{- \frac{r^2}{2} (\frac{1}{m_1} + \frac{1}{m_2})} \nonumber \\
&& = \frac{S_d S_{d-1} }{2 (d-1)} \frac{1}{(2 \pi)^{d}} 
2^{\frac{d-1}{2}} \Gamma(\frac{d+1}{2}) 
\sum_{m_1,m_2 \geq 1 }
\frac{(-1)^{m_1+m_2} e^{(m_1+m_2) \beta {\tilde \mu}}}{m_1^{\frac{1}{2}}
m_2^{\frac{1}{2}}} \frac{1}{(m_1+m_2)^{\frac{d+1}{2}} } \nonumber \\
&& = \frac{S_d S_{d-1} }{2 (d-1)} \frac{1}{(2 \pi)^{d}} 
2^{\frac{d-1}{2}} \int_0^{+\infty} dv v^{\frac{d-1}{2}}
\sum_{m_1,m_2 \geq 1 }
\frac{(-1)^{m_1+m_2} e^{(m_1+m_2) (\beta {\tilde \mu}- v) }}{m_1^{\frac{1}{2}}
m_2^{\frac{1}{2}}} \nonumber \\
&& = \frac{2^\frac{d-3}{2}}{\pi \Gamma(d)} 
\int_0^{+\infty} dv v^{\frac{d-1}{2}} 
[ {\rm Li}_{1/2}(- e^{\beta {\tilde \mu}- v } ) ]^2 \quad , \quad 
e^{\beta \tilde \mu}  = \Upsilon_d^{-1}(1/t^{d/2}) \underset{d=2}{=} e^{1/t} - 1 
\eea 
{This is the formula for $B_d(t)$ which is given in the main text and }which is valid in any dimension $d$
(the restriction $d>1$ was technical). 

At high temperature $t \gg 1$ one finds 
\be 
B_d(t) \simeq \frac{\Gamma(\frac{1+d}{2})}{4 \pi \Gamma(d)} e^{2 \beta \tilde \mu}
\simeq \frac{2^{-d-1}}{\sqrt{\pi }
   \Gamma
   \left(\frac{d}{2}+1\right)^
   2 \Gamma
   \left(\frac{d}{2}\right)} t^{-d} \;.
\ee 
{This gives the $O(R^{d-1})$ correction (i.e. the area correction) to the Poisson formula, which must be added to 
\eqref{Poissoncorrection1} above}
\be \label{Poissoncorrection2}
{\log P(R,T)|_{O(R^{d-1})} = } \frac{\pi ^{d-\frac{1}{2}}}{2
   \Gamma
   \left(\frac{d}{2}\right)} \rho^2 \frac{R^{d-1}}{T^{\frac{d+1}{2}}} \;.
\ee 

In the low temperature limit $t \to 0$, i.e. $e^{\beta \tilde \mu} \to +\infty$,
one has $\tilde \mu \to \mu$ and 
using ${\rm Li}_{1/2}(-e^u) \sim (-2/\sqrt{\pi}) \sqrt{u}$.
Defining $v= \beta  \mu\tilde v$ 
\be \label{BlowT}
B_d(t) \simeq_{t \to 0} \frac{4}{\pi} \frac{2^\frac{d-3}{2}}{\pi \Gamma(d)} 
(\beta \mu)^{\frac{d+3}{2}}
\int_0^{1} d\tilde v \tilde v^{\frac{d-1}{2}} (1- \tilde v)
\simeq { 2} \frac{ (2\beta \mu)^{\frac{d+3}{2}} }{\pi
   ^2 \left(d^2+4 d+3\right)
   \Gamma (d)} \simeq { 2} \frac{ (2/t)^{\frac{d+3}{2}} }{\pi
   ^2 \left(d^2+4 d+3\right)
   \Gamma (d)} \;.
\ee 
\\

\noindent{\bf Remark}. The above calculations are easily modified to obtain $\chi(s)$ from Eqs. (\ref{eq:Widom1})-(\ref{eq:Widom2}). 
One obtains 
\bea 
&& \chi(s) =  - (R \sqrt{T})^d \bigg( - \frac{1}{2^{d/2} \Gamma(1+d/2)} 
( {\rm Li}_{\frac{d}{2}+1} (- e^{\beta \tilde \mu}) 
- {\rm Li}_{\frac{d}{2}+1} (- e^{-s + \beta \tilde \mu}) ) \bigg)  \\
&& + (R \sqrt{T})^{d-1}  
\frac{2^\frac{d-3}{2}}{\pi \Gamma(d)} 
\int_0^{+\infty} dv \, v^{\frac{d-1}{2}} 
[ {\rm Li}_{1/2}(- e^{\beta {\tilde \mu}- v }  )
- {\rm Li}_{1/2}(- e^{-s + \beta {\tilde \mu}- v }  )]^2
\eea 
We see from (\ref{eq:Widom1})-(\ref{eq:Widom2}) that for finite $s$ the logarithm splits into the difference of two contributions, and has the form of a function of $e^{\beta \tilde \mu}$ minus the same function of $e^{-s + \beta \tilde \mu}$. That immediately leads to the above form for the leading term. For the subleading term (second line) the calculation proceeds as follows. 
In \eqref{id_log} we replace $e^{m \beta \tilde \mu} \to e^{m \beta \tilde \mu} - e^{m( \beta \tilde \mu-s)} = \sum_{\sigma=0,1} (-1)^\sigma e^{m(\beta \tilde \mu -  \sigma s)}$.
In the double sum in the first three lines of Eq. \eqref{expl_Bd} we can thus replace
$e^{\beta \tilde \mu(m_1+m_2)}$ by $\sum_{\sigma_1,\sigma_2=0,1} (-1)^{\sigma_1+\sigma_2}  e^{\beta \tilde \mu(m_1+m_2)- s (\sigma_1 m_1 + \sigma_2 m_2)}$. This leads to the above result.


\subsection{Matching from the high temperature regime (i) to the low temperature regime (ii)}

In summary, in the low temperature limit of the high temperature regime, $t=T/T_F \ll 1 $,
with $T_F=k_F^2/2=\mu$, putting together \eqref{AlowT} and \eqref{BlowT}, the expansion \eqref{expansionlogP1} becomes (
denoting $z=k_F R$) 
\bea \label{logPlow}
\log P(R,T) \simeq &-& \frac{z^d}{t} \frac{1}{2^d \Gamma(1+ \frac{d}{2}) \Gamma(2+ \frac{d}{2}) } \left( 1 + \frac{\pi^2 (d+2)}{12} t^2 + O(t^4) \right)
\\
&+&
\frac{z^{d-1}}{t^2}  \frac{8}{\pi^2 \left(d^2+4 d+3\right) \Gamma (d) } 
(1 + o(t)) + o(z^{d-1}) \;.
\eea 
This matches the result of the low temperature calculation, which is performed
in Section \ref{sec:lowT_d}. Indeed, reexpressed in 
the variable $u=2 R T/k_F = z T/T_F = z t$, \eqref{logPlow} reads 
\bea \label{eq:Widom_largeu}
&& \log P(R,T) \simeq - z^{d+1} 
\bigg(   \frac{1}{u} \frac{1}{2^d \Gamma(1+ \frac{d}{2}) \Gamma(2+ \frac{d}{2}) } 
-  \frac{8}{\pi^2(d^2+4d+3)} \frac{1}{\Gamma(d)} \frac{1}{u^2} \bigg) 
+ O(z^{d-1}) 
\eea 
One can check that it reproduces the first two terms in the $1/u$ expansion
at large $u$,
of the low temperature scaling function, denoted $\Phi_{d,+}(u)$, see Eq. 
\eqref{eq:phip_largeu}. 





\subsection{The special case $d=1$}

{The case $d=1$ should a priori be treated separately.}
In $d=1$, the Widom formula for the hole probability in
the interval $[-R,R]$ becomes
\bea 
&& \log P(R,T) \simeq  - 2 R  
 \int_{{\mathbb R}} \frac{dk}{2 \pi}
\log\left(1 + e^{  \beta (\tilde \mu - \frac{k^2}{2})} \right)  +   
\int_{0}^{+\infty} dx\, x  \,\left( \int_{{\mathbb R}}
\frac{d k}{2 \pi} 
\log(1 + e^{ { \beta} 
(\tilde \mu - \frac{k^2}{2})} ) \, e^{i k x}
\right)^2 \\
&& = - R \sqrt{T} A_1(\frac{T}{T_F}) + B_1(\frac{T}{T_F}) 
\quad , \quad T_F= \frac{\pi^2}{2} \rho^2
\eea 
where $t=T/T_F$ is related to $\beta \tilde \mu$ via
$- {\rm Li}_{1/2}(-e^{\beta \tilde \mu})=2/(t^{1/2} \sqrt{\pi})$. One has
\bea 
A_1(t)= \frac{1}{\pi} \int_{{\mathbb R}} dp
\log\left(1 + e^{  \beta
\tilde \mu - \frac{p^2}{2}} \right) = - \sqrt{\frac{2}{\pi}} {\rm Li}_{3/2}(-e^{\beta
\tilde \mu}) 
\eea 
which coincides with specializing the Widom formula in general $d$ in Eq. (\ref{leading_Widom}),
to $d=1$ (setting $V({\sf K}) = 2$ in this case). One also has
\bea 
B_1(t)&=& \int_{0}^{+\infty} dx\, x  \,\left( \int_{{\mathbb R}}
\frac{d p}{2 \pi} 
\log(1 + e^{ \beta\tilde \mu - \frac{p^2}{2})} ) \, e^{i p x}
\right)^2 \\
&=& \frac{1}{2 \pi} \sum_{m_1,m_2 \geq 1} (-1)^{m_1+m_2} \frac{e^{(m_1+m_2) \beta \tilde \mu}}{\sqrt{m_1\,m_2}(m_1+m_2)} \;,
\eea 
where we have used the series expansion (\ref{id_log}) with $d=1$ to perform the integral over $p$ and $x$. Writing $1/(m_1+m_2) = \int_0^\infty e^{-(m_1+m_2)v}\,dv$ and performing explicitly the sums over $m_1$ and $m_2$ one finds
\bea \label{B1}
B_1(t) = \frac{1}{2 \pi} \int_0^\infty dv \left[ {\rm Li}_{1/2}(-e^{\beta \tilde \mu - v})\right]^2 \;,
\eea
which coincides with the expression in (\ref{expl_Bd}) specialized to $d=1$.

\section{From Fredholm determinants to Coulomb gases }

\subsection{The case $d=1$ (sine-kernel)}

In $d=1$ free fermions on the real axis are described by the finite temperature 
version of the sine kernel 
\be \label{def_sineK}
\hat K(x,x')= \int_{-\infty}^{\infty} \frac{dk}{2 \pi} \sigma(k) e^{i k (x-x')}
\quad , \quad 
\sigma(k)= \frac{1}{e^{\beta (\frac{k^2}{2} - \tilde \mu) }+1 } \;,
\ee 
where $\sigma(k)$ is the Fermi factor. The probability
that there are no fermions in $[-R,R]$ (hole probability) 
is given by the Fredholm
determinant
\be \label{holeFD} 
P(R,T) = {\rm Det}( I - \Pi_{[-R,R]} \hat K) \;,
\ee 
where the notation $\Pi_{[-R,R]}$ denotes the projector on the interval $[-R,R]$.

In order to relate this quantity to an observable in a Coulomb gas, it is crucial to use 
the following "duality" relation
\be \label{Duality_SineK}
P(R,T)  = {\rm Det}( I - \Pi_{[-R,R]} \hat K)  
= {\rm Det} (I -  \hat \sigma K_{\rm sine})  
\ee 
where $K_{\rm sine}$ is the standard sine kernel (in momentum space)
and $\hat \sigma$ is given by
\be 
K_{\rm sine}(k,k') = \frac{\sin\left(\pi \left(k-k'\right)  \right)}{\pi(k-k')}
\quad , \quad \hat \sigma(k)= \sigma(\pi k/R) 
= \frac{1}{e^{\beta (\frac{\pi^2 k^2}{2 R^2} - \tilde \mu) }+1 } \;.
\ee
To show this duality one uses the identity 
\be \label{dual2}
\int_{-R}^R \frac{dx}{2 \pi} e^{i (k-k') x} = 
\frac{\sin\left(\left(k-k'\right) R \right)}{\pi(k-k')} \;,
\ee 
and perform the change of variable $k \to \pi k/R$. This
allows to show that, upon integrating first over space variables,
all the traces coincide, i.e.
${\rm Tr} (\Pi_{[0,R]} \hat K)^n = {\rm Tr} (K_{\rm sine} \hat \sigma)^n$ 
where $\hat \sigma(k)=\sigma(\pi k/R)$.

An important consequence of this duality relation is that 
one can now express the hole probability at finite temperature as a linear statistics problem
for free fermions in 1d {\it at zero temperature}. Indeed
using standard properties of Fredholm determinants
and of determinantal point processes one obtains 
\bea  
 P(R,T) = {\rm Det} (I -  \hat \sigma K_{\rm sine})  &=& 
\Bigg \langle \exp\left( \sum_i \log(1 - \hat \sigma(k_i)) \right)  \Bigg \rangle_{
K_{\rm sine}} \\
&=&  \Bigg  \langle \exp\left(  \sum_i \log( 1 + e^{- \beta (\frac{k_i^2}{2 R^2} - \tilde \mu) })  
\right)  \Bigg \rangle_{K_{\rm sine}} \label{linstat} 
\eea 
where the expectation value is over the determinantal sine point process, 
$\{ k_i \}_{i \in \mathbb{N}}$, with kernel $K_{\rm sine}$.
Within this duality the variables $k_i$ play the role
of the positions of fictitious free fermions at zero temperature. It is useful to recall that the mean density of a point process is
given by the diagonal element of its kernel. For the
sine kernel as defined above, this mean density $\rho_0(k)$ is thus equal to unity,
i.e one has 
\be 
\rho_0(k) = \langle \sum_i \delta(k_i-k) \rangle_{K_{\rm sine}}
= K_{\rm sine}(k,k)=1
\ee  

To relate to a Coulomb gas we follow the seminal paper by Dyson~\cite{Dyson_2015}. 
The idea is that the sine kernel describes the universal small scale
correlations of the eigenvalues of random Hermitian matrix models. For
instance, the joint distribution of the eigenvalues $\lambda_i$
of a random matrix of size $N$ belonging to the Gaussian unitary ensemble (GUE)
is given by (up to a normalization) 
\be 
P[\{\lambda_i\}] \sim \prod_{i<j}(\lambda_i-\lambda_j)^2 e^{-  c_N \sum_i \lambda_i^2}
\sim e^{- 2 {\cal E}[\lambda]} \;.
\ee 
The quantity $2 {\cal E}[\lambda]$ can be interpreted as the energy
of a Coulomb gas $2 {\cal E}[\lambda]= 2 \sum_{i<j} \log|\lambda_i-\lambda_j|- c_N \sum_i \lambda_i^2$. To obtain the sine process, we can 
consider the limit $N \to +\infty$ and focus on the region 
near the center of the semi-circle, 
$\lambda=O(1)$. The choice $c_N=\pi^2/(2 N)$ 
leads, in that region, to a sine process with 
unit density, since the edges are far away at $\pm 2N/\pi$.
We can identify $\lambda_i=k_i$ in that region. 
The uniform density $\rho_0(k)=1$ corresponds to the minimal
energy configuration of the Coulomb gas. 
Note that here we define the density without a factor $1/N$.

When studying a linear statistics, i.e an expectation value as \eqref{linstat} over the sine process, 
the energy of the associated Coulomb gas contains an addition potential
term, $V(k)= \log( 1 + e^{- \beta (\frac{\pi^2 k^2}{2 R^2} - \mu) } )$,
which couples to the total density $\rho(k)$.
This total density is different from $\rho_0(k)$ and we define
the difference
\be \label{hatrhodef}
\hat \rho(k) =\rho(k) - \rho_0(k) = \rho(k) - 1 \;.
\ee
The expectation value \eqref{linstat}
can thus be obtained by considering the 
following Coulomb gas 
\be \label{E_deltarho}
2 E = - \int dk dk' \log|k-k'| \hat \rho(k) \hat  \rho(k') 
+ \int dk  (\rho_0(k) + \hat \rho(k) )  \log( 1 + e^{- \beta (\frac{\pi^2 k^2}{2 R^2} - \mu) } ) 
\ee 
where, here, $\rho_0(k) = 1$. The first term (and the total energy) vanishes 
in the absence of a potential, i.e. for $V(k)=0$, as required
by normalization of the expectation value in \eqref{linstat}. By minimizing the energy $2E$ in (\ref{E_deltarho}) with respect to $\hat \rho$ one obtains an approximation of the finite temperature hole probability $P(R,T)$ as
\be  \label{P0_Emin}
P(R,T) \sim e^{-2 E_0} \quad , \quad E_0 = \min_{\hat \rho \geq -1} E \;.
\ee 
We expect this approximation to become exact in the limit of large $R$.
Indeed, at $T=0$, this was explicitly checked by Dyson in \cite{Dyson_2015}.
The validity of this approach at finite temperature is discussed below,
and we will see how to compute explicitly the minimum energy $E_0$. Note that the minimisation in Eq. (\ref{P0_Emin}) must be performed under the constrained $\hat \rho(k) \geq -\rho_0(k)=-1$ -- which comes from the positivity condition $\rho(k) \geq 0$.

\subsection{The case $d>1$ (Bessel kernel)}

To study the case of spatial dimensions $d>1$, 
we start with the formula for the hole probability in each angular sector $\ell$ 
as discussed in the text
\be  \label{Pl_start}
P^{(\ell)}(R,T) = {\rm Det}( I - \Pi_{[0,R]} \hat K_\nu)  
\quad , \quad \nu = \ell + \frac{d}{2} - 1
\ee 
where we recall the expression for the finite temperature Bessel kernel
for $r,r' \in \mathbb{R}^+$
\be  \label{hat_BesselKernel}
\hat K_\nu(r,r') dr' = \sqrt{rr'} dr' \int_0^{+\infty} dk k J_\nu(k r) J_\nu(k r') \sigma(k) \quad , \quad 
\ee 
where we recall that $J_\nu(x)$ is the Bessel function of index $\nu$ and 
$\sigma(k)$ is the Fermi factor defined in Eq.~(\ref{def_sineK}). In Eq. (\ref{Pl_start}) the notation $\Pi_{[0,R]}$ denotes the projector on the interval $[0,R]$. 

Here again, in order to relate to a Coulomb gas we use  
the following "duality" relation
\be \label{Duality_Bessel_FD}
P^{(\ell)}(R,T)  = {\rm Det}( I - \Pi_{[0,R]} \hat K_\nu)  
= {\rm Det} (I -  \bar \sigma K_{\rm Be, \nu})  
\ee 
where $K_{\rm Be, \nu}$ is the standard Bessel kernel 
for $p,p' \in \mathbb{R}^+$
\cite{misprint}
\bea \label{bessel}
 K_{\rm Be,\nu}(p,p') = \frac{1}{4} \int_0^1 dy J_{\nu}(\sqrt{p y}) J_{\nu}(\sqrt{p' y}) &=& \frac{J_{\nu}(\sqrt{p}) \sqrt{p'} J'_{\nu}(\sqrt{p'})
- \sqrt{p} J'_{\nu}(\sqrt{p}) J_{\nu}(\sqrt{p'}) }{2 (p - p') } \\
&=& \frac{J_{\nu}(\sqrt{p}) \sqrt{p'} J_{\nu-1}(\sqrt{p'})
- \sqrt{p} J_{\nu-1}(\sqrt{p}) J_{\nu}(\sqrt{p'}) }{2 (p - p') }
\eea
and where $\bar \sigma(p)$ is defined as
\be 
\bar \sigma(p)= \sigma(\sqrt{p}/R) 
= \frac{1}{e^{\beta (\frac{p}{2 R^2} - \tilde \mu) }+1 } \;.
\ee
To show this duality we use the change of variable
$r = R \sqrt{y}$, $k= \sqrt{p}/R$, $k'= \sqrt{p'}/R$
and the identity
\be \label{dualityBessel}
\int_0^R r dr J_\nu(k r) J_\nu(k' r) = 2 R^2 K_{\rm Be,\nu}(p,p') \;,
\ee 
which allows to show that all the traces coincide, i.e.
${\rm Tr} (\Pi_{[0,R]} \hat K_\nu)^n = {\rm Tr} (K_{\rm Be,\nu} \bar \sigma)^n$ 
where $\bar \sigma(p)=\sigma(\sqrt{p}/R)$. 

As in the case $d=1$, this duality relation allows to express the hole probability $P^{(\ell)}(R,T)$
as a linear statistics problem for the so-called
Bessel determinantal point process (BDPP), which is the 
determinantal point processes on the positive half-line
of kernel $K_{\rm Be,\nu}$. One has 
\be \label{linstatbessel} 
P^{(\ell)}(R,T) =  \Bigg \langle \exp\left( \sum_i \log(1 - \bar \sigma(p_i)) \right)  \Bigg \rangle_{K_{\rm Be,\nu}}
= \Bigg \langle \exp\left(  \sum_i \log( 1 + e^{- \beta (\frac{p_i}{2 R^2} - \tilde \mu) })  
\right)  \Bigg \rangle_{K_{\rm Be,\nu}}
\ee 
where the expectation value is over the BDPP, i.e. the $\{ p_i \}_{i \in \mathbb{N}}$'s, 
with kernel $K_{\rm Be,\nu}$. 

The BDPP arises as the hard edge limit of
complex Wishart-Laguerre (WL) random matrices of size $N \times N$ \cite{Forrester}. 
That ensemble is defined by the
following joint probability distribution function (PDF) for a set of $N$ eigenvalues $\lambda_i$
\be  \label{WL} 
P_{\rm WL}(\vec \lambda) \propto e^{- \sum_{i=1}^N \lambda_i} \prod_{i=1}^N \lambda_i^{\nu} 
\prod_{1 \leq j,k \leq N} (\lambda_j -\lambda_k)^2
= e^{-2 {\cal E}[\lambda]} \;,
\ee
which depends on the continuous parameter $\nu >-1$. 
Again, the quantity $2 {\cal E}[\lambda]$ can be interpreted as the energy
of a Coulomb gas, with $2 {\cal E}[\lambda]= 2 \sum_{i<j} \log|\lambda_i-\lambda_j|- \nu  \sum_i \log \lambda_i - \sum_i \lambda_i$. 
Let us recall that the mean density $\rho_{\rm MP}(\lambda)$
of the eigenvalues $\lambda_i$ is 
given by the Marcenko-Pastur (MP) law
\be 
\rho_{\rm MP}(\lambda) = \frac{1}{N} \frac{ \sqrt{(\lambda - \lambda_-)(\lambda_+-\lambda)}}{2 \pi \lambda} \quad , \quad 
\lambda_\pm = N (\frac{1}{\sqrt{c}} \pm 1)^2 \quad , \quad c = \frac{1}{1 + \frac{\nu}{N}} 
\ee 
which is valid in the bulk, i.e. for $\lambda=O(N)$
and $\nu = O(N)$, and with $\lambda \in [\lambda_-, \lambda_+]$. 

The hard edge scaling is obtained by focusing on 
the eigenvalues near the origin, i.e. $\lambda_i=O(1/N)$, and defining
\be 
\lambda_i = \frac{p_i}{4 N} \;. 
\ee 
Then one can show that the statistics of the $p_i$'s is precisely described 
by the kernel $K_{{\rm Be},\nu}(p,p')$. Here we will focus on the regime of large $R$
in \eqref{linstatbessel}. This regime is controlled
by the behavior of the BPP for large $p_i$
and large $\nu$, with $p_i = O(\nu^2)$. 
In that limit one can show that the exact density
of the BPP takes the form
\be \label{rho0}
\rho_{{\rm Be}, \nu}(p) = \frac{1}{4} \int_0^1 dy J_{\nu}(\sqrt{p y})^2 \quad
\simeq \quad \rho_0(p) = 
\begin{cases}
& 0 \quad, \quad \quad \quad \quad \;\;  \, p < \nu^2 \;,\\
&\\
&{\dfrac{\sqrt{p-\nu^2}}{2 \pi \,p}}  \quad, \quad p \geq \nu^2 \;,
\end{cases}
\ee 
which is the form used below. This form matches the 
the Marcenko-Pastur (MP) density of the WL ensemble
\be 
 \rho_{\rm MP}\left(\frac{p}{4 N}\right) d\left(\frac{p}{4 N}\right)  \simeq  \rho_0(p) dp \;,
\ee 
in the scaling limit $N \to +\infty$, with $p, \nu^2 \gg 1$, 
and $p/\nu^2 = O(1)$.

Again, as for $d=1$, when computing the linear statistics
\eqref{linstatbessel} we need to consider a Coulomb gas
with an additional potential term of total energy 
\be \label{E_deltarho_bessel}
2 E_\nu = - \int_0^{+\infty} dp \int_0^{+\infty} dp' \log|p-p'| \hat \rho(p) \hat  \rho(p') 
+ \int_0^{+\infty} dp  (\rho_0(p)  + \hat \rho(p))
\log( 1 + e^{- \beta  (\frac{p}{2 R^2} - \tilde \mu) } ) 
\ee
with $\hat \rho(p)=\rho(p)-\rho_0(p)$ where $\rho_0(p)$ is given in Eq. (\ref{rho0}). Note that $E_\nu$ depends on $\nu$ only through $\rho_0(p)$. As required, the first term (and the total energy) vanishes 
in the absence of potential. Once again, the
hole probability $P^{(\ell)}(R,T)$ in \eqref{linstatbessel} can be 
approximated as
\be  \label{P0_Eminbessel}
P^{(\ell)}(R,T) \sim e^{-2 E_{0,\nu}} \quad , \quad E_{0,\nu} = \min_{\hat \rho \geq -\rho_0} E_\nu \;.
\ee 
where the energy minimization is under the constraint
that $\hat \rho(p) \geq - \rho_0(p)$ -- which is imposed by the positivity constraint $\rho(p) = \hat \rho(p) + \rho_0(p) \geq 0$. 
Note that $\rho_0(p)$ has support on $[\nu^2,+\infty)$, but, as we will see below, the
optimal density $\rho^*(p)$ has instead support on $[a,+\infty)$ with $a \geq \nu^2$.

\section{Coulomb gas calculations: the case $d=1$}


Let us consider consider the Coulomb gas defined in 
\eqref{E_deltarho}. Let us denote $\rho^*(k)$
and $\hat \rho^*(k)=\rho^*(k) - \rho_0(k)$
its minimizer under the constraint $\rho^*(k) \geq 0$.
The potential $V(k)$ is represented schematically in 
Fig.  We see that it repels the particles away from the origin.
The optimal density $\rho^*(k)$ is thus reduced near the origin, and there
are two cases: (i) high temperature where $\rho^*(k)$
is smooth and strictly positive everywhere (ii) a low temperature regime
where $\rho^*(k)$ vanishes in an interval $[-k_0,k_0]$ around the origin.

More precisely, minimizing $E$ in Eq. (\ref{E_deltarho}) one finds, for $k$ belonging to the support of $\rho^*(k) = 1+ \hat \rho^*(k)$
\bea \label{saddle1}
\frac{\delta E}{\delta \hat \rho(k)} = 0 \Longleftrightarrow 2 \int dk' \log |k-k'| \hat \rho^*(k') = \log( 1 + e^{- \beta (\frac{\pi^2 k^2}{2 R^2} - \tilde \mu) } ) \quad, \quad \hat \rho^*(k) > -1 \;. 
\eea
Taking one more derivative w.r.t. $k$, one finds the necessary condition (again for $k$ in the support of $\rho^*(k)$)
\bea \label{saddle2}
\dashint_{-\infty}^\infty \frac{\hat \rho^*(k')}{k-k'} dk' = - \beta \frac{\pi^2 k}{2 R^2}\frac{1}{1 + e^{\beta (\frac{\pi^2 k^2}{2 R^2} - \tilde \mu)}} \quad, \quad \hat \rho^*(k) \geq -1 \;,
\eea
where the l.h.s is the Hilbert transform of the density.
Below we analyze these equations.

\subsection{High temperature} 

We start with the high temperature regime. In that case the  
 the support of $\hat \rho^*$ is the whole real line
 and the integral equation (\ref{saddle2}) can be solved 
 by inverting the Hilbert transform (which is anti-involutive)
\bea \label{sol_int}
\hat \rho^*(k) = { -} \frac{\beta}{2 R^2} \dashint_{-\infty}^\infty dq \frac{q}{1+e^{\beta (\frac{\pi^2 q^2}{2 R^2} - \tilde \mu)}} \frac{1}{q-k} \quad, \quad \hat \rho^*(k) \geq -1 \;.
\eea
To compute the optimal energy one can first simplify \eqref{E_deltarho}
using the minimization equation \eqref{saddle1}, which 
leads to 
\bea 
\label{E_deltarho_3}
&& 2 E_0 = 2 E_{0,1} + 2 E_{0,2} 
 \quad , \quad  \nonumber \\
&& 2 E_{0,1} = \int_{-\infty}^\infty dk  \log( 1 + e^{- \beta (\frac{\pi^2 k^2}{2 R^2} - \tilde \mu) } ) \quad, \quad 2 E_{0,2} = \frac{1}{2} \int_{-\infty}^\infty dk \hat \rho^*(k) \log( 1 + e^{- \beta (\frac{\pi^2 k^2}{2 R^2} - \tilde \mu) } ) \;.
\eea 
The first term, after the change of variable $k = R p/(\pi \sqrt{\beta})$, recovers precisely the
leading term for large $R$ of Widom's formula, given above in \eqref{leading_Widom} 
(specialized to $d=1$, with $V({\sf K})=2$)
\be  \label{E01}
2 E_{0,1} =
 \sqrt{T} \frac{R}{\pi} \int_{-\infty}^{+\infty} dp 
\log( 1 + e^{\beta \tilde \mu - \frac{p^2}{2} } ) \;.
\ee 
The second term, inserting $\hat \rho^*(k)$
from \eqref{sol_int}, and performing the change of variables 
$q = \sqrt{2} R \tau/( \pi \sqrt{\beta})$,
$k = \sqrt{2} R \tau'/( \pi \sqrt{\beta})$,
is evaluated as
\bea 
2 E_{0,2} =
\frac{-1}{2 \pi^2} \int_{-\infty}^{+\infty} d\tau   \dashint_{-\infty}^\infty d\tau' \frac{\tau}{1+e^{\tau^2-\beta \tilde \mu}} 
\frac{1}{\tau-\tau'}  \log( 1 + e^{- {\tau'}^2 + \beta \tilde \mu}) \;. \label{def_I}
\eea
Taking derivatives with respect to $\beta \tilde \mu$
and using integration by parts in the integral over $\tau'$ and $\tau$, this can be simplified
into 
\bea \label{deriv4}
2 E_{0,2} = -\frac{1}{2 \pi^2} \int_{-\infty}^{\beta \tilde \mu} ds' \left(\int_{-\infty}^{+\infty} d\tau \frac{1}{1+e^{\tau^2-s'}}\right)^2 
= -  \frac{1}{2 \pi} \int_{-\infty}^{\beta \tilde \mu} ds' \left[ {\rm Li}_{1/2}(-e^{s'})\right]^2 \;.
\eea
Remarkably this term coincides with $- B_1(t)$ given in \eqref{B1}
obtained from the Widom's formula. This shows that this Coulomb gas approach
predicts not only the leading term at large $R$, but also the
next to leading. 

This behavior holds until the density $\hat \rho^*(k)$ in Eq. (\ref{sol_int}) is such $\hat \rho^*(k) \geq -1$. Assuming that the minimum value of $\hat \rho^*(k)$ is reached at $k=0$, this solution holds provided $\hat \rho^*(k=0) \geq -1$, i.e.,
\bea \label{no_gap_cond}
\frac{\beta}{R^2} \int_0^\infty \frac{dq}{1+e^{\beta(\frac{\pi^2 q^2}{2R^2}-\tilde \mu)}} \leq 1 \;.
\eea
This inequality can be re-written 
in terms of the variable $u$
\be 
u = \frac{2 R}{\beta k_F}  \geq \frac{2}{\pi k_F} \int_0^\infty \frac{dp}{1+e^{\beta(\frac{p^2}{2}- \tilde \mu)}} \;.
\ee 
In the low temperature scaling limit $R, \beta \to \infty$
with $u$ fixed, $\tilde \mu \to \mu$, this condition becomes
\be 
u \geq u_c = \frac{2}{\pi} \;,
\ee 
which is the critical value mentionned in the text, studied in more details below.


\subsection{Low temperature} 

We now study the low temperature scaling regime 
\be \label{scalingregime}
R , \beta \to +\infty \quad , \quad u = \frac{2 R}{\beta \sqrt{2 \mu}} ~~ \text{fixed} 
\ee 
and we recall that $\tilde \mu \simeq \mu$ in that regime.
In this regime we also denote
\be \label{scalingz} 
z = k_F R = R \sqrt{2 \mu} = \frac{u}{t} 
\ee 
and we will study the limit $z \gg 1$ at fixed $u$.
\\

\noindent {\bf Phase $u>u_c$ (solution without a gap)}. Let us start with the regime $u> u_c=2/\pi$. 
We simply need to take the low temperature scaling limits
of \eqref{E01} and \eqref{deriv4}. 
We substitute 
\be \label{substitute} 
R = \frac{z}{\sqrt{2 \mu}} \quad , \quad \beta = \frac{2 R}{u \sqrt{2 \mu}}
= \frac{z}{u \mu} 
\ee 
one obtains 
\be 
2 E_{0,1} = \int_{-\infty}^\infty dk  \log( 1 + e^{- \frac{z}{u} (\frac{\pi^2 k^2}{z^2} - 1) } ) = \frac{z}{\pi} \int_{-\infty}^\infty dp  \log( 1 + e^{- \frac{z}{u} (p^2 - 1) } ) 
\ee 
In the limit $z \gg 1$ one finds
\be 
2 E_{0,1} \simeq \frac{z^2}{\pi u} \int_{-\infty}^\infty dp (1-p^2)_+ 
= \frac{4 z^2}{3 \pi u} \;.
\ee 

Next from \eqref{deriv4} one has 
\bea \label{deriv4_bis}
2 E_{0,2} =  -  \frac{1}{2 \pi} \int_{-\infty}^{\frac{z}{u}} ds' \left[ {\rm Li}_{1/2}(-e^{s'})\right]^2 
\simeq -  \frac{z^2}{\pi^2 u^2} 
\eea
where we have used the asymptotic behavior ${\rm Li}_{1/2}(-e^{s}) \simeq 2 s^{1/2}/\sqrt{\pi}$ 
for $s \to +\infty$ -- see Eq. (\ref{Liasympt}).

In summary, putting these contributions together, we find that the Coulomb gas predicts in $d=1$
\bea  
&& \log P(R,T) \simeq - (k_F R)^{2} \Phi_{1,+}(u) \quad , \quad u>u_c=2/\pi \\
&& \Phi_{1,+}(u) =  \frac{4}{3 \pi u}  - \frac{1}{\pi^2 u^2} \label{phip_1d}
\eea  
as given in the text. 
\\

\noindent {\bf Phase $u<u_c$ (solution with a gap)}. Let us consider now the regime $u<u_c=2/\pi$.
Let us go back to the minimization equations \eqref{saddle1} 
and \eqref{saddle2} at finite temperature, without at this stage taking the
scaling limit \eqref{scalingregime}. In this case the density $\rho^*(k)$ develops a hole near the origin,
and one has 
\bea
\hat \rho^*(k) = 
\begin{cases}
&- 1 \quad, \quad |k| \leq k_0 \\
& \hat \rho_{0}(k) \quad, \quad |k| \geq k_0 \;,
\end{cases}
\eea
where $\hat \rho_{0}(k)$ still needs to be determined.
The equation for $\hat \rho_0(k)$ can be read off from Eq.  (\ref{saddle2}), namely
\bea
&&\dashint_{|k'|\geq k_0} \frac{\hat \rho_0(k')}{k-k'} dk' = g(k) \quad, \quad k \geq k_0 \label{Trico1} \\
&& g(k) = - \beta \frac{\pi^2 k}{2 R^2}\frac{1}{1 + e^{\beta (\frac{\pi^2 k^2}{2 R^2} - \tilde \mu)}} + \int_{-k_0}^{k_0} \frac{dk'}{k-k'} \quad, \quad k \geq k_0 \;.\label{def_gk}
\eea
Note that, by taking the limit $k \to \infty$ on both sides of Eq. (\ref{Trico1}), one gets
\bea
\dashint_{|k'|\geq k_0} \hat \rho_0(k')dk' = 2 k_0 \Longrightarrow \dashint \hat \rho_0(k')dk'  =  0 \;,
\eea
since we recall that $\hat \rho_0(k') = - 1$ for $|k' |< k_0$. 

Let us first write the general solution of Eq. (\ref{Trico1}) using $\hat \rho_0(k) = \hat \rho_0(-k)$, by symmetry. For this purpose, we introduce 
$\hat \rho_0(k) = \bar \rho_0(K=k^2)$,
such that the integral equation (\ref{Trico1}) can be rewritten as 
\bea \label{Trico2}
\dashint_{k_0^2}^\infty \frac{dQ}{\sqrt{Q}} \frac{\bar \rho_0(Q)}{K-Q} = \frac{1}{\sqrt{K}} g(\sqrt{K}) \;,
\eea
where we have explicitly used the symmetry $\hat \rho_0(k) = \hat \rho_0(-k)$ and 
where the function $g(k)$ is defined in (\ref{def_gk}). We now use the fact that the general solution of the integral equation
\bea \label{Trico_semi} 
\dashint_a^\infty \frac{y(t)}{t-x}\,dt = F(x) \quad, \quad x \geq a \;,
\eea
reads \cite{Handbook} (see formula (50) on p. 229 where we have added the solution of the homogeneous equation $C/\sqrt{x-a}$)
\bea \label{sol_Trico_semi}
y(x) = \frac{C}{\sqrt{x-a}} - \frac{\sqrt{x-a}}{\pi^2}\dashint_a^\infty {dt'}\,  \frac{F(t')}{\sqrt{t'-a}(t'-x)}  \;,
\eea
for some constant $C$. One can thus solve the integral equation for $\bar{\rho}_0(k)$ in Eq. (\ref{Trico2}) and obtain $\hat \rho_0(k)$ as
\bea \label{sol_rhohat0}
\hat \rho_0(k) = C \sqrt{\frac{k^2}{k^2-k_0^2}} + \frac{2}{\pi^2} |k| \sqrt{k^2-k_0^2} \dashint_{k_0}^\infty dq\, \frac{g(q)}{\sqrt{q^2-k_0^2}(q^2-k^2)} \;.
\eea
Using the identity
\bea \label{id1}
\frac{2}{\pi^2} |k| \sqrt{k^2-k_0^2} \dashint_{k_0}^\infty dq\, \frac{g_0(q)}{\sqrt{q^2-k_0^2}(q^2-k^2)} = - 1 \quad {\rm for} \quad g_0(q)= \ln \left(\frac{q+k_0}{q-k_0} \right) \;,
\eea
together with the expression for $g(k)$ in Eq. (\ref{def_gk}) one finds that $\hat \rho_0(k)$ reads for 
$|k| > k_0$
\bea \label{hatrho0}
\hat \rho_0(k) = C \sqrt{\frac{k^2}{k^2-k_0^2}} - 1 - \frac{\beta}{R^2} |k| \sqrt{k^2-k_0^2} \dashint_{k_0}^\infty dq \frac{q}{\sqrt{q^2-k_0^2}(q^2-k^2)} \frac{1}{1 + e^{\beta(\frac{\pi^2 q^2}{2 X^2} - \tilde \mu)}} \;.
\eea
At this stage, the two constants $C$ and $k_0$ remain to be determined. They must be chosen to satisfy the two conditions:
\bea 
\lim_{k \to k_0}  \rho^*(k) = 0 \quad , \quad 
\lim_{k \to \infty} \hat \rho_0(k) = 0 \label{cond2} \;.
\eea
The first condition is natural since the potential 
is pushing smoothly the particles away from the origin. It implies $C=0$. The second condition is also natural since far from
the perturbation $V(k)$ the gas should return to equilibrium. 
This second condition (\ref{cond2}) reads
\bea \label{cond2_bis}
1 =  \frac{\beta}{R^2} \int_{k_0}^\infty \, dq \frac{q}{\sqrt{q^2-k_0^2}} \frac{1}{1+e^{\beta(\frac{\pi^2 q^2}{2 R^2} - \tilde \mu)}} \;,
\eea 
which determines $k_0$. Note that the limit $k_0 \to 0^+$
corresponds precisely to the onset of a solution with a gap
as identified in \eqref{no_gap_cond}. 

We now study the low temperature scaling regime 
\eqref{scalingregime}.
Let us consider equation \eqref{cond2_bis} for the gap $k_0$
setting $\tilde \mu=\mu$. Let us perform again the 
substitution \eqref{substitute}, i.e. 
$R = \frac{z}{\sqrt{2 \mu}}$, $\beta=\frac{z}{u \mu}$,
followed by
the change of variable 
$q= z p/\pi$, $k_0=z p_0/\pi$.
We obtain
\be \label{cond2_bis2}
 \frac{2}{\pi u} \int_{p_0}^\infty \, dp \frac{p}{\sqrt{p^2-p_0^2}} 
 \frac{1}{1+e^{\frac{z}{u}(p^2-1))}} = 1 \;,
\ee
which, in the limit $z \to +\infty$ determines $p_0$ as
\be 
\frac{2}{\pi u} \int_{p_0}^1 \, dp \frac{p}{\sqrt{p^2-p_0^2}}  = 1 
\quad \Longrightarrow \quad p_0^2 = 1 - \frac{\pi^2 u^2}{4} 
\quad \text{for} \quad u < u_c= \frac{2}{\pi} \label{solup0} \;,
\ee 
and we recall that $p_0=0$ for $u> u_c$. 

Similarly we perform the substitution \eqref{substitute}
in the formula for $\hat \rho_0(k)$ given in Eq. (\ref{hatrho0}) with $C=0$.
Then we take the limit $z \gg 1$ and we obtain
\bea 
&& \hat \rho_0(k) =  r(K=\pi k/z) \\
&& r(K)= - 1 - \frac{2}{\pi u} |K| \sqrt{K^2-p_0^2} \dashint_{p_0}^{1} dp \frac{p}{\sqrt{p^2-p_0^2}(p^2-K^2)} 
\quad , \quad |K| > p_0 \label{trico_R}
\eea
and $r(K)=-1$ for $|K| \leq p_0$. Performing the integrals
and using the solution \eqref{solup0} for $p_0$ 
\bea
r(K) =
\begin{cases}
&- 1 \quad, \quad |K| \leq p_0 = 1 - \frac{\pi^2 u^2}{4} \\
&-1 + \frac{2|K|}{\pi u} {\rm tanh}^{-1}\left( \frac{2 \sqrt{K^2-p_0^2}}{\pi u}\right) \quad, \quad p_0 \leq |K| < 1 \\
& - 1 + \frac{2 |K|}{\pi u} 
{\rm tanh}^{-1}\left( \frac{\pi u}{2 \sqrt{K^2 - p_0^2}} \right) \quad, \quad |K| > 1 \;. 
 \end{cases}
\eea
Note that in the limit $K \to \pm 1$, the density $r(K)$ exhibits a logarithmic singularity (see Fig. \ref{plot_rtilde}). 

Let us now compute the optimal energy. We start
again from \eqref{E_deltarho}
and we simplify it using the minimization equation 
\eqref{saddle1} which, however is valid
only for $|k|>k_0$. This leads to 
\be \label{E_deltarho33}
2 E = 
- \int_{|k|<k_0} dk \int_{-\infty}^\infty dk' \log|k-k'| \hat \rho_0(k) \hat  \rho_0(k') +
 2\int_{k_0}^{+\infty}  dk  \left(1+\frac{1}{2}\hat \rho_0(k)\right)   \log( 1 + e^{- \beta (\frac{\pi^2 k^2}{2 R^2} - \mu) } ) 
\ee
where the first term is new as compared to \eqref{E_deltarho_3} and comes from the
exterior of the support of the density. 
In the low temperature limit, performing the substitution \eqref{substitute}
and taking the limit $z \gg 1$ we obtain (using that $\rho_0(k) = -1$ for $|k|< k_0$)
\be \label{E_deltarho34}
2 E = z^2 \bigg(  \frac{1}{\pi^2} 
 \int_{|K|<p_0} dK \int dK' \log|K-K'| r(K') + \frac{2}{\pi u} 
 \int_{p_0}^{1}  dp  (1+\frac{1}{2} r(p) ) (1-p^2) \bigg) 
\ee
\\

The second term is computed setting $y=(1-p^2)/(1-p_0^2)$ and decomposing $r(p)=-1+\tilde r(p)$ one finds 
\be 
\frac{1}{\pi u} \int_{p_0}^1 dp \tilde r(p) (1-p^2) 
= \frac{\pi^2 u^2}{16} \int_{0}^{1} y dy   
{\rm tanh}^{-1}\left( \sqrt{1-y} \right) 
=\frac{\pi^2 u^2}{16} \frac{1}{3} 
\ee 
to which 
one must add
\be 
\frac{1}{\pi u} \int_{p_0}^1 dp (1-p^2) = \frac{1}{\pi u} ( \frac{2}{3} - p_0 + \frac{p_0^3}{3} ) 
\ee 

To analyse the double integral in Eq. (\ref{E_deltarho34}) we first perform an integration by parts (for the integral over $k$) and obtain
\bea \label{first_int_ipp}
\int_{|K|<p_0} dK \int dK' \log|K-K'| r(K') =  2 p_0 \int dK' \ln |p_0-K'| 
r(K')  - \int_{-p_0}^{p_0} dK\, K \,\dashint dK' \frac{1}{K-K'} r(K') 
\eea
The first integral can be evaluated by using the saddle point equation (\ref{saddle1}) in the scaling limit which implies
\bea 
\frac{2}{\pi} \int dK' \log|K-K'| r(K') = \frac{1}{u}(1- K^2)_+
\quad , \quad |K| \geq p_0
\eea 
Evaluated exactly at $K=p_0$ it yields 
\be 
2 p_0 \int dK' \ln |p_0-K'| 
r(K') = 2 p_0 \frac{\pi}{2 u} (1-p_0^2) 
\ee 
To treat the second integral in (\ref{first_int_ipp})
use the identity for $|K|<p_0$ (which is outside
the support)
\bea  
&& \dashint dK' \frac{1}{K-K'} r(K')
= \int_{|K'|>p_0} \frac{1}{K-K'}\bigg( - 1 - \frac{2}{\pi u} |K'| \sqrt{(K')^2-p_0^2} \dashint_{p_0}^{1} dp \frac{p}{\sqrt{p^2-p_0^2}(p^2-(K')^2)} \bigg) 
\\
&& - \dashint_{-p_0}^{p_0} dK' \frac{1}{K-K'} 
\eea  

We use that for $p>p_0$ and $K<p_0$
\bea 
\dashint_{p_0}^{+\infty} dK' K' 
\frac{\sqrt{(K')^2-p_0^2} }{(K^2-(K')^2)(p^2-(K')^2)} 
= \frac{\pi}{2} \frac{\sqrt{p_0^2-  K^2}}{p^2-K^2}
\eea 
and $\dashint_{-\infty}^{\infty} dK' \frac{2K }{K^2-(K')^2} =0$
and the integral becomes elementary and gives 
\bea  
&& \dashint dK' \frac{1}{K-K'} r(K') = - \frac{2 K}{u} 
{\rm tan}^{-1} \left( \frac{ \sqrt{1-p_0^2}}{\sqrt{p_0^2-K^2}} \right)
\eea 
Performing the remaining integral over $K$ we finally obtain
\be 
 - \int_{-p_0}^{p_0} dK\, K \,\dashint dK' \frac{1}{K-K'} r(K') = \frac{\pi}{3 u}  \left(2
   p_0^3+\sqrt{1-p_0^2} \,p_0^2+2
   \sqrt{1-p_0^2}-2\right)
\ee 

Putting all terms together one finds, in the same order as
in \eqref{E_deltarho34} 
\bea 
2 E = z^2 \bigg( 
-\frac{p_0^3}{3 \pi  u}+\frac{\sqrt{1-p_0^2} p_0^2}{3
   \pi  u}+\frac{p_0}{\pi  u}+\frac{2 \sqrt{1-p_0^2}}{3
   \pi  u}-\frac{2}{3 \pi  u} + \frac{\frac{p_0^3}{3}-p_0+\frac{2}{3}}{\pi 
   u}+\frac{\pi ^2 u^2}{48}
\bigg) 
\eea 
Using \eqref{solup0} to eliminate $p_0$ the expression drastically
simplifies into 
\be
2 E = z^2 ( \frac{1}{2}-\frac{\pi ^2 u^2}{48} ) 
\ee
In summary the Coulomb gas predicts in $d=1$
\bea  \label{phim_sine}
&& \log P(R,T) \simeq - (k_F R)^{2} \Phi_{1,-}(u) \quad , \quad u<u_c=2/\pi \nn \\
&& \Phi_{1,-}(u) =  \frac{1}{2}-\frac{\pi ^2 u^2}{48} 
\eea  
as given in the text. One sees that defining $v=u-u_c$
\bea
&& \Phi_{1,+}(u) =  \frac{4}{3 \pi u}  - \frac{1}{\pi^2 u^2} = \frac{5}{12}-\frac{\pi  v}{12}-\frac{\pi ^2
   v^2}{48}+\frac{\pi ^3 v^3}{24}+O\left(v^4\right) \quad , \quad v>0 \\
&& \Phi_{1,-}(u) = \frac{5}{12} - \frac{\pi v}{12}  - \frac{\pi^2 v^2}{48}  \quad , \quad -2/\pi < v<0 \\
\eea 
Hence near the transition one has
\be  \label{delta_phi_sine}
\Phi_{1,+}(u)-\Phi_{1,-}(u) \simeq  \frac{\pi ^3 (u-u_c)^3}{24} \theta(u-u_c)
\ee 
which shows that it is a third order transition. The vicinity of the transition is discussed below in Eqs. (\ref{hole_TW}) and (\ref{third_order}).
\\

\noindent {\bf Remark}. One can ask for an interpretation in terms of the original (physical) momentum variables $k$ appearing in~\eqref{def_sineK}. At $T=0$ the positions $x_i$ of the infinite free fermion gas form a DPP with sine kernel of Fermi momentum $k_F$. The hole probability of the interval $[-R,R]$ in \eqref{holeFD} has a dual representation. 
{Thanks to Eq. \eqref{Duality_SineK} (and the manipulations around this equation) it is equal to the hole probability on the interval $[-k_F,k_F]$ for a dual DPP describing momenta $k_i$ with sine kernel of "Fermi momentum" $R$ (see Eq. (\ref{dual2}))}. Hence at $T=0$ it is natural that the gap of the Coulomb gas studied above is exactly $k_F$ \cite{Dyson_2015}. At finite $T$ this duality remains, and is now expressed as \eqref{linstat} in rescaled variables $k'=k R/\pi$ (noted $k$ above). 
Hence the gap $|K|= \pi |k'|/z=p_0$ obtained above in the low temperature regime corresponds, in the original momenta variable, to a momentum gap 
\be  \label{kgap1} 
k_{\rm gap}= k_F \sqrt{(1- \frac{\pi^2 u^2}{4})_+} \;.
\ee

\subsection{Connection with the results of Ref. \cite{Xu2025}} 

It is interesting to compare our results obtained here by the Coulomb gas method with recent rigorous work, using quite different Riemann-Hilbert techniques, in the mathematics literature \cite{Xu2025}. 

Let us recall that we are studying the probability that there are no fermions in $[-R,R]$, which is given by the Fredholm determinant
\be 
P(R,T) = {\rm Det}( I - \Pi_{[-R,R]} \hat K) \quad {\rm where} \quad \hat K(y,y')= \int_{-\infty}^{\infty} \frac{dk}{2 \pi} \sigma(k) e^{i k (y-y')} \quad {\rm and} \quad \sigma(k)= \frac{1}{e^{\beta (\frac{k^2}{2} - \tilde \mu) }+1 } \;.
\ee 
where the notation $\Pi_{[-R,R]}$ denotes the projector on the interval $[-R,R]$.
In \cite{Xu2025} the following quantity is introduced in Eqs. (1.4)-(1.7)
(with the notations used there) 
\be 
D(x,s)= {\rm Det}(1 - \Pi_{[-\frac{x}{\pi},\frac{x}{\pi}]} K(\lambda,\lambda')) \quad , \quad 
K(\lambda,\lambda') = \int_0^{+\infty} d\tau \cos(\pi \tau (\lambda-\lambda')) \frac{1}{1+ e^{\tau^2-s}} \;.
\ee 
Using the change of variables $\tau= k \sqrt{\beta/2}$, 
$\pi \tau \lambda = k y$ and $\pi \sqrt{\beta/2} \lambda = y$ 
and $d \lambda K(\lambda,\lambda') = dy  \hat K(y,y') $, we
find that the correspondence between the two sets of notations is 
\be \label{corr_Xu}
P(R,T)  = D(x= R \sqrt{2/\beta} , s= \beta \tilde \mu ) 
\ee 

{\bf (i) High temperature regime}. In Theorem 1 in \cite{Xu2025} it is
proved that for $x \to +\infty$, with either $s$ fixed, or $s \ll x^2$
one has (see also \cite{Its1990})
\bea  \label{const_Xu}
\log D(x,s) = - \frac{2 x}{\pi} \int_{\mathbb{R}} \frac{\tau^2 d\tau}{1 + e^{\tau^2 - s}} 
+ \frac{1}{2 \pi^2} \int_{-\infty}^s ds' \left( 
\int_{\mathbb{R}} \frac{d\tau}{1 + e^{\tau^2 - s'}}
\right)^2 + O(e^{-c x^2}) 
\eea 

We can check that this expression is exactly the same as the one obtained above from Widom's formula, and from the Coulomb gas formula. Indeed, in our notations, the first term becomes
\be 
\log P(R,T) = - \frac{\beta}{\pi} R \int_{\mathbb{R}} \frac{dk k^2}{1 + e^{ \beta ( \frac{k^2}{2} - \tilde \mu)}} 
\ee 
which identifies with \eqref{expansionlogP1}-\eqref{leading_Widom} setting $d=1$, upon
an integration by part. For the second term in (\ref{const_Xu}), we can check that it coincides precisely with the result obtained by the Coulomb gas method in Eq. (\ref{deriv4}). As discussed above, it also coincides with $- B_1(t)$ given in \eqref{B1}
obtained from the Widom formula.

{\bf (ii) Low temperature regime}.

In Theorem 2 and Remark 2 in \cite{Xu2025}, the limit 
$x,s \to +\infty$ with 
$\ell = \frac{\pi}{2} x/\sqrt{s}$ fixed is considered. 
In our notations $s x^2 = z^2$ and $u=2 \ell/\pi$.
Hence this is what we call the low temperature regime (ii) 
in the present paper. The formula (2.7) and (2.8) are in perfect
agreement with $\Phi_{1,+}(u)$ and $\Phi_{1,-}(u)$
respectively, obtained above in \eqref{phi_dp} and \eqref{phi_dm}. 

{\bf Critical region}. In Theorem 3 of \cite{Xu2025}
the critical regime is studied. The scaling variable $y$ is defined 
through $\ell = 1 + (\frac{\pi}{4 s})^{2/3} y$. 
In our notations their Eq. (2.13) is equivalent to writing  
\bea \label{hole_TW}
P(R,T) \simeq   G_1(y) e^{- z^2 \Phi_+(u) } \quad {\rm with}  \quad G_1(y) = F_2(y) 
\eea 
in the critical region, where 
$u = \frac{2}{\pi} (1 + (\frac{1}{2 z})^{2/3} y)$ 
and $y=O(1)$. The function $F_2(y)$ is the cumulative GUE Tracy Widom 
distribution given  by \cite{TW1994}
\bea 
\log F_2(y) = - \int_y^{+\infty} d\tau (\tau-y) q^2(\tau) + O(z^{-1/6}) \;,
\eea 
where $q(\tau)$ satisfies the Painlev{\'e} II equation, $q''(\tau)=\tau q(\tau) + 2 q^3(\tau)$ with $q(\tau) \sim {\rm Ai}(\tau)$ for $\tau \to \infty$. We recall the asymptotic behaviors $\log F_2(y) \sim - |y|^3/(12)$ as $y \to - \infty$, while $\log F_2(y) \to 0$ as $y \to +\infty$.
One can check the matching on the side $u<u_c$, between the regime 
$u_c-u \sim z^{-2/3} $ and the regime $u_c-u= O(1)$, which reads
\bea \label{third_order}
- \log F_2(y) = - \log F_2((2 z)^{2/3} \frac{\pi}{2} (u-u_c)) 
\simeq  \frac{z^2}{24} \pi^3 (u_c-u)^3 \simeq 
z^2 ( \Phi_-(u)- \Phi_+(u) ) \;,
\eea 
for $u$ near $u_c$. This coincides with the singularity obtained from the Coulomb gas in \eqref{delta_phi_sine}.

\section{Coulomb gas calculations: the case $d>1$}

We recall from the text that the 
the total probability $P(R,T)$ factorizes over the different $\ell$-sectors and one has 
\be \label{holeproduct}
P(R,T) = \prod_{\ell \geq 0} P^{(\ell)}(R,T)^{g_d(\ell)} \quad, \quad g_d(\ell) = \frac{(2 \ell + d-2) \Gamma(\ell+d-2)}{\Gamma(\ell+1) \Gamma(d-1)} \quad , \quad \ell \geq 1 \quad {\rm and} \quad  g_d(0)= 1 \;.
\ee
It is useful to define
\bea \label{def_nu_sm}
\nu = \ell + \frac{d}{2} - 1 \;.
\eea

In each sector with a given $\nu$ we will
consider a Coulomb gas with total energy given in
\eqref{E_deltarho_bessel}
\be \label{E_deltarho_bessel2}
2 E_\nu = - \int_0^{+\infty} dp \int_0^{+\infty} dp' \log|p-p'| \hat \rho(p) \hat  \rho(p') 
+ \int_0^{+\infty} dp  (\rho_0(p)  + \hat \rho(p))
\log( 1 + e^{- \beta  (\frac{p}{2 R^2} - \tilde \mu) } ) 
\ee
with $\hat \rho(p)=\rho(p)-\rho_0(p)$ where we recall
that $\rho_0(p)= \dfrac{\sqrt{(p-\nu^2)_+}}{2 \pi \,p}$ is the equilibrium density in the absence of
potential. The
hole probability $P^{(\ell)}(R,T)$ in the corresponding sector is then  
approximated as
\be  \label{P0_Eminbessel2}
P^{(\ell)}(R,T) \sim e^{-2 E_{0,\nu}} \quad , \quad E_{0,\nu} = \min_{\hat \rho \geq -\rho_0} E_\nu \;.
\ee 
where we recall that the energy minimization is under the constraint
that $\hat \rho(p) \geq - \rho_0(p)$ -- which is imposed by the positivity constraint $\rho(p) = \hat \rho(p) + \rho_0(p) \geq 0$. 
 
Below we compute $E_{0,\nu}$ separately in three different regimes

\begin{enumerate}

\item   $\nu$ and $\beta \tilde \mu$, and large $R$, which is of interest
for the finite temperature Bessel process of index $\nu$.

\item fixed $\beta \tilde \mu$ and both $\nu \sim R$ large. This regime is
of interest to analyze the hole probability in $d$ dimensions in the fixed temperature regime.

\item $\nu \sim \beta \mu \sim z= R \sqrt{2 \mu} \to + \infty$.
This regime is
of interest to analyze the hole probability in $d$ dimension in the low temperature regime.

\end{enumerate}
In the last two regimes, the product $P(R,T)$ in \eqref{holeproduct} 
is dominated by terms with $\ell \sim \nu = O(z)$ for large $z$.
\\

The main results of this section, obtained via the CG method, and detailed below, are as follows. The asymptotic behavior of the hole probability $P^{(\ell)}(R,T)$ for the Bessel process at fixed $\nu,\ell$ is given 
(a) in the high temperature regime in \eqref{res_Pl}, in agreement with \eqref{Basor_exp} (b) in the low temperature regime in \eqref{lowThighTBessel} and \eqref{lowThighTBessel2}. 
The asymptotic behavior of the hole probability $P^{(\ell)}(R,T)$ for the Bessel process for $\nu,\ell \sim R$ is given 
(a) in the high temperature regime in \eqref{double_scaling}, \eqref{func_A},
\eqref{func_B},
and (b) in the low temperature regime in \eqref{phiexpression}, \eqref{resultlambda}.
Finally, the results for free fermions in spatial dimension $d$ 
obtained by summing over $\ell$, are given (a) in the high temperature regime 
around \eqref{P2_Fnu}, recovering the results of Section \ref{sec:widom}, and (b) 
in the low temperature regime in Section \ref{sec:ddimlowT}.


\subsection{First regime: fixed $\nu,\beta \tilde \mu$ and large $R$}

\subsubsection{Optimal density for fixed $\nu$}

Let us denote $\rho^*(p)$
and $\hat \rho^*(p)=\rho^*(p) - \rho_0(p)$
the minimizer of the total energy $E_\nu$ in Eq. \eqref{E_deltarho_bessel}
under the constraint $\rho^*(p) \geq 0$.
Here we assume that $\rho^*(p)$ has a support over $[a, + \infty)$, with $a \geq \nu^2$ since the effective potential $\log( 1 + e^{- \beta (\frac{\pi^2 p}{2 R^2} - \tilde \mu) } )$ is repulsive. One finds, for $p$ belonging to the support of $\rho^*(p)$, i.e. $p \geq a$
\bea \label{saddle1nu}
\frac{\delta E_\nu}{\delta \hat \rho(p)} = 0 \Longleftrightarrow 2 \int_0^{+\infty} dp' \log |p-p'| \hat \rho^*(p') = \log( 1 + e^{- \beta (\frac{ p}{2 R^2} - \tilde \mu) } ) \quad, \quad \hat \rho^*(p) > -1 \;. 
\eea
Taking one more derivative w.r.t. $p$, one finds the necessary condition 
for $p \in [a, +\infty)$ (since $\rho^*$ has support over $[a, + \infty)$)
\be \label{sp_a2}
 \dashint_a^{+\infty} dp' \frac{\hat \rho^*(p')}{p-p'} = - \frac{\beta}{4 R^2} \frac{1}{e^{\beta (\frac{ p}{2 R^2}-\tilde \mu)}+1} + \dashint_{\nu^2}^{a} dp' \frac{ \rho_0(p')}{p-p'} \;,
\ee
which is valid for $p \geq a$ and we recall that  
$\rho^*(p) = \rho_0(p) + \hat \rho^*(p) \geq  0$. The 
second term in the r.h.s originates from the Coulomb interaction term,
taking into account that $\hat \rho^*(p) = 0$ for $p \in [0,\nu^2]$, and that $\hat \rho^*(p) = -  \rho_0(p) = - \sqrt{p-\nu^2}/(2 \pi p)$ for $p \in [\nu^2, a]$.  Note also that, by taking the limit $p \to \infty$ on both sides of Eq. (\ref{sp_a2}) one finds the conservation law 
\begin{eqnarray} \label{conservation}
\int_{\nu^2}^\infty \hat \rho^*(p')\, dp' = 0 \;.
\end{eqnarray}
As above in Eqs. (\ref{Trico_semi})-(\ref{sol_Trico_semi}), this integral equation (\ref{sp_a2}) can be solved explicitly. Using the identity
\bea \label{id_pps}
\dashint_a^{\infty} dp' \frac{1}{\sqrt{p'-a}} \frac{1}{p'-p} \frac{1}{p'-p''}  = \frac{\pi}{p'' - p} \frac{1}{\sqrt{a-p''}} \quad, \quad p>a \quad, \quad p''<a \;,
\eea
one obtains 
\bea \label{Trico_half3}
\hat \rho^*(p) = \frac{C}{\sqrt{p-a}} &-& \frac{\sqrt{p-a}}{4 \pi^2 R^2} \beta  \dashint_{a}^\infty dp' \frac{1}{\sqrt{p'-a}} \frac{1}{p'-p} \frac{1}{e^{\beta (p'/(2 R^2)-\tilde \mu)}+1} \\
&+& \frac{\sqrt{p-a}}{\pi} \int_{\nu^2}^a dp'' \frac{1}{p''-p} \frac{1}{\sqrt{a-p''}} \frac{\sqrt{p'' - \nu^2}}{2\pi p''} \quad, \quad p \geq a \nn 
\eea
%
It is easy to see on (\ref{Trico_half3}) that, for large $p$, $\hat \rho(p)$ behaves as $\hat \rho(p) \sim D/\sqrt{p}$. Since we expect $\hat \rho(p) \ll \rho_0(p) \sim 1/(2 \pi \sqrt{p})$ as $p \to \infty$, which imposes the condition $D=0$. This yields
\bea \label{eqD0}
C + \frac{\beta}{4 \pi^2 R^2} 
\dashint \int_a^\infty \frac{1}{\sqrt{p'-a}} \frac{1}{e^{\beta (p'/(2 R^2)-\tilde \mu)}+1} - \frac{1}{\pi} \int_{\nu^2}^a dp''\,\frac{1}{\sqrt{a-p''}} \frac{\sqrt{p''-\nu^2}}{2 \pi p''} = 0 \;,
\eea
which fixes the value of the constant $C$. Inserting back this expression for $C$ in Eq. (\ref{Trico_half3}) one finds
\bea \label{Trico_half4}
\hat \rho^*(p) = - \frac{\beta}{4 \pi^2 R^2} \frac{1}{\sqrt{p-a}} \dashint_a^{\infty} dp' \frac{\sqrt{p'-a}}{p'-p} \frac{1}{e^{\beta (p'/(2 R^2)- \tilde \mu)}+1} - \frac{1}{\pi} \frac{1}{\sqrt{p-a}} \int_{\nu^2}^a dp'' \frac{\sqrt{p''-\nu^2}}{2\pi p''} \sqrt{a - p''} \frac{1}{p'' - p}  \;.
\eea
In fact the second integral can be computed explicitly, which gives
finally for $p>a$
\bea \label{total_density}
\rho^*(p) = - \frac{\beta}{4 \pi^2 R^2} \frac{1}{\sqrt{p-a}} \dashint_a^{\infty} dp' \frac{\sqrt{p'-a}}{p'-p} \frac{1}{e^{\beta (p'/(2 R^2) -\tilde \mu)}+1} + \frac{1}{2 \pi p} \frac{p - \nu \sqrt{a}}{\sqrt{p-a}} \;.
\eea
Note that this solution is correct only if it is positive for all $p \geq a$.

We still have to obtain the value of $a$. 
One notices that this expression for the density $\rho^*(p)$ diverges as $p \to a$ as $\hat \rho(p) \sim H(a)/\sqrt{p-a}$. Since it is reasonable to assume that $H(a) = 0$, we obtain an equation determining $a$, which reads
\bea \label{eq_for_a}
1 - \frac{\nu}{\sqrt{a}}  = \frac{\beta}{2 \pi R^2} \int_{a}^\infty dp' \frac{1}{\sqrt{p'-a}} \frac{1}{e^{\beta (p'/(2 R^2) - \tilde \mu)}+1} = - 
\sqrt{ \frac{\beta}{2 \pi}} \frac{1}{R} {\rm Li}_{1/2}(-e^{\beta (\tilde \mu -a/(2 R^2))})\;.
\eea
For large $R$ at fixed $\nu$, we see that $a \to \nu^2$. 
In summary, the optimal density $\rho^*(p)$ is determined by Eq. (\ref{Trico_half4}) together with Eq. (\ref{eq_for_a}). 

\subsubsection{Hole probability for fixed $\nu$}

The optimal energy, from Eq. (\ref{E_deltarho_bessel2}), and using the minimization equation 
\eqref{saddle1nu} is then given by
\bea  
2 E_{0, \nu} &=&  2 E_{\nu, 0} + 2E_{\nu, 1} + 2 E_{\nu, 2} \label{energy_sp_nu2} \\
2 E_{\nu,0} &=& \int_{\nu^2}^a dp \int_{\nu^2}^\infty dp' \ln|p-p'|  \rho_0(p) \hat \rho^*(p') - \int_{\nu^2}^{a} dp \, \rho_0(p) \log( 1 + e^{- \beta  (\frac{p}{2 R^2} - \tilde \mu) } ) \label{def_E0} \\
2 E_{\nu, 1} &=& \int_{\nu^2}^\infty dp \rho_0(p) 
\log( 1 + e^{- \beta  (\frac{p}{2 R^2} - \tilde \mu) } )\;, \label{E1_new} \\
2E_{\nu, 2} &=& \frac{1}{2} \int_{a}^\infty dp\, \hat \rho^*(p) \log( 1 + e^{- \beta  (\frac{p}{2 R^2} - \tilde \mu) } )  \;. \label{E2_new}
\eea
This set of relations (\ref{total_density})-(\ref{E2_new}) can now be analysed in the various limits described above, see 1,2,3 below \eqref{P0_Eminbessel2}.
\\


Consider now the regime 1, i.e. the limit of large $R$ at $\beta \tilde \mu$ and $\nu$ fixed. In this limit, assuming that $a = O(1)$,
changing variable $p'=2 R^2 x$ in \eqref{eq_for_a} we see that the equation for $a$ becomes 
\bea \label{eq_for_a_new}
1 - \frac{\nu}{\sqrt{a}}  \simeq  - 
\sqrt{ \frac{\beta}{2 \pi}} \frac{1}{R}  {\rm Li}_{1/2}(-e^{\beta \tilde \mu})\;
\quad \Rightarrow \quad a= \nu^2 - \sqrt{\frac{2 \beta}{\pi}} \frac{\nu^2}{R}  {\rm Li}_{1/2}(-e^{\beta \tilde \mu}) + o(R^{-1} )  
\eea
In this limit, the density $\hat \rho^*(p)$ takes the scaling form
\bea \label{rho_tilde}
\hat \rho^*(p) \approx \frac{1}{2 \tilde \mu R^2} 
\tilde \rho\left(\frac{p}{2 \tilde \mu R^2}\right) \quad, \quad \tilde \rho(\tilde p) = 
- \frac{\beta \tilde \mu}{2 \pi^2} \frac{1}{\sqrt{\tilde p}}\dashint_0^\infty d \tilde q \frac{\sqrt{\tilde q}}{(\tilde q-\tilde p)} \frac{1}{e^{\beta \tilde \mu(\tilde q-1)}+1}
\eea
where we have used that $\frac{1}{2 \pi p} 
\frac{p - \nu \sqrt{a}}{\sqrt{p-a}}- \rho_0(p) \sim p^{-3/2} 
\sim R^{-3} \tilde p$ at large $p$.

Let us now analyze the optimal energy. 
It is easy to see that the term $2 E_{\nu,0}$ in Eq. (\ref{def_E0}) 
behaves as $\sim (a-\nu^2)^{3/2} \sim R^{-3/2}$ at large $R$.
More precisely one can show that (using the minimization equation (\ref{saddle1nu}) at $p=a$)
\bea \label{E0_final}
2 E_{\nu,0}  = - \frac{1}{6 \pi \nu^2} (a - \nu^2)^{3/2} \log\left(1+e^{- \beta  (\nu^2/(2 R^2)-\tilde \mu)} \right) + o(1/R^{3/2}) \;.
\eea
Next one has
\bea  \label{E1}
2 E_{\nu , 1} = \int_{\nu^2}^\infty dp \frac{\sqrt{p-\nu^2}}{2 \pi p} \log( 1 + e^{- \beta  (\frac{p}{2 R^2} - \tilde \mu) } ) \;,
\eea 
which we write as
\bea 
2E_{\nu,1} &=& \frac{1}{2 \pi} \int_0^\infty \frac{dp}{\sqrt{p}} \log( 1 + e^{- \beta  (\frac{p}{2 R^2} - \tilde \mu) } ) - \frac{1}{2 \pi} \int_0^{\nu^2} \frac{dp}{\sqrt{p}} \log( 1 + e^{- \beta  (\frac{p}{2 R^2} - \tilde \mu) } ) \\
&+& \frac{1}{2 \pi} \int_{\nu^2}^\infty dp \left( \frac{\sqrt{p-\nu^2}}{p}-\frac{1}{\sqrt{p}}\right)  \log( 1 + e^{- \beta  (\frac{p}{2 R^2} - \tilde \mu) } )
\eea 
To extract the large $z$ asymptotics, we perform the change of variable $q = p/z^2$ in the first integral and we immediately see that this first term is of order $O(z)$. On the other hand, in the second and third integrals one can directly take the limit $z \to \infty$ in the integrands (without any rescaling). This gives
\bea \label{E1_2}
2 E_{\nu,1} &=& \frac{R}{2 \pi} \sqrt{2 \tilde \mu} \int_0^\infty \frac{dq}{\sqrt{q}} \log( 1 + e^{- \beta \tilde \mu ({q} - 1) } ) + \frac{1}{2 \pi}\log( 1 + e^{\beta \tilde \mu} ) \left(  \int_{\nu^2}^\infty dp \left( \frac{\sqrt{p-\nu^2}}{p}-\frac{1}{\sqrt{p}}\right) - \int_0^{\nu^2} \frac{dp}{\sqrt{p}} \right) + O(1/R)\; \nn \\
&=& \frac{R}{2 \pi} \sqrt{2 \tilde \mu} \int_0^\infty \frac{dq}{\sqrt{q}} \log( 1 + e^{- \beta \tilde \mu ({q} - 1) } ) - \frac{\nu}{2} \log( 1 + e^{\beta \tilde \mu} ) + O(1/R)
\eea 
where we have used that $\int_{\nu^2}^\infty  dp ( \frac{\sqrt{p-\nu^2}}{p}-\frac{1}{\sqrt{p}}) = 2 \nu - \pi \nu$. 
Finally, the analysis of $E_{\nu,2}$ in Eq. (\ref{E2_new}) can be performed by injecting the scaling form (\ref{rho_tilde}) in Eq. (\ref{E2_new}), leading to (after the change of variables $q= \beta p/(2 R^2)$ and $q'= \beta p'/(2 R^2)$)
\bea \label{E2_new2}
2E_{\nu,2} = - \frac{1}{4 \pi^2} \int_{0}^\infty dq\, \log( 1 + e^{ \beta \tilde \mu - q}  ) 
\int_{0}^\infty \frac{dq'}{q'-q} 
\frac{ \sqrt{q'} }
{\sqrt{q}} 
 \frac{1}{e^{-\beta \tilde \mu+ q'}+1}  + o(1) \;.
\eea   
Finally, combining (\ref{E1_2}) and  (\ref{E2_new2}) yields our final result from the Coulomb gas in the limit of large $R$,
for fixed $\beta \tilde \mu$ and $\nu$ 
\bea \label{res_Pl}
\log P^{(\ell)}(R,T) = -2 E_{\nu,0} -2 E_{\nu,1}- 2 E_{\nu,2} &=& 
- \frac{R}{2 \pi} \sqrt{2 \tilde \mu} \int_0^\infty \frac{dq}{\sqrt{q}} \log( 1 + e^{- \beta \tilde \mu ({q} - 1) } ) + \frac{\nu}{2} \log( 1 + e^{\beta \tilde \mu} ) \\
&+&  \frac{1}{4 \pi^2} \int_{0}^\infty dq\, \log( 1 + e^{ \beta \tilde \mu - q}  ) 
\dashint_{0}^\infty \frac{dq'}{q'-q} 
\frac{ \sqrt{q'} }
{\sqrt{q}} 
 \frac{1}{e^{-\beta \tilde \mu+ q'}+1}  + o(1) \nonumber \;.
\eea
Note that the leading term can be computed explicitly as $\frac{z}{2 \sqrt{\pi}} \frac{1}{\sqrt{\beta \tilde \mu}} {\rm Li}_{3/2}(- e^{\beta \tilde \mu})$.

Let us now compare our results with those of Basor and Ehrhardt 
from Ref. \cite{BE2003}. 
To make contact with our notations one must choose in their paper 
the functions 
\bea  
&& a(v)= - \frac{1}{1 + e^{\beta (\frac{v^2}{2} - \tilde \mu) }} \\
&& b(v) = \log(1+a(v)) = - \log( 1 + e^{\beta (\tilde \mu - \frac{v^2}{2}) })  \\
&& \hat b(u) = \frac{1}{\pi} \int_0^{+\infty} dv b(v) \cos(u v)
\eea  
Their result given in Eq. (5) of Ref. \cite{BE2003}, translated in our notations, then reads 
\be \label{Basor_exp}
P^{(\ell)}(R,T) 
\sim \exp\left( R \, \hat b(0) - \frac{\nu}{2} b(0) 
+ \frac{1}{2} \int_0^{+\infty} dx  x (\hat b(x))^2 
\right) 
\ee 
The first term reads
\bea \label{first_term}
R \, \hat b(0) = - R \int_0^{+\infty} dv \log(1  + e^{\beta (\tilde \mu- v^2/2) } )  
= - \frac{R \sqrt{2 \tilde \mu}}{ 2\pi} \int_0^{+\infty} \frac{dq}{\sqrt{q}} \log(1  + e^{-\beta \tilde \mu(q-1)}  ) 
\eea 
where in the last equality we performed the change of variable $v = \sqrt{2 \tilde \mu q}$. It coincides with our first term in \eqref{res_Pl}. 
The second term in (\ref{Basor_exp}) obviously coincides 
with our second term in \eqref{res_Pl}. 

The third term is more complicated. Let us write it explicitly
\bea \label{third_term}
&&\frac{1}{2} \int_0^{+\infty} dx  x (\hat b(x))^2 = \frac{1}{2 \pi^2} \int_0^\infty dx \, x \int_0^\infty dv\, \cos{(x v)} \int_0^\infty \cos{(x w)} \log\left(1 + e^{\beta (\tilde  \mu - v^2/2) } \right)\log\left(1 + e^{\beta 
(\tilde  \mu - w^2/2)} \right) \nonumber \\
&=& \frac{1}{2 \pi^2} \int_0^\infty dy\, y \int_0^\infty dv' \int_0^\infty dw' \cos{(y v')} \cos{(y w')} \log\left(1 + e^{\beta \tilde \mu - (v')^2} \right) \log\left(1 + e^{\beta \tilde \mu - (w')^2} \right)\;,
\eea 
where we performed the changes of variables $v' = v \sqrt{\beta/2}$, $w' = w \sqrt{\beta/2}$ and $y = x/\sqrt{\beta/2} $. 
We now perform a first integration by parts on the integral over $w'$ (integrating $y \cos(y w')$). This yields (returning back to the notations $v$ and $w$ instead of $v'$ and $w'$)
\bea \label{third_term_2}
&&\frac{1}{2} \int_0^{+\infty} dx  x (\hat b(x))^2 = \frac{1}{2 \pi^2} \int_0^\infty dy \int_0^\infty dv \, \cos{(y v)} \int_0^\infty dw\, \sin{(y w)} \frac{2 w}{1+ e^{-\beta \mu +w^2}}\log\left(1 + e^{\beta \mu -v^2} \right) \;.
\eea
We can now perform the integral over $y$ using the identity
\bea \label{id_cos}
\int_0^\infty dy \cos{(y v)} \sin{(y w)} = - \frac{w}{v^2-w^2} \;.
\eea
Using this identity in Eq. (\ref{third_term}) one finds
\bea  \label{third_term_2}
\frac{1}{2} \int_0^{+\infty} dx  x (\hat b(x))^2 = - \frac{1}{\pi^2} \int_0^\infty dv \dashint_0^\infty dw \frac{w^2}{v^2-w^2} \frac{1}{1 + e^{-\beta \mu + w^2}} \log \left( 1+ e^{\beta \mu - v^2}\right) \;.
\eea 
Finally, performing the changes of variables $q=v^2$, $q' = w^2$ one arrives precisely at the last term in \eqref{res_Pl}. It is quite
remarkable that our Coulomb gas calculation recovers both the
order $O(R)$ term as well as the constant term, for any
fixed $\nu$. 


\subsection{The case $\beta \tilde \mu$ fixed, and $\nu \sim R$ large}

In this subsection we first obtain a new result for the hole probability of the 1D Bessel kernel in the regime of $\beta \tilde \mu$ fixed and $\nu \sim R$, which goes beyond the results in the mathematics literature, such as \cite{BE2003}, valid only for fixed $\nu$. In a second stage we perform a summation over $\ell$ (i.e. $\nu$) which gives the hole probability in the high temperature regime (i) of the main text. Interestingly, as we show here, this calculation reproduces the results obtained using the $d$-dimensional Widom's formula (see Section \ref{sec:widom}). 

\subsubsection{Hole probability associated to the 1D Bessel kernel of index $\nu \sim R$}

In this regime, the parameters $a \sim \nu^2 \sim R^2$ are
all large, and $\beta \tilde \mu$ is fixed. The equation (\ref{eq_for_a_new})
which determines $a$, is thus modified into (since we find $a \sim R^2$)
\bea \label{eq_for_a_new2}
 a- \nu^2 = - \sqrt{\frac{2 \beta}{\pi}} \frac{\nu^2}{R}  {\rm Li}_{1/2}(-e^{\beta (\tilde \mu- \frac{a}{2 R^2} ) }) + o(R) \;, 
\eea
where we recall that $\nu = O(R)$, hence $a \sim R^2$. Thus we 
see that $a- \nu^2 = O(R)$ in that regime.

The goal is now to obtain the asymptotics of $\log P^{(\ell)}(R,T)$
in that regime. One shows below that in the limit 
$\beta \tilde \mu$ fixed, and $\nu \sim R$ large 
\bea \label{double_scaling}
\log P^{(\ell=\nu-\frac{d}{2} + 1)}(R,T) = \log {\rm Det}( I - \Pi_{[0,R]} \hat K_\nu)
= R \sqrt{2 \tilde \mu} \, {\cal A}\left(\frac{\nu}{R \sqrt{2 \tilde \mu}}, \beta \tilde \mu \right) + {\cal B}\left(\frac{\nu}{R \sqrt{2 \tilde \mu}},  \beta \tilde \mu \right) + o(1) 
\eea
where the function ${\cal A}(x,\beta \tilde \mu)$ and ${\cal B}(x,\beta \tilde \mu)$ are given by
\bea \label{func_A}
&&{\cal A}(x, \beta \tilde \mu) = -\int_{x^2}^\infty \frac{dq}{2 \pi q} \sqrt{q-x^2} \log\left(1 + e^{-\beta \tilde \mu(q-1)} \right) \\
&&{\cal B}(x, \beta \tilde \mu) = \frac{\beta \tilde \mu}{4 \pi^2} \int_{x^2}^\infty 
\frac{dp}{\sqrt{p-x^2}} \log\left(1+e^{\beta \tilde \mu(1-p)} \right) \int_{x^2}^\infty dq \frac{\sqrt{q-x^2}}{q-p} \frac{1}{1+ e^{\beta \tilde \mu(q-1)}} \label{func_B}
\eea

To derive this result we note that 
there are three contributions to the optimal energy
denoted $2 E_{\nu,j}$, $j=0,1,2$, and
given in Eqs. \eqref{def_E0}, \eqref{E1_new} and \eqref{E2_new}. 
The hole probability is obtained from the sum  
$\log P^{\ell}(R,T) = - 2 (E_{\nu,0}+ E_{\nu,1}+ E_{\nu,2})$.

We first analyze the term $2 E_{\nu, 0}$ from Eq. (\ref{def_E0}). From the large $R$ asymptotics in (\ref{E0_final}), which is also valid for $\nu \sim R$,
we see that $2 E_{\nu,0} = O(\nu/R^{3/2})= O(R^{-1/2})$ in that regime.
Hence this term is subdominant since here we
are interested in the terms $O(R)$ and $O(1)$.

The second contribution, $2 E_{\nu,1}$, is given in Eq. \eqref{E1_new} and reads
\bea
2 E_{\nu, 1} &=&  \int_{\nu^2}^\infty dp \frac{\sqrt{p-\nu^2}}{2\pi p} 
\log( 1 + e^{- \beta  (\frac{p}{2 R^2} - \tilde \mu) } )
\eea
Performing the change of variable $p= 2 \tilde \mu R^2 q$ one obtains
the first term ${\cal A}$ in \eqref{double_scaling}. 

The last contribution comes from $2E_{\nu, 2}$ in \eqref{E2_new} which reads
\bea \label{Enu2_2}
2E_{\nu, 2} &=& \frac{1}{2} \int_{a}^\infty dp\, \hat \rho^*(p) \log( 1 + e^{- \beta  (\frac{p}{2 R^2} - \tilde \mu) } )
\eea
where we recall that 
\bea \label{total_density4}
\hat \rho^*(p) = - \frac{\beta}{4 \pi^2 R^2} \frac{1}{\sqrt{p-a}} \dashint_a^{\infty} dp' \frac{\sqrt{p'-a}}{p'-p} \frac{1}{e^{\beta (p'/(2 R^2) -\tilde \mu)}+1} + \frac{1}{2 \pi p} 
( \frac{p - \nu \sqrt{a}}{\sqrt{p-a}} - \sqrt{p-\nu^2} )  \;.
\eea
which is correct only if it is positive for all $p \geq a$.
Performing the change of variable $p=a (1 + y/\nu)$ one finds
that at large $\nu$ the integral involving the second term
in \eqref{total_density4} is of order $O(1/\sqrt{\nu})=O(R^{-1/2})$
which we can discard here. In the first term we approximate 
$a \simeq \nu^2$ to the leading order and perform the
change of variable $p \to 2 \tilde \mu R^2 p$,
$p' \to 2 \tilde \mu R^2 q$ which leads to the term ${\cal B}$
in \eqref{func_B}. 

We can now transform the function ${\cal B}$ to obtain a simpler
expression. For this one writes ${\cal B}(x, \beta \tilde \mu)$
as a function of $x^2$ and take a derivative w.r.t. $x^2$.
One has 
\bea
&& \partial_{x^2} {\cal B}(x, \beta \tilde \mu) =  \frac{\beta \tilde \mu}{4 \pi^2} \partial_{x^2} \int_{0}^\infty 
\frac{dp}{\sqrt{p}} \log\left(1+e^{\beta \tilde \mu(1-p-x^2)} \right) \int_{0}^\infty dq \frac{\sqrt{q}}{q-p} \frac{1}{1+ e^{\beta \tilde \mu(q+x^2-1)}} \\
&& =  
- \frac{(\beta \tilde \mu)^2}{4 \pi^2}  
 \int_{0}^\infty 
\frac{dp}{\sqrt{p}} \frac{1}{1+e^{\beta \tilde \mu(x^2+p-1)}} \int_{0}^\infty dq \frac{\sqrt{q}}{q-p} \frac{1}{1+ e^{\beta \tilde \mu(q+x^2-1)}} \\
&& + \frac{\beta \tilde \mu}{4 \pi^2} 
 \int_{0}^\infty 
\frac{dp}{\sqrt{p}} \log\left(1+e^{\beta \tilde \mu(1-p-x^2)} \right) \int_{0}^\infty dq \frac{\sqrt{q}}{q-p} \partial_{x^2}  \frac{1}{1+ e^{\beta \tilde \mu(q+x^2-1)}}
\eea 
We focus on the second integral and note that $\partial_{x^2}$
can be replaced by $\partial_q$. Performing an integration by part over $q$ using
the identity 
\bea \label{trivial2}
\partial_q \frac{\sqrt{q}}{q-p} = - \frac{(p+q)}{2(p-q)^2 \sqrt{q}} \;.
\eea
We obtain
\bea 
&& \partial_{x^2} {\cal B}(x, \beta \tilde \mu) = - \frac{(\beta \tilde \mu)^2}{4 \pi^2} 
 \int_{0}^\infty 
\frac{dp}{\sqrt{p}} \frac{1}{1+e^{\beta \tilde \mu(x^2+p-1)}} \int_{0}^\infty dq \frac{\sqrt{q}}{q-p} \frac{1}{1+ e^{\beta \tilde \mu(q+x^2-1)}} \\
&& + \frac{\beta \tilde \mu}{4 \pi^2} 
 \int_{0}^\infty 
\frac{dp}{\sqrt{p}} \log\left(1+e^{\beta \tilde \mu(1-p-x^2)} \right) \int_{0}^\infty dq 
\frac{(p+q)}{2(p-q)^2 \sqrt{q}}\frac{1}{1+ e^{\beta \tilde \mu(q+x^2-1)}}
\eea 
Integrating now by parts over $p$ and using 
\bea \label{trivial3}
\partial_p \frac{\sqrt{p}}{(q-p) \sqrt{q} } =  \frac{(p+q)}{2(p-q)^2 \sqrt{p q}} \;
\eea
one gets 
\bea 
&& \partial_{x^2} {\cal B}(x, \beta \tilde \mu) = 
- \frac{(\beta \tilde \mu)^2}{4 \pi^2}  
 \int_{0}^\infty 
\frac{dp}{\sqrt{p}} \frac{1}{1+e^{\beta \tilde \mu(x^2+p-1)}} \int_{0}^\infty dq \frac{\sqrt{q}}{q-p} \frac{1}{1+ e^{\beta \tilde \mu(q+x^2-1)}} \\
&& + \frac{(\beta \tilde \mu)^2}{4 \pi^2} 
 \int_{0}^\infty 
dp 
\frac{1}{1+e^{\beta \tilde \mu(x^2+p-1)}} 
\int_{0}^\infty dq 
\frac{\sqrt{p}}{(q-p) \sqrt{q} } \frac{1}{1+ e^{\beta \tilde \mu(q+x^2-1)}}
\eea 
Combining both terms using that $\frac{\sqrt{q}}{\sqrt{p}} - \frac{\sqrt{p}}{\sqrt{q}}
= \frac{q-p}{\sqrt{p q}}$ we see that the last two integrals decouple
and one obtains 
\bea 
&& \partial_{x^2} {\cal B}(x, \beta \tilde \mu) 
= - \frac{(\beta \tilde \mu)^2}{4 \pi^2} 
 \bigg[ \int_{0}^\infty 
\frac{dp}{\sqrt{p}} \frac{1}{1+e^{\beta \tilde \mu(x^2+p-1)}} \bigg]^2 
= - \frac{\beta \tilde \mu}{4 \pi} \bigg[ {\rm Li}_{1/2}(- e^{\beta \tilde \mu(1-x^2)})
\bigg]^2 
\eea 

Hence we obtain
\be 
{\cal B}(x, \beta \tilde \mu) = \frac{\beta \tilde \mu}{4 \pi} 
\int_{x^2}^{+\infty} dw   \left( {\rm Li}_{1/2}(- e^{\beta \tilde \mu(1-w)}) \right)^2 \;.
\ee 


\subsubsection{Hole probability for $d>1$ at large $R$, and the 
$d$-dimensional Widom formula}

The goal is now to obtain the asymptotics of $P(R,T)$ in $d$ dimensions from the angular decomposition, which reads
\be  \label{P2_Fnu}
\log P(R,T) = \log P^{(\ell=0)}(R,T) + \sum_{\ell \geq 1} g_d(\ell) \log P^{(\ell)}(R,T) \quad , \quad \nu = \ell + \frac{d}{2} - 1 \;,
\ee 
It turns out that the sum is dominated by large values of $\nu$, hence we will need the 
asymptotics 
we can write
\be \label{gdasymptnu}
g_d(\ell)  = \frac{2 \nu  \Gamma \left(\frac{d}{2}+\nu
   -1\right)}{\Gamma (d-1) \Gamma
   \left(-\frac{d}{2}+\nu +2\right)} = \frac{2 \nu^{d-2}}{\Gamma(d-1)} 
  ( 1-\frac{(d-4) (d-3) (d-2)}{24 \nu ^2}+ O(\nu^{-4})  ) 
\ee 
We use this formula as an analytic continuation for $g_d(\ell)$ (we restrict here to $d>1$),
and note that it is analytic at $\ell=1$ but not at $\ell=0$.

We now consider separately the cases $d>2$ and $d=2$. Let us define the 
functions $f_{\rm EML}(\ell)$ and $\tilde f_{\rm EML}(\ell)$
as 
\bea 
f_{\rm EML}(\ell) = - 2 (E_{\nu,0} + E_{\nu,1} + E_{\nu,2} ) \quad , \quad 
\tilde f_{\rm EML}(\ell) = g_d(\ell) f_{\rm EML}(\ell) \quad , \quad 
\nu = \ell + \frac{d}{2} - 1 
\eea 
where $E_{\nu,j}$, $j=0,1,2$ are defined in Eqs. \eqref{def_E0}, \eqref{E1_new}, \eqref{E2_new}.

\vspace*{0.5cm}
\noindent{\bf The case $d>2$.} In this case we use the following Euler-Mac Laurin formula to transform the discrete sum over $\ell \geq 1$ in Eq. (\ref{P2_Fnu}) into an integral. 
This formula, for a function $\tilde f_{\rm EML}(\ell)$ which vanishes at infinity, as well as its derivatives, reads
\bea \label{P_EML1}
\sum_{\ell \geq 1}^\infty \tilde f_{\rm EML}(\ell) =  \int_1^\infty \tilde f_{\rm EML}(\ell) d\ell + \frac{1}{2} \tilde f_{\rm EML}(1)-  \sum_{k=1}^\infty \frac{B_{2k}}{(2k)!} \tilde f^{(2k-1)}_{\rm EML}(1) \;,
\eea
where $B_{k}$'s are the Bernoulli numbers, with $B_2 = 1/6$, $B_4 = -1/30$, and so on.
Let us start by the evaluation of the integral in \eqref{P_EML1}.
It turns out that this integral is dominated by the region $\nu \sim R$ hence we
can simply insert the scaling form \eqref{double_scaling}. This leads to 
\bea  \label{intAB}
&& \int_1^\infty f_{\rm EML}(\ell) d\ell 
= \frac{2}{ \Gamma(d-1)} \bigg( (R \sqrt{2 \tilde \mu})^d
\int_{0}^{+\infty} dy y^{d-2} 
 {\cal A}\left(y, \beta \tilde \mu \right) + (R \sqrt{2 \tilde \mu})^{d-1}
 \int_{0}^{+\infty} dy y^{d-2} {\cal B}\left(y,  \beta \tilde \mu \right)  \bigg) + o(R^{d-1}) \nn  
\eea 
We have used that (i) 
the correction term in \eqref{gdasymptnu} leads to a subdominant contribution
(ii) there is an additional correction $O(R)$
from the lower bound in the integral in the first term (which is $d/(2 R \sqrt{2 \tilde \mu})$), which is again subdominant for $d>2$. In addition we must estimate the boundary terms: the first is the term with $\ell=0$ in \eqref{P2_Fnu} and the second is the 
$\ell=1$ term in the Euler Mac Laurin formula \eqref{P_EML1}. 
From the results at large $R$ but fixed $\ell$ in \eqref{res_Pl} 
we see that they are both $O(R)$, hence also subdominant for $d>2$.
Finally the last term in \eqref{P_EML1} is also subdominant.

\vspace*{0.5cm}
\noindent{\bf The case $d=2$.} In this case one has $g_d(0)=1$
and $g_d(\ell \geq 1)=2$ so we can use another version of the Euler-Mac Laurin formula 
which reads
\bea \label{P_EML3}
f_{\rm EML}(0) + 2\sum_{\ell \geq 1}^\infty f_{\rm EML}(\ell) =  2\int_0^\infty f_{\rm EML}(\ell) d\ell -  2\sum_{k=1}^\infty \frac{B_{2k}}{(2k)!} f^{(2k-1)}_{\rm EML}(0) \;.
\eea
One can again show that the integral is dominated by the values of $\nu = \ell \sim R$
leading to the same result as in \eqref{intAB} setting $d=2$, and that 
the correction term (the last term in \eqref{P_EML3}) is $o(1)$. 

In conclusion we have shown that for $d \geq 2$ the hole probability
behaves at large $R$ and fixed $\beta \tilde \mu$ as
\bea \label{logP_final}
&& \log P(R,T) = \frac{2}{ \Gamma(d-1)} \bigg( (R \sqrt{2 \tilde \mu})^d I_{\cal{A}}
 + (R \sqrt{2 \tilde \mu})^{d-1} I_{\cal{B}} \bigg) + o(R^{d-1})  \\
&& I_{\cal{A}} = \int_{0}^{+\infty} dy y^{d-2} {\cal A}\left(y, \beta \tilde \mu \right) 
\quad , \quad I_{\cal{B}} = \int_{0}^{+\infty} dy y^{d-2} {\cal B}\left(y, \beta \tilde \mu \right) 
\eea 
which gives the leading and subleading terms in the large $R$ expansion,
where ${\cal A}$ and ${\cal B}$ are given in 

We now show that they coincide with the leading and subleading terms in the large $R$ expansion of the Widom formula (setting $s=+\infty$ in \eqref{eq:Widom1}, \eqref{eq:Widom2}), which
were studied in detail in Section \ref{sec:widom}. 

Let us start with the leading term. One has 
\bea 
&& I_{\cal{A}} 
 = - \int_{0}^{+\infty} dy \, y^{d-2} 
 \int_{y^2}^\infty \frac{dq}{2 \pi q} \sqrt{q-y^2} \log\left(1 + e^{-\beta \tilde \mu(q-1)} \right) \\
&& = - \frac{\sqrt{\pi }  \Gamma
   \left(\frac{d-1}{2}\right)}{4 \Gamma
   \left(\frac{d}{2}+1\right)} 
\int_{0}^\infty  \frac{dq}{2 \pi q}  q^{d/2} \log(1 + e^{-\beta \tilde \mu(q-1)} ) 
\eea 
where we have interchanged the order of integration
and used the identity to perform the integration over $y$
\be 
\int_{0}^{\sqrt{q}} dy y^{d-2} \sqrt{q-y^2}  = \frac{\sqrt{\pi } q^{d/2} \Gamma
   \left(\frac{d-1}{2}\right)}{4 \Gamma
   \left(\frac{d}{2}+1\right)} 
\ee 
Performing the change of variable $q=k^2/(2 \tilde \mu \beta)$
the leading term of the hole probability becomes 
\bea 
\log P(R,T) = - \frac{2^{-d+1}}{\Gamma \left(\frac{d}{2}+1\right)
   \Gamma \left(\frac{d}{2}\right)} (R \sqrt{T})^d 
\int_0^{+\infty} dk k^{d-1} 
\log\left(1 + e^{ \beta \tilde \mu-\frac{k^2}{2}} \right) + O(R^{d-1}) 
\eea 
which agrees with the leading term of the Widom formula given in Eqs. \eqref{expansionlogP1} and \eqref{leading_Widom} 
using $V_d=\pi^{d/2}/\Gamma(1+d/2)$ and $S_d=d V_d$.

Let us consider now the subleading term in (\ref{logP_final}). One has
\be 
I_{\cal{B}} = \frac{\beta \tilde \mu}{4 \pi}
\int_{0}^{+\infty} dy y^{d-2}  
\int_{y^2}^{+\infty} dw   \left( {\rm Li}_{1/2}(- e^{\beta \tilde \mu(1-w)}) \right)^2 \; \\
= \frac{1}{4 \pi (d-1) (\beta \tilde \mu)^{\frac{d-1}{2}}}
 \int_{0}^{+\infty} dv v^{\frac{d-1}{2}} 
 \left( {\rm Li}_{1/2}(- e^{\beta \tilde \mu-v}) \right)^2
\ee 
where in the second equality we have performed an integration by part,
and the change of variable $w= v/(\beta \tilde \mu)$. Hence we obtain that
the subleading term in \eqref{logP_final} becomes 
\bea \label{logP_final}
&& \log P(R,T)|_{\rm subleading}  
= \frac{2^\frac{d-3}{2}}{\pi \Gamma(d)} (R \sqrt{T} )^{d-1}
 \int_{0}^{+\infty} dv v^{\frac{d-1}{2}} 
 {\rm Li}_{1/2}(- e^{\beta \tilde \mu-v} )^2
\eea 
We see that this expression is identical to Eqs. \eqref{expansionlogP1} and (\ref{expl_Bd}) obtained from Widom's formula.

\subsection{Low temperature regime for the Bessel process: fixed $\nu=O(1)$,
$u=z/(\beta \mu)$, $\beta \mu \sim  z= R \sqrt{2 \mu} \to + \infty$}\label{sec:lowT_d}

We now analyse the low temperature regime, for $z = R \sqrt{2 \mu}$
and $\beta \mu \to \infty$, keeping $u=z/(\beta \mu)$ fixed. 
There are two subcases (i) $\nu$ fixed studied in the present section, and (ii) for $\nu \to \infty$ 
keeping $\nu/z$ fixed, studied in the next section.
In the whole subsection since we study
the low temperature regime we can replace everywhere $\tilde \mu=\mu$.
We will use the variable $z= R \sqrt{2 \mu}$ below instead of $R$.

\subsubsection{Optimal density for the Bessel process at fixed $\nu$: low temperature regime}

Let us start with the equation which determines $a$, the edge of the support \eqref{eq_for_a}. In the limit $\beta \mu \to +\infty$ the Fermi factor
becomes a step function and the equation simplifies into
\bea \label{eq_for_a32}
1 - \frac{\nu}{\sqrt{a}} = \frac{\beta \mu}{\pi z^2} \int_{a}^\infty dp' \frac{1}{\sqrt{p'-a}} \frac{1}{e^{\beta \mu(p'/z^2 - 1)}+1} 
\to 
\frac{\beta \mu}{\pi z^2} \int_{a}^{z^2} dp' \frac{1}{\sqrt{p'-a}} 
\eea
Let us denote $z = \beta \mu u$, with $u=O(1)$, i.e. both $z$ and $\beta \mu$ large, 
and recall that $a \geq \nu^2$.
The value of $a$ is thus determined by the equation
\be \label{eqanew}
\frac{2}{\pi u} \sqrt{1 - \frac{a}{z^2}} = 1 - \frac{\nu}{\sqrt{a}} \quad , \quad 
u = \frac{z}{\beta \mu}
\ee 

From now on, we focus in this subsection on the case $\nu=O(1)$. 
There is a change of behavior in the solution to \eqref{eqanew}
as a function of the parameter $u$. 
Let us first assume $a=O(1)$ in which case one finds  
\be 
a = \nu^2 (\frac{\pi u}{\pi u - 2})^2  \quad , \quad u > \frac{2}{\pi} 
\ee 
However, for $u < \frac{2}{\pi}$ this solution fails and
one finds that $a$ scales as $a \sim z^2$. One
then obtains 
\be  
\frac{a}{z^2} = 1 - \frac{\pi^2 u^2}{4} \quad , \quad u < \frac{2}{\pi}
\ee 
This implies that in the limit $z \to \infty$
\begin{eqnarray} \label{alowT}
\tilde a = \frac{a}{z^2} = 
\begin{cases}
& 1 - \dfrac{\pi^2 u^2}{4} \quad, \quad u < \frac{2}{\pi} \\
& \\
& 0 \quad, \quad \quad \hspace*{0.9cm} u > \frac{2}{\pi} \;,
\end{cases}
\end{eqnarray}
which yields the result for the support of the density announced in the text. 

Furthermore, the density profile is given in Eq. (\ref{total_density}).
Taking the limit of large $\beta \mu$ and replacing the Fermi factor,
we obtain, in the scaled variables, for $p \geq a$
\bea \label{total_density2}
\rho^*(p) \simeq -\frac{\beta \mu}{2 \pi^2 z^2} \frac{1}{\sqrt{p-a}} \dashint_a^{z^2} dp' \frac{\sqrt{p'-a}}{p'-p}  + \frac{1}{2 \pi p} \frac{p - \nu \sqrt{a}}{\sqrt{p-a}} 
\eea
Let us consider this function in the limit $z \to \infty$
and in the region $p = \tilde p\, z^2$ with $\tilde p$ fixed.
Using the formula 
\begin{eqnarray} \label{id_integral}
\dashint_{\tilde a}^1 dp' \frac{\sqrt{p'-\tilde  a}}{p'-\tilde p}
=
\begin{cases}
&2 \sqrt{1-\tilde  a} - 2 \sqrt{\tilde p - \tilde a} \, {\rm tanh}^{-1}\left( \sqrt{\frac{\tilde p - \tilde a}{1-\tilde a}}\right) \quad, \quad \tilde a < \tilde p < 1 \\
& 2 \sqrt{1-\tilde  a} - 2 \sqrt{\tilde p - \tilde a} \, {\rm tanh}^{-1}\left( \sqrt{\frac{1 - \tilde a}{\tilde p - \tilde a}}\right) \quad, \quad \tilde p > 1
\end{cases}
\end{eqnarray}
one finds that $\hat \rho^*(p)= \rho^*(p)- \frac{1}{2 \pi p} \sqrt{(p-\nu^2)_+}$
takes the scaling form
\begin{eqnarray} \label{rho_hat_explicit}
\hat \rho^*(p) \sim \frac{1}{z} \hat r\left( \frac{p}{z^2}\right) \quad, \quad \hat r(\tilde p) = 
\begin{cases}
&- \frac{1}{2\pi \sqrt{\tilde p}} \quad, \quad \hspace*{3.8cm} 0<\tilde p < \tilde a \\
& \\
&- \frac{1}{2\pi \sqrt{\tilde p}} + \frac{1}{\pi^2 u} {\rm tanh}^{-1}\left( \sqrt{\frac{\tilde p-\tilde a}{1-\tilde a}}\right) \quad, \quad \tilde a < \tilde p < 1\\
& \\
& - \frac{1}{2\pi \sqrt{\tilde p}} + \frac{1}{\pi^2 u} {\rm tanh}^{-1}\left( \sqrt{\frac{1-\tilde a}{\tilde p-\tilde a}}\right) \quad, \quad \tilde p > 1
\end{cases}
\end{eqnarray}
where the above expression for $\hat r(\tilde p)$ is valid for $u < \frac{2}{\pi}$,
and we have used the equation \eqref{alowT} for $u < \frac{2}{\pi}$.

For $u> 2/\pi$ one has $a=O(1)$ and one can use instead \eqref{rho_tilde}.
Taking the limit $\beta \tilde \mu \to +\infty$ in that equation
and using \eqref{id_integral} for $\tilde a=0$
one finds
\begin{eqnarray} \label{rho_hat_explicit22}
 \hat r(\tilde p) = 
\begin{cases}
&- \frac{1}{\pi^2 u \sqrt{\tilde p}} + \frac{1}{\pi^2 u} \, {\rm tanh}^{-1}\left( \sqrt{\tilde p}\right) \quad, \quad 0 < \tilde p < 1\\
& \\
& - \frac{1}{\pi^2 u \sqrt{\tilde p}} + \frac{1}{\pi^2 u} \, {\rm tanh}^{-1}\left( \sqrt{\frac{1}{\tilde p}}\right) \quad, \quad \tilde p > 1
\end{cases}
\end{eqnarray} 
The two expressions match correctly at $u=2/\pi$ with $\tilde a=0$.
The equations \eqref{rho_hat_explicit} and \eqref{rho_hat_explicit22}
for the scaled function $r(\tilde p)$ of the density $\rho^*(p)$ are summarized in the main text.

\subsubsection{Hole probability for the Bessel process for fixed $\nu$ at low temperature}

We now compute the optimal energy from \eqref{energy_sp_nu2}, which reads in the scaled 
variables
\be  \label{energy_sp_nu22}
2 E_{0,\nu}  = \int_{\nu^2}^a dp \int_{\nu^2}^\infty dp' \ln|p-p'|  \rho_0(p) \hat \rho^*(p') +  \int_{a}^\infty dp \left( \rho_0(p) + \frac{1}{2}\hat \rho^*(p)\right) \log( 1 + e^{- \beta \mu (\frac{p}{z^2} - 1) } ) 
\ee
Let us recall that $\int_{\nu^2}^\infty dp' \hat \rho(p')=0$ (see Eq. (\ref{conservation})).
Consider now the limit $\beta \mu =z/u \to +\infty$, at fixed $u$.
One can rescale $p = z^2 \tilde p$ and $p' = z^2 \tilde p'$ 
and replace 
$\log( 1 + e^{- \beta \mu (\frac{p}{z^2} - 1) } ) \to \frac{z}{u} (1-\tilde p)_+$.
Using that the density takes the scaling form \eqref{rho_hat_explicit}, 
and replacing $\rho_0(p) \simeq \frac{1}{2 \pi z \sqrt{\tilde p}}$ for $\tilde p>0$,
we 
obtain 
\bea \label{2E_scaling_1}
2 E_{0,\nu} &\simeq& z^2 \bigg( \frac{1}{2 \pi} \int_{0}^{\tilde a} d\tilde p \int_{0}^\infty d\tilde p' \ln|\tilde p-\tilde p'| 
\frac{\hat r(\tilde p')}{\sqrt{\tilde p}} 
+  \frac{1}{u}  \int_{\tilde a}^{1} d\tilde p ( \frac{1}{2 \pi \sqrt{\tilde p}} + \frac{1}{2} \hat r(\tilde p)) 
(1-\tilde p)     \quad , \quad u < 2/\pi \label{um_Bessel}\\
&\simeq&   \frac{z^2}{u}  \int_{0}^{1} d\tilde p \left( \frac{1}{2 \pi \sqrt{\tilde p}} + \frac{1}{2} \hat r(\tilde p)\right) 
(1-\tilde p)    \quad , \quad u > 2/\pi  \label{up_Bessel}
\eea 


\vspace*{0.5cm}
\noindent{\bf The case $u>2/\pi$}. Let us first consider the case $u> 2/\pi$. The integrals in (\ref{up_Bessel}) can be easily evaluated and one finds the low temperature scaling 
form for the hole probability 
\bea \label{lowThighTBessel}
&& \log P^{(\ell)}(R,T) = - 2 E_{0,\nu} \simeq  - z^2 \Phi_{\rm Be,+}(u) \\
&& \Phi_{\rm Be,+}(u) = \frac{2}{ 3 \pi u} - \frac{1}{2 \pi^2 u^2}
\eea  
One notes that $\Phi_{\rm Be,+}(u)= \frac{1}{2} \Phi_{1,+}(u)$
where $\Phi_{1,+}(u)$ was obtained in \eqref{phip_1d}.
This result is probably related to the fact that the Bessel process
is a half-axis 1$d$ DPP, and the dependence in $\nu$, when $\nu=O(1)$ appears
to be subdominant in that low temperature regime. 

One can check that this matches the low temperature limit of 
the high temperature expression obtained in \eqref{res_Pl}.
Indeed in the low temperature limit \eqref{res_Pl}
becomes (upon rescaling $q=\beta \mu \tilde q$,$q'=\beta \mu \tilde q'$)
and $\beta \tilde \mu=z/u \gg 1$
\bea 
&& \log P^{(\ell)}(R,T)  \simeq  -\frac{z^2}{u} \frac{1}{2\pi} \int_0^1 d\tilde q\, \frac{1-\tilde q}{\sqrt{\tilde q}}
+\frac{z^2}{4 \pi^2 u^2}\int_0^1 d\tilde q (1-\tilde q) 
\dashint \frac{d \tilde q'}{\tilde q'-\tilde q} \frac{\sqrt{\tilde q'}}{\sqrt{\tilde q}}
\eea 
Performing the integrals one recovers \eqref{lowThighTBessel}.

\vspace*{0.5cm}
\noindent{\bf The case $u \leq 2/\pi$}. We start from the expression in \eqref{2E_scaling_1}.
Let us evaluate the first term. 
Upon integration by part over $\tilde p$ one has
\bea \label{alternative1}
&& \frac{1}{2 \pi} \int_{0}^{\tilde a} d\tilde p \int_{0}^\infty d\tilde p' \ln|\tilde p-\tilde p'| 
\frac{\hat r(\tilde p')}{\sqrt{\tilde p}} \\
&& = \frac{1}{\pi} \bigg( \sqrt{\tilde a} \int_{0}^\infty d\tilde p' \ln|\tilde a-\tilde p'| \hat r(\tilde p') 
- \int_{0}^{\tilde a} d\tilde p \sqrt{\tilde p} \dashint_{0}^\infty d\tilde p' 
\frac{\hat r(\tilde p')}{\tilde p- \tilde p'}  
\bigg) \nn \\
&& = \frac{1}{\pi} \bigg( \sqrt{\tilde a} (1- \tilde a)_+ \frac{1}{2 u} 
- \int_{0}^{\tilde a} d\tilde p \sqrt{\tilde p} \dashint_{0}^\infty d\tilde p' 
\frac{\hat r(\tilde p')}{\tilde p- \tilde p'}  
\bigg)  \nn
\eea 
We have used the saddle point equation \eqref{saddle1nu},
which in the low temperature limit becomes,
for $\tilde p \geq \tilde a$,
\bea \label{sp_a03}
  \int_{0}^\infty d\tilde p' \ln|\tilde p-\tilde p'| \hat r(\tilde p') = 
\frac{1}{2 u} (1- \tilde p)_+ 
\eea
and we have used it here for $\tilde p=\tilde a$ to evaluate the first integral in the second line of Eq. (\ref{alternative1}).

To proceed it is convenient to go back to \eqref{total_density2} 
and take the low temperature limit. This yields the 
following integral representation for $\hat r(\tilde p)$
and $\tilde p > \tilde a$
\bea 
\hat r(\tilde p)= 
\frac{1}{2 \pi \sqrt{\tilde p-\tilde a}}  \bigg( 1 - \frac{1}{\pi u} 
\dashint_{\tilde a}^1 d \tilde p' 
\frac{ \sqrt{\tilde p'- \tilde a}}{\tilde p'- \tilde p} \bigg) 
- \frac{1}{2 \pi \sqrt{\tilde p}} \quad , \quad \tilde p > \tilde a
\eea 
valid for $u \leq 2/\pi$, but which is also valid for $u>2/\pi$ upon setting $\tilde a=0$.
Note that for $\tilde p<\tilde a$ one has $\hat r(\tilde p)=-1/(2 \pi \sqrt{\tilde p})$.

We now evaluate the integral in the third line of \eqref{alternative1}.
One needs to compute for $\tilde p < \tilde a$
\bea 
&& \dashint_{0}^\infty d\tilde p' 
\frac{\hat r(\tilde p')}{\tilde p- \tilde p'}  
= - \frac{1}{2 \pi} \dashint_{0}^{\tilde a} d\tilde p' 
\frac{1}{(\tilde p- \tilde p') \sqrt{\tilde p'}} 
+ \int_{\tilde a}^\infty d\tilde p' 
\frac{\hat r(\tilde p')}{\tilde p- \tilde p'} \\
&& = - \frac{1}{2 \pi \sqrt{\tilde p}} \log( \frac{\sqrt{\tilde a} + \sqrt{\tilde p}}{\sqrt{\tilde a} - \sqrt{\tilde p}}) 
+ \int_{\tilde a}^\infty d\tilde p' 
\frac{\hat r(\tilde p')}{\tilde p- \tilde p'} \label{firstfirst}
\eea 

\bea 
\int_{\tilde a}^\infty d\tilde p' 
\frac{\hat r(\tilde p')}{\tilde p- \tilde p'} 
= \int_{\tilde a}^\infty d\tilde p' 
\frac{1}{\tilde p- \tilde p'}  ( \frac{1}{2 \pi \sqrt{\tilde p'-\tilde a}} - 
\frac{1}{2 \pi \sqrt{\tilde p'}} ) 
- \frac{1}{\pi u} \int_{\tilde a}^\infty d\tilde p' 
\frac{1}{\tilde p- \tilde p'} 
\frac{1}{2 \pi \sqrt{\tilde p'-\tilde a}}  
\dashint_{\tilde a}^1 d \tilde p'' 
\frac{ \sqrt{\tilde p''- \tilde a}}{\tilde p''- \tilde p'} 
\eea 
The first term simplifies with the first term in \eqref{firstfirst} and one gets
\bea 
&& \dashint_{0}^\infty d\tilde p' 
\frac{\hat r(\tilde p')}{\tilde p- \tilde p'}  
= - \frac{1}{2 \sqrt{\tilde a - \tilde p}} - \frac{1}{\pi u} \int_{\tilde a}^\infty d\tilde p' 
\frac{1}{\tilde p- \tilde p'} 
\frac{1}{2 \pi \sqrt{\tilde p'-\tilde a}}  
\dashint_{\tilde a}^1 d \tilde p'' 
\frac{ \sqrt{\tilde p''- \tilde a}}{\tilde p''- \tilde p'}  
\eea     
In the second term we first perform the integral over $\tilde p'$.
One obtains, for $\tilde p<\tilde a<\tilde p''$
\bea 
\int_{\tilde a}^\infty d\tilde p' 
\frac{1}{\tilde p- \tilde p'} 
\frac{1}{\sqrt{\tilde p'-\tilde a}}  
\frac{ 1}{\tilde p''- \tilde p'}   = 
\frac{\pi}{\sqrt{\tilde a - \tilde p} (\tilde p - \tilde p'')} 
\eea 
This leads to 
\bea \label{rhat1}
\int_{0}^\infty d\tilde p' 
\frac{\hat r(\tilde p')}{\tilde p- \tilde p'} 
=   - \frac{1}{2 \sqrt{\tilde a - \tilde p}} - \frac{1}{2 \pi u} 
\int_{\tilde a}^1 d \tilde p'' 
\sqrt{\tilde p''- \tilde a}    \frac{1}{ \sqrt{\tilde a - \tilde p} (\tilde p - \tilde p'')} 
\eea 
We then perform the final integral over $\tilde p''$ for $\tilde p < \tilde a$
\bea  \label{rhat2}
\int_{\tilde a}^1 d \tilde p'' 
\sqrt{\tilde p''- \tilde a}    \frac{1}{  (\tilde p - \tilde p'')} = 
2 \sqrt{\tilde a-\tilde p} \tan^{-1}\left(\sqrt{\frac{1-\tilde a}{\tilde a-\tilde p}}\right)-2 \sqrt{1-\tilde a} 
\eea 
Finally, combining Eqs. (\ref{rhat1}) and (\ref{rhat2}), and using the relation between $u$ and $\tilde a$ in \eqref{alowT} which
leads to a cancellation,
one finds
\bea \label{rhat3}
\int_{0}^\infty d\tilde p' 
\frac{\hat r(\tilde p')}{\tilde p- \tilde p'} = - \frac{1}{\pi u}\tan^{-1}\left(\sqrt{\frac{1-\tilde a}{\tilde a-\tilde p}}\right)= - \frac{1}{\pi u} {\sin}^{-1}\left(\sqrt{\frac{1-\tilde a}{1-\tilde p}}\right)  
\eea

Injecting this in Eq. (\ref{alternative1}) and performing the remaining integral over $\tilde p$ one finds 
\bea \label{alternative2}
\frac{1}{2 \pi} \int_{0}^{\tilde a} d\tilde p \int_{0}^\infty d\tilde p' \ln|\tilde p-\tilde p'| 
\frac{\hat r(\tilde p')}{\sqrt{\tilde p}} = \frac{1}{\pi u} \left[\frac{\sqrt{\tilde a}}{2}(1-\tilde a) + \frac{1}{6} \left( 2 \tilde a^{3/2} - 2 + 2\sqrt{1-\tilde a} + \tilde a \sqrt{1-\tilde a}\right) \right]
\eea
with $u = (2/\pi) \sqrt{1 - \tilde a}$. 


The second term in the energy in Eq. (\ref{um_Bessel}) reads
\begin{eqnarray}\label{E_second}
\frac{1}{u}  \int_{\tilde a}^{1} d\tilde p ( \frac{1}{2 \pi \sqrt{\tilde p}} + \frac{1}{2} \hat r(\tilde p)) 
(1-\tilde p)  = \frac{\pi  \left(\sqrt{\tilde a} (\tilde a-3)+2\right) u+(\tilde a-1)^2}{6 \pi ^2 u^2} \;,
\end{eqnarray}
where we have used the explicit expression for $\hat r(\tilde p)$ for $\tilde a <\tilde p < 1$ given in the second line of Eq. (\ref{rho_hat_explicit}). Finally, summing the two contributions to the energy given in Eqs. (\ref{alternative2}) and (\ref{E_second}) one finds from Eq. (\ref{2E_scaling}), using $a = 1-\pi^2 u^2/4$
 the low temperature scaling 
form for the hole probability of the Bessel process at fixed $\nu=O(1)$
\bea \label{lowThighTBessel2}
&& \log P^{(\ell)}(R,T) = - 2 E_{0,\nu} \simeq  - z^2 \Phi_{\rm Be,-}(u) \\
&& \Phi_{\rm Be,-}(u) = \frac{1}{4} - \frac{\pi^2 u^2}{96}
\eea  
Remarkably, as noted above, these expressions for $\Phi_-(u)$ in Eq. (\ref{lowThighTBessel2}) and for $\Phi_+(u)$ in Eq. (\ref{lowThighTBessel}) found here for the finite temperature Bessel kernel are exactly half of their counterpart found for the finite temperature sine kernel given respectively in Eqs. (\ref{phim_sine}) and (\ref{phip_1d}).

\subsection{Low temperature regime for the Bessel process: $\nu,z,\beta \mu \to + \infty$, with
$\nu/z = \lambda$ and $u=z/(\beta \mu)$ fixed and free fermions in $d$ dimensions}  
\label{sec:lowT_d}

\subsubsection{Optimal density for the Bessel process for $\nu = \lambda z \to +\infty$: low temperature regime}

We start again with the equation \eqref{eqanew} which determines $a$ in the limit 
$z , \beta \mu \to \infty$ with $u=z/(\beta \mu)$ fixed, which we recall 
here 
\be \label{eqanew2}
\frac{2}{\pi u} \sqrt{1 - \frac{a}{z^2}} = 1 - \frac{\nu}{\sqrt{a}} \quad , \quad 
u = \frac{z}{\beta \mu}
\ee 
Now we study the case where $\nu = \lambda z$ with fixed $\lambda$. 
Denoting as above $\tilde a = a/z^2$, one finds that 
in the scaling limit, $\tilde a$ is determined by the following equation,
for given $(u,\lambda)$,
with $\lambda^2 \leq \tilde a \leq 1$ 
\be  \label{rel_a_u}
\frac{2}{\pi u} \sqrt{1 - \tilde a}  = 1 - \frac{\lambda}{\sqrt{\tilde a}} 
\ee 
It is a quartic equation for $\tilde a$, which can be solved but is too bulky
to analyze. 
Instead we note that it can be written as
\be
u = \frac{2}{\pi} \frac{ \sqrt{\tilde a(1-\tilde a)} }{\sqrt{\tilde a}-\lambda} 
\ee 
A detailed analysis shows that the r.h.s is a strictly decreasing function of $\tilde a \in [\lambda^2,1]$ for $\lambda \in [0,1]$. Since it diverges for $\tilde a=\lambda^2$
and vanishes for $\tilde a=1$), for each $u$ in $[0,+\infty[$ there is a unique root $0<\tilde a=\tilde a(u,\lambda)<1$. To be more precise, it is convenient to rewrite the equation as follows 
\be 
\lambda = f_u(\tilde a) = \sqrt{\tilde a} - \frac{2 \sqrt{\tilde a(1-\tilde a)}}{\pi u} \label{lambdaa} \;,
\ee 
with $f_u(0)=0$ and $f_u(1)=1$, which implies $\tilde a(u,0)=0$
and $\tilde a(u,1)=1$, and $f_u(\tilde a) \simeq (1- \frac{2}{\pi u}) \sqrt{\tilde a}$ for small $\tilde a$.
Plotting the function for various values of $u$, see Fig. \ref{Fig_fu}, one sees that
for $u>2/\pi$ the function $f_u(\tilde a)$ is increasing, hence 
$\tilde a(u,\lambda)$ increases from $0$ to $1$ as $\lambda$ increases from $0$ to $1$.
For $u<2/\pi$ one sees that $f_u(\tilde a)$ is negative at small $\tilde a$, with
a negative minimum for some $\tilde a \in [0,1]$ and becomes 
positive for $\tilde a>\tilde a_c(u)$ with 
\bea  \label{ac_of_u}
\tilde a_c(u) = 
\begin{cases}
&1 - \frac{\pi^2 u^2}{4} \quad, \quad u < \frac{2}{\pi} \;, \\
& \\
&0 \quad, \quad \quad \quad \quad u > \frac{2}{\pi} \;.
\end{cases}
\eea 
One then finds that $\tilde a(u,\lambda)$ is again an increasing function of $\lambda$,
but now it varies from $\tilde a(u,\lambda=0)=\tilde a_c(u)$ to $\tilde a(u,\lambda=1)=1$
as $\lambda$ increases from $0$ to $1$.
\begin{figure}[t]
\includegraphics[width = 0.7 \linewidth]{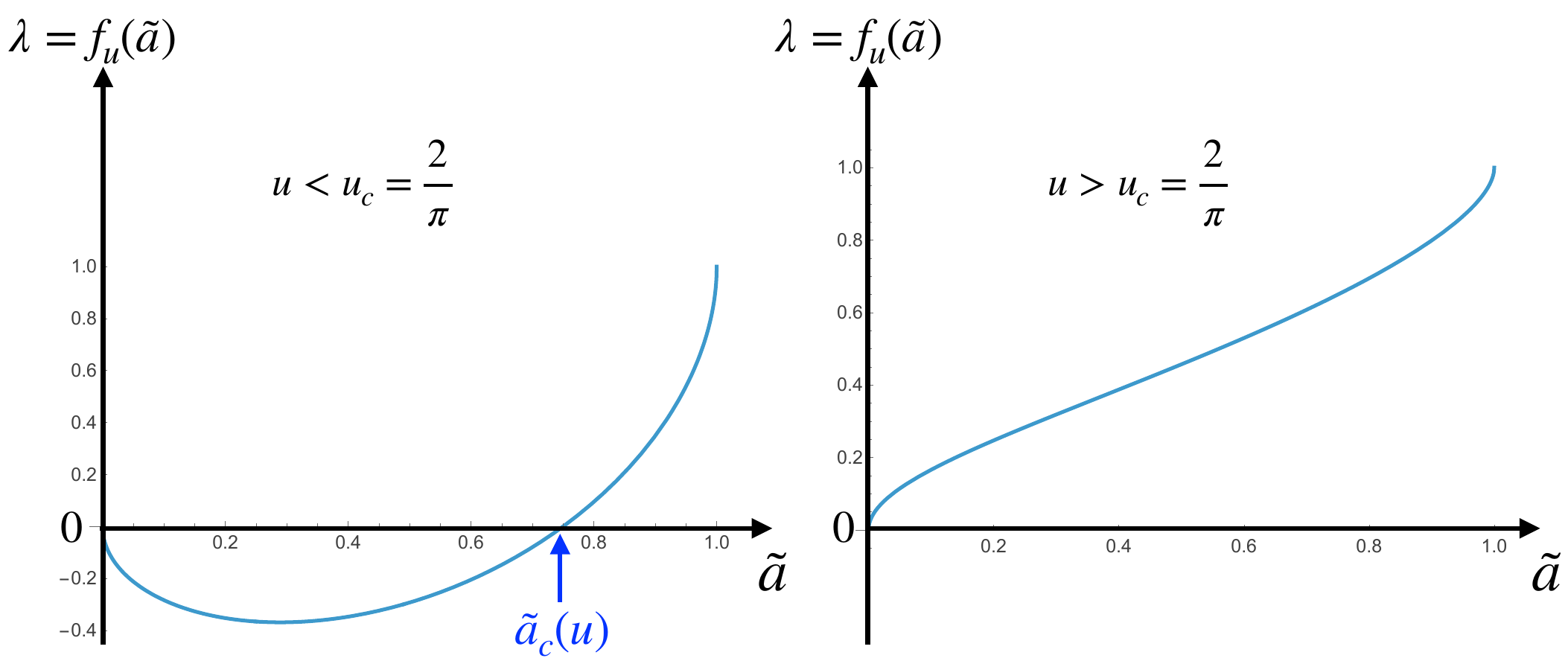}
\caption{Plot of $f_u(\tilde a)$ vs $\tilde a$ for two different values of $u$, both $u<u_c$ (left panel) and $u>u_c$ (right panel). This shows that $\tilde a=\tilde a(u,\lambda)$, the root of $\lambda=f_u(\tilde a)$, behaves differently in each phase.}\label{Fig_fu}
\end{figure}


We start again from the optimal density in Eq. \eqref{total_density2} for $p> a$
\bea \label{total_density3}
\rho^*(p) = -\frac{\beta \mu}{2 \pi^2 z^2} \frac{1}{\sqrt{p-a}} \dashint_a^{\infty} dp' \frac{\sqrt{p'-a}}{p'-p} \frac{1}{e^{\beta \mu(p'/z^2-1)}+1} + \frac{1}{2 \pi p} \frac{p - \nu \sqrt{a}}{\sqrt{p-a}} \;.
\eea
Replacing $\nu = \lambda z$, $a = z^2 \tilde a$ and taking $\beta \mu = z/u \to +\infty$,
at fixed $\lambda,u$ it takes the scaling form
\bea  \label{rho_hat_explicit2}
&& \hat \rho^*(p) \sim \frac{1}{z} \hat r\left( \frac{p}{z^2}\right) \nn \\
&& \hat r(\tilde p) = \theta(\tilde p-\tilde a) 
\bigg( \frac{1}{2 \pi \tilde p} \frac{\tilde p - \lambda \sqrt{\tilde a}}{\sqrt{\tilde p-\tilde a}}
- \frac{1}{2 \pi^2 u \sqrt{\tilde p-\tilde a}}   
\dashint_{\tilde a}^1 d \tilde p' 
\frac{ \sqrt{\tilde p'- \tilde a}}{\tilde p'- \tilde p} \bigg) 
- \theta(\tilde p-\lambda^2) \frac{\sqrt{\tilde p -  \lambda^2}}{2 \pi \tilde p}  
\eea  
Performing the integral using \eqref{id_integral} one obtain the explicit expression

\begin{eqnarray} \label{rho_hat_explicit3}
\hat r(\tilde p) = 
\begin{cases}
&- \frac{\sqrt{\tilde p - \lambda^2}}{2 \pi \tilde p}  \quad, \quad \hspace*{3.8cm} \lambda^2<\tilde p < \tilde a \\
& \\
& \frac{1}{2 \pi \tilde p} (\frac{ \lambda}{\sqrt{\tilde a}} \sqrt{\tilde p - \tilde a} 
- \sqrt{\tilde p-  \lambda^2} )
+ \frac{1}{\pi^2 u} {\rm tanh}^{-1}\left( \sqrt{\frac{\tilde p-\tilde a}{1-\tilde a}}\right) \quad, \quad \tilde a < \tilde p < 1 \\
& \\
& \frac{1}{2 \pi \tilde p} (\frac{ \lambda}{\sqrt{\tilde a}} \sqrt{\tilde p - \tilde a} 
- \sqrt{\tilde p-  \lambda^2} )
 + \frac{1}{\pi^2 u} {\rm tanh}^{-1}\left( \sqrt{\frac{1-\tilde a}{\tilde p-\tilde a}}\right) \quad, \quad \tilde p > 1
\end{cases}
\end{eqnarray}
This leads to the equations for the optimal density given in the main text.

\subsubsection{Hole probability for the Bessel process for $\nu= \lambda z \to +\infty$ at low temperature}

We now compute the optimal energy from \eqref{energy_sp_nu22} 
\be  \label{energy_sp_nu222}
2 E_{0,\nu} = \int_{\nu^2}^a dp \int_{\nu^2}^\infty dp' \ln|p-p'|  \rho_0(p) \hat \rho^*(p') +  \int_{a}^\infty dp \left( \rho_0(p) + \frac{1}{2}\hat \rho^*(p)\right) \log( 1 + e^{- \beta \mu (\frac{p}{z^2} - 1) } ) 
\ee
Let us recall that $\int_{\nu^2}^\infty dp' \hat \rho^*(p')=0$ (see Eq. (\ref{conservation})). Consider now the low temperature, large $z$ limit,
$\beta \mu =z/u \to +\infty$, at fixed $u$. 
Upon setting $\nu = \lambda z$, $a = z^2 \tilde a$
and rescaling $p = z^2 \tilde p$ and $p' = z^2 \tilde p'$,
we can replace 
$\log( 1 + e^{- \beta \mu (\frac{p}{z^2} - 1) } ) \to \frac{z}{u} (1-\tilde p)_+$.
Using that the density takes the scaling form \eqref{rho_hat_explicit2}, 
and replacing $\rho_0(p) \simeq \frac{1}{2 \pi z \sqrt{\tilde p}}$ for $\tilde p>0$,
we 
obtain 
\bea \label{2E_scaling}
2 E_{0,\nu} &\simeq& 
z^2 \bigg( \frac{1}{2 \pi} \int_{\lambda^2}^{\tilde a} d\tilde p \int_{\lambda^2}^\infty d\tilde p' \ln|\tilde p-\tilde p'| 
\frac{\hat r(\tilde p') \sqrt{\tilde p - \lambda^2}}{\tilde p}
+  \frac{1}{u}  \int_{\tilde a}^{1} d\tilde p ( \frac{\sqrt{\tilde p - \lambda^2}}{2 \pi \tilde p} + \frac{1}{2} \hat r(\tilde p)) 
(1-\tilde p) 
\bigg)  
\eea 
Note that this is valid for any value of $u$.

Let us first study the first term. Upon integration by part over $\tilde p$ it can be transformed into
\bea 
&& \frac{1}{2 \pi} \int_{\lambda^2}^{\tilde a} d\tilde p \int_{\lambda^2}^\infty d\tilde p' \ln|\tilde p-\tilde p'| 
\frac{\hat r(\tilde p') \sqrt{\tilde p - \lambda^2}}{\tilde p}  = \frac{z^2}{2 \pi} \int_{\lambda^2}^\infty d\tilde p' 
\hat r(\tilde p')  ( 2 \sqrt{\tilde a - \lambda^2} - 2 \lambda {\rm tan}^{-1}( 
\frac{\sqrt{\tilde a - \lambda^2}}{\lambda} ) \log|\tilde p' - \tilde a| \nn \\
&& 
-  \frac{z^2}{2 \pi} 
\int_{\lambda^2}^{\tilde a} d\tilde p  
( 2 \sqrt{\tilde p - \lambda^2} - 2 \lambda {\rm tan}^{-1}( \frac{\sqrt{\tilde p - \lambda^2}}{\lambda} )
\dashint_{\lambda^2}^\infty d\tilde p' 
\frac{\hat r(\tilde p')}{\tilde p- \tilde p'}   \\
&& = \frac{z^2}{2 \pi} ( 2 \sqrt{\tilde a - \lambda^2} - 2 \lambda \, {\rm tan}^{-1}( 
\frac{\sqrt{\tilde a - \lambda^2}}{\lambda} ) (1- \tilde a) \frac{1}{2 u}  -\frac{z^2}{2 \pi} 
\int_{\lambda^2}^{\tilde a} d\tilde p  
( 2 \sqrt{\tilde p - \lambda^2} - 2 \lambda \,{\rm tan}^{-1}( \frac{\sqrt{\tilde p - \lambda^2}}{\lambda} )
\dashint_{\lambda^2}^\infty d\tilde p' 
\frac{\hat r(\tilde p')}{\tilde p- \tilde p'} \nn \label{last_1}\\
\eea 
We have used the saddle point equation \eqref{saddle1nu}, which states that for any $p \in [a,+\infty[$
\bea \label{sp_a02}
2 \int_{\nu^2}^\infty dp' \ln|p-p'| \hat \rho(p') = \log \left(1 + e^{- \beta \mu(p/z^2-1)} \right) \;.
\eea
In the present low temperature scaling regime it becomes for $\tilde p \geq \tilde a$
\bea \label{sp_a03}
  \int_{\lambda^2}^\infty d\tilde p' \ln|\tilde p-\tilde p'| \hat r(\tilde p') = 
\frac{1}{2 u}\,(1- \tilde p)_+  
\eea
and above in Eq. (\ref{last_1}) we have used this relation for $\tilde p=\tilde a$. 
Let us now study the integral which appears in the second term in \eqref{last_1}.
Using Eq. \eqref{rho_hat_explicit3} we obtain for $\lambda^2 < \tilde p < \tilde a$
\bea 
&& \dashint_{\lambda^2}^\infty d\tilde p' 
\frac{\hat r(\tilde p')}{\tilde p- \tilde p'}  
= - \frac{1}{2 \pi} \dashint_{\lambda^2}^{\tilde a} d\tilde p' 
 \frac{\sqrt{\tilde p' - \lambda^2}}{(\tilde p- \tilde p') \tilde p'} 
+ \int_{\tilde a}^\infty d\tilde p' 
\frac{\hat r(\tilde p')}{\tilde p- \tilde p'} \\
&& = \frac{1}{\pi p} ( \lambda \,{\rm tan}^{-1} (\frac{\sqrt{\tilde a - \lambda^2}}{\lambda})
- \sqrt{\tilde p- \lambda^2} \, {\rm tanh}^{-1}(\frac{\sqrt{\lambda^2 - p}}{\sqrt{\lambda^2 - \tilde a}})
+ \int_{\tilde a}^\infty d\tilde p' 
\frac{\hat r(\tilde p')}{\tilde p- \tilde p'}
\eea 
Inserting the integral representation of $\hat r(\tilde p')$ 
from \eqref{rho_hat_explicit2}, the last term can be decomposed into three parts.
One has for $\lambda^2 < \tilde p < \tilde a$,
\bea 
&& \int_{\tilde a}^\infty d\tilde p' 
\frac{\hat r(\tilde p')}{\tilde p- \tilde p'} 
= B_1 + B_2 + B_3 \quad , \quad B_1= - \frac{1}{2 \pi^2 u} \int_{\tilde a}^\infty d\tilde p' 
\frac{1}{\tilde p- \tilde p'}
\frac{1}{\sqrt{\tilde p'-\tilde a}}   
\dashint_{\tilde a}^1 d \tilde p'' 
\frac{ \sqrt{\tilde p''- \tilde a}}{\tilde p''- \tilde p'} \\
&&  
B_2= \int_{\tilde a}^\infty d\tilde p' 
\frac{1}{\tilde p- \tilde p'} 
\frac{1}{2 \pi \tilde p'} \frac{\tilde p' - \lambda \sqrt{\tilde a}}{\sqrt{\tilde p'-\tilde a}}
\quad , \quad 
B_3= - \frac{1}{2 \pi} \int_{\tilde a}^{\infty} d\tilde p' 
 \frac{\sqrt{\tilde p' - \lambda^2}}{ (\tilde p- \tilde p') \tilde p'} 
\eea 
Using the following identity, valid for $\tilde p<\tilde a<\tilde p''$
\bea 
\int_{\tilde a}^\infty d\tilde p' 
\frac{1}{\tilde p- \tilde p'} 
\frac{1}{\sqrt{\tilde p'-\tilde a}}  
\frac{ 1}{\tilde p''- \tilde p'}   = 
\frac{\pi}{\sqrt{\tilde a - \tilde p} (\tilde p - \tilde p'')} 
\eea 
we obtain the explicit expressions for the above integrals 
\bea 
&& B_1= - \frac{1}{2 \pi u \sqrt{\tilde a - \tilde p}} \int_{\tilde a}^1 d \tilde p'' 
\frac{ \sqrt{\tilde p''- \tilde a}}{\tilde p -\tilde p''} 
= \frac{1}{\pi u} \left( \frac{\sqrt{1- \tilde a}}{\sqrt{\tilde a - \tilde p}} - 
{\rm tan}^{-1}( \frac{\sqrt{1- \tilde a} }{\sqrt{\tilde a - \tilde p} } )\right) \\
&& B_2 = \frac{\lambda \sqrt{\tilde a} - \lambda \sqrt{\tilde a-\tilde p} - \tilde p}{2 \tilde p \sqrt{\tilde a-\tilde p}} \\
&& B_3 = \frac{\lambda}{2 p} - \frac{\lambda}{\pi p} \,{\rm tan}^{-1} ( \frac{\sqrt{\tilde a- \lambda^2}}{\lambda} )
+ \frac{\sqrt{p - \lambda^2}}{\pi p} {\rm tanh}^{-1} ( \frac{\sqrt{p- \lambda^2}}{\sqrt{\tilde a- \lambda^2}} )
\eea 
This allows to obtain, after some simplifications
\bea 
&& \dashint_{\lambda^2}^\infty d\tilde p' 
\frac{\hat r(\tilde p')}{\tilde p- \tilde p'}  
= \frac{\lambda}{2 \tilde p \sqrt{\tilde a} } \sqrt{\tilde a-\tilde p} - \frac{1}{\pi u} \,
{\rm tan}^{-1}( \frac{\sqrt{1- \tilde a} }{\sqrt{\tilde a - \tilde p} } )
\eea 
Inserting this result into Eq. \eqref{last_1} and performing the 
remaining integral over $\tilde p$ we finally obtain the first term 
in \eqref{2E_scaling} as 
\bea \label{2E1_expl}
&& \frac{1}{2 \pi} \int_{\lambda^2}^{\tilde a} d\tilde p \int_{\lambda^2}^\infty d\tilde p' \ln|\tilde p-\tilde p'| 
\frac{\hat r(\tilde p') \sqrt{\tilde p - \lambda^2}}{\tilde p}  \nonumber \\
&& = \frac{z^2}{2 \pi} \left( 2 \sqrt{\tilde a - \lambda^2} - 2 \lambda \,{\rm tan}^{-1}( 
\frac{\sqrt{\tilde a - \lambda^2}}{\lambda} )\right) (1- \tilde a) \frac{1}{2 u}  - \frac{z^2}{2 \pi} \Bigg[\frac{\pi  \lambda \left(\sqrt{\tilde a}-\lambda\right)^2}{2 \sqrt{\tilde a}} + \pi  \lambda^2 \left(-\frac{\lambda}{\sqrt{\tilde a}}-\frac{\ln \tilde a}{2}+\ln \lambda+1\right) \nonumber \\
&& + \frac{\lambda^2 \left(2 \sqrt{\tilde a-\lambda^2}+3 \sqrt{1-\tilde a}-2 \sqrt{1-\lambda^2}\right)-\tilde a \left(2
   \sqrt{\tilde a-\lambda^2}+\sqrt{1-\tilde a}\right)-2 \sqrt{1-\tilde a}+2 \sqrt{1-\lambda^2}}{3 u} \nonumber\\
&& \frac{\lambda \left(\lambda \left(-\sqrt{\tilde a-\lambda ^2}-2 \sqrt{1-\tilde a}+\sqrt{1-\lambda ^2}\right)-\tilde a \tan
   ^{-1}\left(\frac{\lambda }{\sqrt{\tilde a-\lambda ^2}}\right)+\frac{\pi  \tilde a}{2}+\sqrt{(1-\tilde a) \tilde a}-\sin
   ^{-1}\sqrt{\tilde a}+\sin ^{-1}\lambda \right)}{u} \Bigg] \nonumber \\
\eea

Let us now consider the second term in \eqref{2E_scaling}, which we denote $2 E_2$. Inserting
the explicit expression of $\hat r(\tilde p)$ in \eqref{rho_hat_explicit3} we obtain
\bea  \label{2E_scaling2}
&& 2 E_2 \simeq   
\frac{z^2}{u}  \int_{\tilde a}^{1} d\tilde p \bigg(
\frac{\sqrt{\tilde p - \lambda^2}}{4 \pi \tilde p} + 
 \frac{1}{4 \pi \tilde p} \frac{\lambda}{\sqrt{\tilde a}} \sqrt{\tilde p - \tilde a} 
+ \frac{1}{2} \frac{1}{\pi^2 u} {\rm tanh}^{-1}\left( \sqrt{\frac{\tilde p-\tilde a}{1-\tilde a}}\right) \bigg) 
(1-\tilde p) \\
&& = \frac{z^2}{12 \pi u}  \bigg( 
2 \left(\tilde a-\lambda
   ^2\right)^{3/2}-6
   \sqrt{\tilde a-\lambda ^2}+6
   \lambda  \,{\rm cot}^{-1}\left(\frac{\lambda
   }{\sqrt{\tilde a-\lambda
   ^2}}\right) \\
   && -\frac{3
   (1-\tilde a)^{3/2} \lambda
   }{\sqrt{\tilde a}}+6 \lambda
   \sqrt{\frac{1}{\tilde a}-1}
    -6 \lambda  {\rm cos}^{-1}\left(\sqrt{\tilde a}\right)+
   (1-\tilde a)^{3/2}-2
   \left(1-\lambda
   ^2\right)^{3/2}+6
   \sqrt{1-\lambda ^2}-6
   \lambda  {\rm cos}^{-1}(\lambda
   ) \bigg) 
\eea

Combining this result with Eq. (\ref{2E1_expl}) 
major simplifications arise and we finally obtain
\bea
&&2E_{0,\nu} \simeq z^2 \phi(\lambda, \tilde a(u,\lambda)) \quad, \quad {\rm for} \quad  (1- \frac{\pi^2 u^2}{4})_+ = a_c(u) \leq \lambda \leq 1 \\
&& \phi(\lambda, \tilde a) := \frac{1}{24} \left(\frac{3 (3 \tilde a-1) \lambda^2}{\tilde a}+6 \lambda \left(\lambda \log \left(\frac{\tilde a}{\lambda^2}\right)+\frac{2
   \left(\lambda-\sqrt{\tilde a}\right) {\rm cos}^{-1}\left(\sqrt{\tilde a}\right)}{\sqrt{((1-\tilde a) \tilde a)}}\right)-\frac{2 (5 \tilde a+1)
   \lambda}{\sqrt{\tilde a}}+\tilde a+5\right) \label{phiexpression}
\eea 
where we recall that $\tilde a=\tilde a(u,\lambda)$ is the root of Eq. \eqref{lambdaa}.

Hence we have obtained the asymptotics of the hole probability of the Bessel process
in the low temperature, large $\nu \sim z$ limit
\bea \label{resultlambda}
\log P^{(\ell)}(R,T) \sim e^{- z^2 \phi(\lambda, \tilde a(u,\lambda))}
\quad , \quad u=\frac{\beta \mu}{z} = O(1) \quad , \quad 
\ell \simeq \nu = \lambda z \quad , \quad z = R \sqrt{2 \mu} \to +\infty
\eea 

We can check this formula in two limits. 
In the limit $\lambda \to 0$ one must consider separately two cases. For $u<2/\pi$
one has $\tilde a(u,\lambda) \to \tilde a_c(u)= 1 - \frac{\pi^2 u^2}{4}$ and
one finds 
\be 
\lim_{\lambda \to 0} \phi(\lambda, \tilde a(u,\lambda))
= \frac{1}{24} (5+\tilde a_c(u)) = \frac{1}{4} - \frac{\pi^2 u^2}{96} \quad , \quad u<2/\pi 
\ee 
which agrees with $\Phi_{\rm Be,-}(u)$ given in \eqref{lowThighTBessel2}.
For $u>2/\pi$ one must use that $\tilde a \simeq \lambda^2/(1- \frac{2}{\pi u})^2$
as $\lambda \to 0$ and one finds 
\be 
\lim_{\lambda \to 0} \phi(\lambda, \tilde a(u,\lambda))
 = \frac{2}{3 \pi u} - \frac{1}{2\pi^2 u^2} \quad , \quad u>2/\pi 
\ee 
which agrees with $\Phi_{\rm Be,+}(u)$ given in \eqref{lowThighTBessel}.

A second check is to consider the zero temperature limit, $u\to 0$. 
In that case one sees from \eqref{lambdaa} that $\tilde a =\tilde a(0,\lambda) \to 1$ 
and one has 
\be \label{eq:phi_lambda1}
\log P^{(\ell= \lambda z)}(R,T=0) \sim e^{- z^2 \phi(\lambda,1) } 
\quad , \quad  
\phi(\lambda,1)  
   = \theta(1-\lambda) \left( - \frac{\lambda^2}{2} \log \lambda + \frac{3}{4} \lambda^2 - \lambda + \frac{1}{4} \right) 
\ee
This result agrees with Eq. (16) in \cite{Gabriel_gap}.
It was originally derived in the context of lattice QCD in \cite{GW1980,Wadia},
not as a hole probability but as a partition function, by a completely
different Coulomb gas technique. It was also proved later in the study of the longest increasing subsequence of random permutations \cite{Joh1998} (see also \cite{KC2010}).
It is quite remarkable that it is recovered here by a different method. 
\\

\noindent{\bf Remark}. As in the remark above \eqref{kgap1}, let us again give an interpretation in the original (physical) momentum variables $k$. For a fixed angular sector $\ell$, at $T=0$ the radii $r_i$
of the free fermi gas form a well studied DPP on a half axis with "Fermi momentum" $k_F$ (in the variables $r_i^2$ it forms a Bessel DPP) \cite{BLACT2018}. Again the hole probability of the interval $[0,R]$ for this DPP identifies with the hole probability on $[0,k_F]$ of a dual DPP with 
"Fermi momentum" $R$, thanks to Eqs. \eqref{Duality_Bessel_FD} and \eqref{dualityBessel}.
At finite $T$ this extends to a linear statistics for that dual DPP, see Eq. \eqref{linstatbessel}.
In the (physical) momentum variables $k_i= \sqrt{p_i}/R$ the gap is thus, at fixed $\ell \sim \nu$
\bea 
&& k_{\rm gap}  \simeq \frac{\nu}{R (1 - \frac{2}{\pi u}) } \quad , \quad u >2/\pi \\
&& k_{\rm gap}  \simeq \sqrt{1 - \frac{\pi^2 u^2}{4}} k_F \quad , \quad u < 2/\pi
\eea 
and for $\ell \sim \nu = \lambda z$
\be
k_{\rm gap}  \simeq \sqrt{\tilde a(u,\lambda)} k_F  \;.
\ee
In that latter case the transition at $u=2/\pi$ is whether 
the gap vanishes or not (in order $O(k_F)$) as $\lambda \to 0$.

\subsection{The hole probability in dimension $d>1$ for 
$z , \beta \mu \to \infty$ with $u=z/(\beta \mu)$ fixed} \label{sec:ddimlowT}

We now return to the formula for the hole probability in dimension $d>1$.
It is obtained by summation over the angular sectors
\be  \label{P2_Fnu2}
\log P(R,T) = \log P^{(\ell=0)}(R,T) + \sum_{\ell \geq 1} g_d(\ell) \log P^{(\ell)}(R,T) \quad , \quad \nu = \ell + \frac{d}{2} - 1 \;,
\ee
where $g_d(\ell)$ is given in Eq. (\ref{holeproduct}) -- we recall that $g_d(\ell) \simeq 2\,\ell^{d-2}/\Gamma(d-1)$ for $\ell \gg 1$. 

In the low temperature, large $z= R \sqrt{2 \mu}$ regime, i.e.
$z , \beta \mu \to \infty$ with fixed $u=\beta \mu/z$,
the summation over $\ell$ is dominated by values $\ell \simeq \nu \simeq z$
and at leading order one can approximate the sum by an integral. Using
the results of the previous subsection, and replacing $\ell^{d-2} d\ell \simeq \nu^{d-2} d\nu \simeq z^{d-1} \lambda^{d-2} d\lambda$, it reads 
\be 
- \log P(R,T) \simeq  \frac{2}{\Gamma(d-1)} z^{d+1} \int_0^1 d \lambda \lambda^{d-2} 
\phi(\lambda, \tilde a(\lambda,u)) \quad , \quad 
\quad , \quad 
\lambda = f_u(\tilde a) := \sqrt{\tilde a} \left(1 - \frac{2 \sqrt{1-\tilde a}}{\pi u}\right) \;, 
\ee 
where $\phi(\lambda, \tilde a)$ is given in \eqref{phiexpression}.

One can obtain an alternative formula as follows. Since $\tilde a =\tilde a(u,\lambda)$ as a function of $\lambda$ 
is an increasing function, see Fig. \ref{Fig_fu}, one can
change variable and one obtains an 
integral over $\tilde a$ on $\tilde a \in [\tilde a_c(u),1]$, where $\tilde a_c(u)$ is given in Eq. (\ref{ac_of_u}), and one obtains the integral representation for $d>1$
\be \label{eq:Pd}
- \log P(R,T) \simeq  \frac{2}{\Gamma(d-1)} z^{d+1} \int_{\tilde a_c(u)}^1 d \tilde a f_u'(\tilde a) [f_u(\tilde a)]^{d-2} 
\phi(f_u(\tilde a), \tilde a) \quad , \quad \tilde a_c(u) = (1- \frac{\pi^2 u^2}{4})_+
\ee 
There are thus two branches and one obtains
\be \label{eq:def_phipm}
- \log P(R,T) = z^{d+1} \begin{cases} \Phi_{d,+}(u) \quad , \quad u>2/\pi  \\
\Phi_{d,-}(u) \quad , \quad u<2/\pi
\end{cases} 
\ee 
where we recall that $z= R \sqrt{2 \mu}$ and $u= z T/\mu$.

A first check can be performed on the form \eqref{P2_Fnu2} and one can verify that it recovers the known result at $T=0$, i.e. for $u=0$. In that case 
$\tilde a(\lambda,u=0)=1$ for all $\lambda \in [0,1]$
and one can use the result \eqref{eq:phi_lambda1}. Integrating over $\lambda$
we recover in general $d$
\be
\Phi_{d,-}(0)= \kappa_d=\frac{2}{(d+1)^2 \Gamma (d+1)} \;,
\ee 
in agreement with \cite{Gabriel_gap}.


We can now evaluate the integrals in \eqref{eq:Pd} in fixed dimension $d$. Surprisingly
the indefinite integrals can be computed \cite{UsToBePublished}. We give them
there only in $d=1,2,3$. The
explicit formula read

\begin{itemize} 

\item

In $d=2$ one obtains 
\bea \label{Phi_d=2}
\Phi_{2,+}(u) 
   &=& -\frac{8}{15 \pi ^2 u^2}+\frac{1}{12 \pi ^2 u^3}+\frac{1}{8 u} \\
\Phi_{2,-}(u) 
&=&  \frac{\pi  \sqrt{4-\pi ^2 u^2} \left(324 + 28 \pi ^2 u^2 - \pi ^4 u^4 \right) u+120 \left(3 \pi ^2
   u^2+2\right) {\rm sin}^{-1}\left(\frac{\pi  u}{2}\right)-768 \pi  u}{1440 \pi ^3 u^3}  \\
&=& \sqrt{4-\pi ^2 u^2} \left(-\frac{\pi ^2
   u^2}{1440}+\frac{9}{40 \pi ^2
   u^2}+\frac{7}{360}\right)-\frac{8}{15 \pi ^2
   u^2}+\left(\frac{1}{6 \pi ^3 u^3}+\frac{1}{4 \pi 
   u}\right) {\rm sin}^{-1}\left(\frac{\pi  u}{2}\right) 
\eea 

One finds that in $d=2$ the transition is of order $9/2$, i.e.
one has for $u \lesssim u_c=2/\pi$
\be 
\Phi_{2,-}(u) - \Phi_{2,+}(u) = -\frac{32}{945} \sqrt{2}
   y^{9/2} \left(1 + \frac{117 y}{44} + O(y^2) \right) 
   \quad , \quad y = 1 - \frac{\pi u}{2} >0 
\ee 

The expansion around zero temperature, $u= z T/\mu=0$, reads 
\be 
\Phi_{2,-}(u) = \frac{1}{9}-\frac{\pi ^2
   u^2}{240}+\frac{\pi ^4
   u^4}{13440}+\frac{\pi ^6
   u^6}{774144}+\frac{\pi ^8
   u^8}{16220160}+O\left(u^9\right)
\ee 
in agreement with the $T=0$ result, with $\Phi_{2,-}(0)=1/9$ 
in $d=2$~\cite{Gabriel_gap}. 

\item 
In $d=3$ one finds 
\bea \label{Phi_d=3}
\Phi_{3,+}(u) &=& -\frac{1}{6 \pi ^2
   u^2}+\frac{16}{105 \pi ^3
   u^3}-\frac{1}{18 \pi ^4
   u^4}+\frac{4}{45 \pi  u} \\
\Phi_{3,-}(u) &=& \frac{\pi ^4
   u^4}{53760}-\frac{\pi ^2
   u^2}{1440}+\frac{1}{48} \nn
\eea 
One finds that in $d=3$ the transition is of order $6$, i.e.
one has for $u \lesssim u_c=2/\pi$
\be 
\Phi_{3,-}(u) - \Phi_{3,+}(u) = \frac{y^6}{180} \left(  1+\frac{25 y}{7}+\frac{467 y^2}{56}+O(y^3) \right) 
   \quad , \quad y = 1 - \frac{\pi u}{2} >0 
\ee 
which has opposite sign as in $d=2$. The result for $\Phi_{3,-}(u)$
agrees with the $T=0$ result, with $\Phi_{3,-}(0) = 1/48$ in $d=3$~\cite{Gabriel_gap}.



\end{itemize}

We find that these integrals can also be done explicitly in higher 
integer dimensions. For any $d$, $\Phi_{d,+}(u)$ is a polynomial of the variable
$1/u$ of degree $d+1$, with alternating signs. In any $d$, $\Phi_{d,-}(u)$ is formally
an even function of $u$. For odd $d$, $\Phi_{d,-}(u)$ is a polynomial of the variable
$u$ of degree $d+1$, which furthermore is a polynomial in $u^2$. 
For even $d$, $\Phi_{d,-}(u)$ has the following form
\bea 
\Phi_{d,-}(u) = P_d\left(\frac{1}{u}\right) + u^d \sqrt{4-\pi ^2 u^2}\, Q_d\left(\frac{1}{u} \right)
+ R_d\left(\frac{1}{u}\right) {\rm sin}^{-1}\left(\frac{\pi  u}{2}\right) 
\eea 
where $P_d(X)$ has degree $d$, $Q_d(X)$ has degree $2d$ and $R_d(X)$ has degree $d+1$.
Finally we find that the degree of the singularity is $\frac{3}{2}(d+1)$, i.e. to leading order one has 
\bea \label{deltaPhi_SM}
\Phi_{d,-}(u) - \Phi_{d,+}(u) \simeq \widetilde C_d \left(\frac{2}{\pi} -u\right)^{\frac{3}{2}(d+1)} \quad, \quad u \to (2/\pi)^- \;.
\eea 
where the amplitude has the explicit form valid in any $d$ \cite{UsToBePublished} 
\bea
\widetilde C_d =
(-1)^{d-1}
\frac{\pi^{\frac{3(d+1)}{2}}}{2^{d-2}}
\frac{\Gamma\!\bigl(\frac{d-1}{2}\bigr)}
{9(d+1)(9d^2-1)\Gamma\!\bigl(\frac{3(d-1)}{2}\bigr)}
\eea

Finally let us note that the present low temperature results match the ones obtained
in the high temperature regime from the Widom formula, namely Eq. \eqref{eq:Widom_largeu}.
For instance one finds that in $d=1,2,3$
\bea  \label{eq:phip_largeu}
\Phi_{d,+}(u) = \begin{cases}  \frac{4}{3 \pi  u}-\frac{1}{\pi ^2 u^2} + O(u^{-3})  \quad , \quad d=1 \\
 \frac{1}{8 u}-\frac{8}{15 \pi ^2 u^2}  + O(u^{-3})  \quad , \quad d=2 \\
 \frac{4}{45 \pi  u}-\frac{1}{6 \pi ^2 u^2} + O(u^{-3}) \quad , \quad d=3 
\end{cases} 
\eea  
More generally the first two terms in the $1/u$ expansion of
$\Phi_+(u)$ can be obtained from the Widom formula in any $d$ (which
however is valid only in the high temperature regime). 
\\

{\bf Critical regime}. Given that we found that the transition occurs at $u_c=2/\pi$ independently of $\ell, \nu$ (and of the
dimension $d$) there is no shift in the critical point when summing over $\ell$. 
Hence we can conjecture that in any $d$, we will have in the critical regime
\bea \label{hole_TW_txt2}
P(R,T) \simeq   G_d(y) e^{- z^{d+1} \Phi_{d,+}(u) } \;,
\eea 
where the critical scaling variable $y = \frac{\pi}{2} (2 z)^{2/3}(u-u_c)$
will be the same as in $d=1$. The scaling function $G_d(y)$, however, will depend on $d$.
By matching the asymptotics within the critical regime for $y \to - \infty$ with the singularity at $u=2/\pi$ in the regime of fixed $u<2/\pi$, we must have 
\bea 
- \log G_d(y) = - \log G_d((2 z)^{2/3} \frac{\pi}{2} (u-u_c)) 
\simeq  z^{d+1} ( \Phi_{d,-}(u)- \Phi_{d,+}(u) )
\simeq \tilde C_d (z^{2/3} (u_c-u))^{\frac{3}{2} (d+1)}
\eea 
for $u$ near $u_c=2/\pi$, where in the last equation we have used
\eqref{deltaPhi_SM}. This matching argument is consistent with the scaling exponent being $2/3$ in any $d$. Furthermore, it allows to obtain the asymptotics 
\be 
- \log G_d(y) \simeq g_d |y|^{\frac{3}{2}(d+1)} \quad , \quad g_d = \tilde C_d (\frac{2}{\pi^3})^{\frac{1}{2} (d+1)} = (-1)^{d-1} 2^{(5-d)/2} \frac{\Gamma(\frac{d-1}{2})}{9(d+1)(9d^2-1)\Gamma(\frac{3(d-1)}{2})} \;.
\ee 
One can check that $\lim_{d \to 1} g_d = 1/12$ in agreement with the left tail of the Tracy-Widom distribution, which holds in $d=1$ \cite{Xu2025}.

\subsection{Comparison with Refs. \cite{Ruzza2025} and \cite{XuBessel}, and extensions: a non trivial
check of our Coulomb gas result}

In a recent mathematical work \cite{Ruzza2025} the Fredholm determinant based 
on the finite temperature Bessel kernel was studied. 
In our notation it reads
\be \label{PellSMR}
P^{(\ell)}(R,T)  = {\rm Det}( I - \Pi_{[0,R]} \hat K_\nu)  
\quad , \quad \nu = \ell + \frac{d}{2}-1 
\ee 
involving the finite $T$ Bessel kernel $\hat K_\nu$ given in Eq. (\ref{hat_BesselKernel}). In Ref. \cite{Ruzza2025} the variables $x,t, \alpha$ are used. We will follow the convention of \cite{XuBessel} and perform the change $t \to -t$ everywhere. With this convention (after this change) and the dictionary with our notations reads 
\be 
u = x/\sqrt{t} \quad , \quad z= x \sqrt{t} \quad , \quad \nu = \alpha 
\ee 
where $z=k_F R$ where $k_F=\sqrt{2 \tilde \mu}$ and $u=2 R T/k_F$. We thus replace $\alpha$ by $\nu$ below.
Beware 
that in this section the variable $t$ of \cite{Ruzza2025} denotes $t=\beta \tilde \mu$
and not $T/T_F$ as in the remainder of the paper.

One of the main result of \cite{Ruzza2025} is Theorem 1.2. Expressing the
observable in the variable $x,t$ via the definition 
$Q(x,t)  := P^{(\ell)}(R,T)$, this theorem states that 
the function
\be \label{defvxt}
v(x,t) := - \partial_x Q(x,t) - \frac{1}{2} t x 
- \frac{4 \nu^2 -1}{8 x}
\ee 
satisfies the equation
\bea \label{Ruzza0} 
(2 \partial_x v + t) (\partial_t v)^2 + \frac{1}{4} 
(\partial_x \partial_t v)^2 - \frac{1}{2} (\partial_x^2 \partial_t v)
(\partial_t v) - \frac{\nu^2}{4} = 0 
\eea 
This is a necessary condition obeyed by the function $Q(x,t)$ and the Theorem 1.2 contains
an additional (complicated) construction which allows to obtain $Q(x,t)$
and which we will not consider here. In the following we will use \eqref{Ruzza0} to put some constraints on the behavior of $P^{(\ell)}(R,T)$ in the low temperature regime. This will provide some highly non-trivial checks for our Coulomb gas calculation.

We now analyze Ruzza's equation \eqref{Ruzza0} in our low temperature regime (ii), which corresponds to $x,t \to +\infty$ with $u=x/\sqrt{t}$ fixed.
We will consider separately the case $\ell,\nu$ fixed, and the case $\ell = \lambda z \to + \infty$
at fixed $\lambda$.

\begin{itemize}

\item
{\bf case $\nu= O(1)$}. Let us write everywhere the ratio $x/\sqrt{t}=u$.
Then \eqref{defvxt} becomes 
\be \label{defv}
v(x,t) = - \partial_x \log Q(x,t) - \frac{1}{2} u t^{3/2} 
- \frac{4 \nu^2 -1}{8 u t^{1/2}}
\ee 

To search for a scaling solution of Ruzza's equation, we first recall what we expect from our study 
\be \label{seriesPl}
- \log P^{(\ell)}(R,T) = z^2 \Phi(u) + o(z^2) \;,
\ee 
where $o(z^2)$ is a function of $z$ and $u$, which, for large $z$, is uniformly 
subdominant w.r.t. $z^2$ at fixed $u$ (see below).
It implies, taking into account that $u=x/\sqrt{t}$ and $z=x \sqrt{t}$, that at large $t$
\be \label{der1}
- \partial_x \log Q(x,t) = t^{3/2} (2 u \Phi(u) + u^2 \Phi'(u)) + 
o(t^{3/2}) \;.
\ee  
Hence it is natural to search for a solution of Ruzza's equation
as a large $t$ expansion of the form
\bea \label{ansatz1}
v(x,t) = t^{3/2} U\left(\frac{x}{\sqrt{t}}\right) 
+ o(t^{1/2}) \quad {\rm where} \quad 
U(u)= 2 u \Phi(u) + u^2 \Phi'(u) - \frac{u}{2} 
\eea 

Injecting \eqref{ansatz1} into \eqref{Ruzza0} and expanding at large $t$, the leading term 
is $O(t^2)$ and leads to the equation
\be 
\left(2 U'(u)+1\right) \left(u U'(u)-3 U(u)\right)^2 = 0
\ee 
There are two solutions. The first one corresponds to our high temperature branch
\be 
U'(u)= - \frac{1}{2} \quad \Rightarrow \quad U(u)= c_+ - \frac{1}{2} u 
\quad \Rightarrow \quad 2 u \Phi(u) + u^2 \Phi'(u) = c_+
\quad \Rightarrow \quad \Phi(u)= \frac{c_+}{u} + \frac{c_1}{u^2} 
\ee 
where the constants $c_+$ and $c_1$ are a priori undetermined (one would need
additional boundary conditions to Ruzza's equation which we do not discuss here).
We simply remark that this first solution is consistent with our result $\Phi(u)=\Phi_+(u)=\frac{2}{3 \pi u} 
- \frac{1}{2 \pi^2 u^2}$, i.e. $c_+= \frac{2}{3 \pi}$, $c_1=- \frac{1}{2 \pi^2}$.

The second solution is 
\bea 
&& u U'(u)-3 U(u) = 0 \quad \Rightarrow \quad U(u)= c_- u^3 \\
&& 
\quad \Rightarrow \quad 2 u \Phi(u) + u^2 \Phi'(u) - \frac{u}{2} = c_- u^3 
\quad \Rightarrow \quad \Phi(u)= \frac{1}{4} + \frac{c_- u^2}{4} 
+ \frac{c'_1}{u^2}  
\eea 
Although $c_-$ and $c'_1$ are again undetermined, imposing that $\Phi(u)$
has a well defined $T=0$ limit implies $c'_1=0$, and we note that the
$T=0$ is then fully determined, as $\Phi(0)=1/4$. Again the
second solution is consistent with our result for the lower branch, $\Phi(u)= \Phi_-(u) = \frac{1}{4} - \frac{\pi^2 u^2}{96}$, leading to $c_-=- \frac{\pi^2}{24}$. In summary the existence
of these two solutions is consistent with the existence of a transition
at $u=u_c$, but fixing the value at $u_c=2/\pi$ requires extra knowledge.

We have looked at the equation \eqref{Ruzza0} to the two next orders in
an expansion at large $t$. It gives interesting constraints if one assumes some form for the expansion, for instance assuming
$- \log P^{(\ell)}(R,T) = z^2 \Phi(u) + z \Psi(u) + \gamma(u) + o(1)$
it gives some information on the functions $\Psi(u)$ and $\gamma(u)$,
which take different values in the two branches. However one also finds that logarithmic corrections to each of the terms in the expansions of $- \log P^{(\ell)}(R,T)$ are possible. Overall these subleading terms are compatible with the form of the low temperature limit of the high temperature regime given in
\eqref{limits2}, which actually exhibit logarithmic corrections.

{\bf Critical behavior}. 
In a recent work \cite{XuBessel} the authors focus on the critical region, i.e. large $t$, fixed $x/\sqrt{t}=u$ and $u$ very close to $u=u_c=2/\pi$. Denoting 
\be \label{us} 
u=\frac{2}{\pi} (1 - \frac{1}{2} (\frac{\pi}{t})^{2/3} s)
\ee
they show that at fixed $s$ and large $t$, the function
$v(x,t)$ defined in \eqref{defvxt} takes the form 
\be \label{asymptxu}
v(x,t) = - \frac{x t}{2} + \frac{2}{3 \pi} t^{3/2} + \pi^{1/3} t^{1/6} (-s^2/4 + \sigma_\nu(s)) + O(t^{-1/6}) 
\ee
where the function $\sigma(s)=\sigma_\nu(s)$ solves the $\sigma$-form of the second Painlev\'e equation
\bea 
(\sigma'')^2 + 4 (\sigma')^3 - 4 s (\sigma')^2 + 4 \sigma' \sigma - \nu^2 =0  
\eea 
with 
\bea \label{asymptsigma}
\sigma_\nu(s) = \begin{cases}  
&- \nu \sqrt{s} + O(1/s) \quad 
s \to +\infty \\
& s^2/4 + \frac{4 \nu^2 -1}{8 s} 
+ O(s^{-5/2}) \quad , \quad s \to -\infty \end{cases} 
\eea 
One can check that the critical behavior in \eqref{asymptxu} is compatible with the large $t$, fixed $u$ expansion studied above of Ruzza's equation \eqref{Ruzza0}
\bea  
&& v(x,t) \simeq t^{3/2} U_-(u) + t^{1/2} V_-(u) \quad , \quad 
U_-(u)=- \frac{\pi^2}{24} u^3 \quad , \quad V_-(u)= - \nu \sqrt{1- \frac{\pi^2 u^2}{4}} \quad , \quad u<2/\pi \\
&& v(x,t) \simeq t^{3/2} U_+(u) + t^{1/2} V_+(u) \quad , \quad U_+(u)=\frac{2}{3 \pi} - \frac{u}{2}  \quad , \quad V_+(u)= 0 \quad , \quad u>2/\pi 
\eea  
where we have specified the integration constants to agree with our result for $\Phi_\pm(u)$ in \eqref{lowThighTBessel} and 
\eqref{lowThighTBessel2}, and added a possible choice of the
subdominant term. Indeed using the asymptotic behavior
\eqref{asymptsigma} for $\sigma_\nu(s)$, respectively for $s \to + \infty$
($-$ branch, low temperature) and $s \to - \infty$
($+$ branch, high temperature), one can check 
the matching between the critical regime and the fixed $u$ regime.
To show this, one replaces $u$ using \eqref{us} and
expand at large $t$ with $x=u \sqrt{t}$.
This is also 
compatible with the form near criticality
\be 
P^{(\ell)}(R,T) \sim \tilde G_\nu(s) e^{ - z^2 \Phi_+(u)} 
\ee 
where $\partial_s \log \tilde G_\nu(s)= \sigma_\nu(s)-s^2/4$
and $\Phi_+(u)= \frac{2}{3 \pi u} - \frac{1}{2 \pi^2 u^2}$ was computed in \eqref{lowThighTBessel}. This is consistent
with the asymptotics $\tilde G_\nu(s) \sim e^{-s^3/12}$ for $s \to +\infty$ and
$\tilde G_\nu(s) = O(1)$ for $s \to +\infty$. 

\item 
{\bf case $\nu = \lambda z$, $z \to +\infty$, $\lambda=O(1)$}. We expect from our Coulomb gas study that at large $z$, fixed $u,\lambda$ the hole probability takes the form
\be \label{seriesPlambda}
- \log P^{(\ell)}(R,T) = - \log Q(x,t) = z^2 \Phi(u=\frac{x}{\sqrt{t}},\lambda= \frac{\nu}{x \sqrt{t}}) + o(z^2 )  \quad , \quad z = x \sqrt{t} \gg 1
\ee 
where the function $\Phi$ was obtained in \eqref{phiexpression}.
When computing $v(x,t)$ in \eqref{defv},
the derivative w.r.t. $x$ should be done at fixed $\nu$. This 
implies 
\bea 
&&  - \partial_x \log Q(x,t) 
\simeq t^{3/2} u ( 2  \Phi(u,\lambda) + u \partial_u \Phi(u,\lambda) 
- \lambda \partial_\lambda \Phi(u,\lambda)) \\
&& 
v(x,t) = - \partial_x \log Q(x,t)  - \frac{1}{2} (1 + \lambda^2 ) u t^{3/2}    
+  \frac{1}{8 u t^{1/2}} 
\eea 
We will thus search for a solution to Ruzza's equation of the form
\bea \label{ansatz3}
v(x,t) = t^{3/2} U(u=\frac{x}{\sqrt{t}}, \lambda=\frac{\nu}{x \sqrt{t}}) + o(t^{3/2})
\eea  
which generalizes \eqref{ansatz1}, and for now we do not write subleading terms.
The function $U(u,\lambda)$ and $\Phi(u,\lambda)$ are thus related via
\be \label{fromPhitoUlambda}
U(u,\lambda)= u \left(  ( 2   + u \partial_u 
- \lambda \partial_\lambda ) \Phi(u,\lambda) - \frac{1}{2} (1 + \lambda^2 ) 
\right)
\ee 
Injecting \eqref{ansatz3} into \eqref{Ruzza0} 
and expanding at large $t$, the leading term 
is $O(t^2)$ and now leads to the equation for $U(\lambda)$
\bea 
\label{partiallambda}
\left(u + 2 ( u \partial_u  - \lambda \partial_\lambda) U  \right) 
\left(- 3 U + (u \partial_u + \lambda \partial_\lambda) U \right)^2
-\lambda ^2 u^3 = 0
\eea 
We will not attempt to solve this (complicated) equation in full generality (although it can be studied using the method of characteristics), 
but we will check that the function 
$\Phi(u,\lambda)$ obtained in \eqref{resultlambda} is indeed a solution.
This is a highly non-trivial check of our Coulomb gas method, as we now discuss.

Let us recall that we obtained $\Phi(u,\lambda)$ in the following parametric form
by introducing the variable $\tilde a$ 
\be \label{ansatzPhi}
\Phi(u,\lambda)=\phi(\lambda,\tilde a) \quad , \quad 
u= \frac{2}{\pi} \frac{\sqrt{\tilde a (1-\tilde a)}}{\sqrt{\tilde a}-\lambda} \;,
\ee
where $\phi(\lambda,\tilde a)$ is the not so simple function displayed in 
\eqref{phiexpression}. Let us introduce some compact notations
and denote $D_\pm=u \partial_u \pm \lambda \partial_\lambda$. 
Introducing the shorthand function $F(u)$ such that 
\be
U = u F \quad , \quad F=(2 + D_-) \Phi - \frac{1}{2} (1+ \lambda^2) \;.
\ee
The relation in Eq.~\eqref{partiallambda} becomes 
\be \label{neweq} 
 (1 + 2 F + 2 D_- F) (-2 F + D_+ F)^2 = \lambda^2
\ee
Now we want to try a solution of the form \eqref{ansatzPhi}.
We thus need to change variables from $u,\lambda$ to $\lambda,\tilde a$.
The transformation rules for the derivatives read
\bea 
&& D_\pm = u \partial_u \pm \lambda \partial_\lambda 
= M_\pm(\lambda,\tilde a) \partial_{\tilde a} \pm \lambda \partial_\lambda \quad , \quad  M_\pm(\lambda,\tilde a) =  \frac{u}{\frac{du}{d\tilde a}|_\lambda} \pm \frac{\lambda}{\frac{d\lambda}{d\tilde a}|_u} \\
&& M_+(\lambda,\tilde a) = \frac{2 (1-\tilde a) \tilde a \left(2 \lambda- \sqrt{\tilde a}
   \right)}{\tilde a^{3/2}-2 \tilde a \lambda +\lambda } \quad , \quad 
 M_-(\lambda,\tilde a) = - \frac{2 (1-\tilde a) \tilde a^{3/2}}{\tilde a^{3/2}-2 \tilde a \lambda +\lambda }
\eea 
Amazingly, our expression \eqref{phiexpression} for $\phi(\lambda,\tilde a)$ ($\tilde a \to a$) leads to
great simplifications and one finds
\bea 
&& F = \frac{1}{6} \left(-\frac{2 (2 \tilde a+1) \lambda}{\sqrt{\tilde a}}+\tilde a-1\right) \\
&& (1 + 2 F + 2 D_- F) = \tilde a \quad , \quad  -2 F + D_+ F = \frac{\lambda }{\sqrt{\tilde a}} 
\eea 
Hence the equation \eqref{neweq} is obeyed. 
\\

\end{itemize}

\section{Variance (low temperature), cumulants and entropy (high temperature)}
\label{app:var_scaling}


\subsection{High T expression of cumulants from Widom's formula}

From \eqref{chisFD}-\eqref{eq:Widom2}, Widom's formula gives to leading order for a large domain
\be \label{chiscum}
\chi(s) 
= (R \sqrt{T})^d V({\sf K}) 
 \int_{{\mathbb R}^d} \frac{d^d {\bf k}}{(2 \pi)^d}
\left( \log (1 + e^{ \beta \tilde \mu - \frac{k^2}{2} - s} ) 
- \log (1 + e^{ \beta \tilde \mu - \frac{k^2}{2}} )
 \right)
\ee
Performing the integrals for a spherical domain, with 
$V({\sf K}) = V_d=S_d/d= \pi^{d/2}/\Gamma(1+d/2)$,
after similar manipulations as in Section \ref{sec:widom},
we obtain the $n$-th cumulant as 
\bea  
\langle {\cal N}_R^n \rangle^c = (- \partial_s)^n|_{n=0} \chi(s) 
\simeq  - (R \sqrt{T})^d \frac{1}{2^{d/2} \Gamma(1+d/2)} 
{\rm Li}_{\frac{d}{2}+1-n} 
(-e^{\beta \tilde \mu})
\eea  
where we have used that $(- \partial_s) (- {\rm Li}_a(A e^{-s})= (- {\rm Li}_{a-1}(A e^{-s}))$. This is valid to leading order for $R \sqrt{T} \gg 1$ at fixed $T=O(T_F)$. 
The variance is obtained setting $n=2$. 

In the low temperature limit, $T/T_F \ll 1$, i.e. $\beta \mu \to +\infty$, we can use the large argument asymptotics
\eqref{Upsilonasympt}, \eqref{Liasympt}, 
and one obtains, with $z=k_F R$ and $u=z/\beta \mu$
\bea
\langle {\cal N}_R^n \rangle^c \simeq \frac{1}{2^{d} \Gamma(1+d/2) \Gamma(2-n+d/2)} 
z^{d+1-n} u^{n-1} 
\label{widomcumlowT}
\eea 
Note that in even dimension $d$ the amplitude vanishes for $n \geq 2+d/2$.
Again the variance is obtained setting $n=2$, hence behaves as $z^{d-1} u$.
\\
 
{\bf Remark}. The leading order of Widom's formula \eqref{chiscum} for $\chi(s)$ can be obtained from simple thermodynamics. We recall that in the grand canonical ensemble $T \log Z_{gc}(\tilde \mu,T) \simeq p(\tilde \mu,T) V$ to leading order in the system volume $V$, where $Z_{gc}(\tilde \mu,T)$ is the grand canonical partition function. Hence 
one has, to leading order in the large $R$ limit 
\be \label{thermowidom}
\langle e^{-s {\cal N}_R} \rangle = \frac{Z_{gc}(\tilde \mu- s/\beta,T)}{Z_{gc}(\tilde \mu,T)} \simeq \exp\left(  V_d R^d \beta \left( p(\tilde \mu - s/\beta,T)
- p(\tilde \mu,T)\right)  + O(R^{d-1}) \right) 
\ee
since $- s/\beta$ can be interpreted as a shift in the chemical potential within the sphere of radius $R$. On the other hand for a free Fermi gas 
\be 
\beta p(\tilde \mu,T) = 
 \int_{{\mathbb R}^d} \frac{d^d {\bf k}}{(2 \pi)^d}
\log (1 + e^{ \beta (\tilde \mu - \frac{k^2}{2}}) 
\ee 
which leads to \eqref{chiscum}.

\subsubsection{Variance at low temperature}

We start from the general formula 
\be  \label{var1} 
 \langle {\cal N}_R^2 \rangle^c = \int_{|x|<R} d^d {\bf x} \left( g_1(0) 
 -  \int_{|y|<R} d^d {\bf y} 
g_1(|{\bf x}-{\bf y}|)^2 \right) \quad , \quad  g_1(|x|)= 
\int_{{\mathbb R}^d} \frac{d^d {\bf q}}{(2 \pi)^d} 
\frac{ e^{ i {\bf q} \cdot {\bf x}} }{1 + e^{\beta  (\frac{q^2}{2} - \tilde \mu)}}
\ee 
Let us rescale ${\bf x} \to {\bf x}/k_F$, ${\bf y} \to {\bf y}/k_F$, ${\bf q} \to k_F {\bf q}$, and denote $z= k_F R$ and $\beta \tilde \mu=z/u$. We also denote the Fermi factor after rescaling as $n_{z/u}({\bf q}) := 1/(1+ e^{\frac{z}{u} (q^2-1)})$. Using the
identity
\bea 
\int_{|x|<z} d^d {\bf x} e^{i {\bf q} \cdot {\bf x}}=
  (2 \pi)^{d/2} z^{d/2} q^{-d/2} J_{d/2}(z q) 
\eea 
Eq. \eqref{var1} becomes 
\bea 
\langle {\cal N}_R^2 \rangle^c = V_d z^d 
\int \frac{d^d {\bf q}}{(2 \pi)^d} n_{z/u}({\bf q}) 
- z^d \int \frac{d^d {\bf q}}{(2 \pi)^d}  
\int d^d {\bf p} \, n_{z/u}({\bf q}) n_{z/u}({\bf p}) 
 |{\bf q}-{\bf p}|^{-d} (J_{d/2}(z |{\bf q}-{\bf p}|))^2 
\eea 
Upon the change of variable ${\bf q} \to {\bf q} + \frac{1}{2} {\bf k}$,
${\bf p} \to {\bf q} - \frac{1}{2} {\bf k}$ in the second term, we can
rewrite the variance as 
\bea \label{varD} 
\langle {\cal N}_R^2 \rangle^c = \frac{1}{(2 \pi)^d} z^d \int d^d {\bf k} 
k^{-d} (J_{d/2} (k))^2 D(\frac{{\bf k}}{z})
\eea 
where we have defined the function
\be \label{defD}
D({\bf k}) := \int d^d {\bf q}  \left(  n_{z/u}({\bf q}) - n_{z/u}({\bf q} + \frac{1}{2} {\bf k})
n_{z/u}({\bf q} - \frac{1}{2} {\bf k}) \right) 
\ee
performed the rescaling ${\bf k} \to {\bf k}/z$,
and used the identity $V_d = \int d^d {\bf k} k^{-d} (J_{d/2} (k))^2$. 
Until now this is exact.

Now we consider the low temperature limit, i.e. the limit $z \gg 1$ at fixed $u$. Let us change variables in \eqref{defD} as ${\bf q}={\bf n} (1 + u \tilde q/z)$ where ${\bf n}$
is a unit vector, i.e. ${\bf n}^2=1$, and $\tilde q$ a real number. Expanding
inside the Fermi factors we obtain to leading order at large $z$ and fixed $u>0$
\bea 
D(\frac{{\bf k}}{z}) \simeq \frac{u}{z} \int_{-\infty}^{+\infty}
d \tilde q  \int_{{\cal S}_{d}} d \Omega_{\bf n}  \left( \frac{1}{1 + e^{2 \tilde q}}
- \frac{1}{1 + e^{2 \tilde q+ \frac{1}{u} {\bf n} \cdot {\bf k}}} 
\frac{1}{1 + e^{2 \tilde q - \frac{1}{u} {\bf n} \cdot {\bf k}}} \right) 
= \frac{1}{2 z} \int_{{\cal S}_{d}} d \Omega_{\bf n} ({\bf n} \cdot {\bf k}) \coth( \frac{1}{u} {\bf n} \cdot {\bf k}) 
\eea
where ${\cal S}_{d}$ is the unit sphere in dimension $d$ and $d \Omega_{\bf n}$ denotes the solid angle element. 

This formula does not allow to obtain the $T=0$ result. Indeed setting $u=0$
formally in the last expression in the r.h.s. gives $D_0(\frac{{\bf k}}{z}) \simeq 
\frac{1}{2 z} |{\bf n} \cdot {\bf k}|$. Inserted into \eqref{varD} 
it gives formally $z^{d-1}$ times a log divergent integral. Hence one needs
to go back to \eqref{defD}, replace $n_{z/u}(q)=\theta(1-q^2)$ 
and keep $z$ as a cutoff in the integral \eqref{varD}, 
to produce the standard result $\langle {\cal N}_R^2 \rangle^c 
\propto z^{d-1} \log z$. We will not follow that route here
since we are primarily interested in finite temperature. Instead we will
use that the result at $T=0$ is known, including the $O(z^{d-1})$ term
\cite{Castin2007,SmithLMS2021}, and reads
\bea \label{varT0}
\langle {\cal N}_R^2 \rangle^c|_{T=0}
= \frac{1}{\pi^2 \Gamma(d)} z^{d-1} \left( \log z + b_d + o(1) \right) \quad , \quad 
b_d = 2 \log 2 - \frac{\gamma_E}{2} + 1 - \frac{3}{2} \psi^{(0)}(\frac{d+1}{2}) 
\eea 
We can now substract it from the finite temperature variance, and
in the low $T$ limit we replace $D_0({\bf k}) \simeq V_{d-1} k$
which leads to 
\be
\langle {\cal N}_R^2 \rangle^c - \langle {\cal N}_R^2 \rangle^c|_{T=0} \simeq 
\frac{z^{d-1}}{(2 \pi)^d} \int \frac{d^d{\bf k}}{ k^d}
J_{d/2}(k)^2 \left( \frac{1}{2}  
\int_{{\cal S}_{d}} d \Omega_{\bf n}  ({\bf n} \cdot {\bf k}) \coth( \frac{1}{u} {\bf n} \cdot {\bf k}) 
- V_{d-1} k \right) 
\ee
where the integral over $k$ is now convergent. We can thus rewrite the result 
for the variance in the low temperature regime $z \gg 1$ with fixed $u$, as
\be \label{varscaling} 
\langle {\cal N}_R^2 \rangle^c - \langle {\cal N}_R^2 \rangle^c|_{T=0}  =
z^{d-1} {\cal V}_d(u)  \quad , \quad z= k_F R \quad , \quad  u = \frac{2 T R}{k_F}
\ee
where the scaling function ${\cal V}_d(u)$ takes several equivalent forms.
In $d=1$ one finds 
\be \label{V1} 
{\cal V}_1(u) = \frac{2}{\pi} \int_0^{+\infty} dk \frac{J_{1/2}(k)^2}{e^{2k/u}-1} 
= \frac{1}{\pi^2} \sum_{m=1}^{+\infty} \log(1 + \frac{u^2}{m^2}) = \frac{1}{\pi^2} \log \left(\frac{\sinh(\pi u)}{\pi u} \right)
= \begin{cases} \frac{u^2}{6} - \frac{\pi^2 u^4}{180} + O(u^4) \\
\frac{u}{\pi} - \frac{1}{\pi^2} \log(2 \pi u) + o(1) \end{cases} 
\ee 
Putting together with \eqref{varT0} we see that in $d=1$ in the low temperature regime for $z \gg 1$ and $u$ fixed
\be \label{vard1}
\langle {\cal N}_R^2 \rangle^c = 
\frac{1}{\pi^2}  \left( \log \left( z  \frac{\sinh(\pi u)}{\pi u}\right)  + 
b_1 \right)  + o(1) \quad , \quad 
b_1 = 2 \log 2 + \gamma_E + 1  
\ee

For $d>1$ one can use ${\bf k} \cdot {\bf n}= k t$ with $t=\cos \theta$, as well as
\bea 
\int_{{\cal S}_{d}} d \Omega_{\bf n}  = S_{d-1} \int_{-1}^{+1} dt (1-t^2)^{\frac{d-3}{2}} 
\quad , \quad  V_{d-1}= \frac{S_{d-1}}{d-1} = 
S_{d-1} \int_{0}^{+1} dt t (1-t^2)^{\frac{d-3}{2}}
\eea 
which lead to the two equivalent expressions
\bea 
&& {\cal V}_d(u) = \frac{1}{(2 \pi)^d} 2 S_{d} S_{d-1} 
\int_0^1 dt \, t (1-t^2)^{\frac{d-3}{2}} 
\int_0^{+\infty} dk \frac{ J_{d/2}(k)^2}{e^{2 k t /u }-1} \\
&& = \frac{1}{(2 \pi)^d} 
\frac{2}{\pi} S_{d} S_{d-1}  
\sum_{m=1}^{+\infty} 
\int_0^1 dt \, t (1-t^2)^{\frac{d-3}{2}}  
Q_{\frac{d-1}{2}}(1 + 2 \frac{m^2 t^2}{u^2}) 
\eea 
where $Q_\alpha$ is the Legendre of second kind
and $S_d=d \pi^{d/2}/\Gamma[1+d/2]$. 
The asymptotics for general $d$ are 
\be 
{\cal V}_d(u) \simeq \begin{cases} \frac{2^{1-d}}{d \Gamma(d/2)^2} u \quad , \quad u \gg 1 \\
 \frac{2^{-d} \sqrt{\pi}}{3 \Gamma(d/2) \Gamma(\frac{d+3}{2}) } u^2 \quad , \quad u \ll 1
\end{cases}
\ee 
The large $u$ asymptotics matches precisely the low $T$ limit of the result \eqref{widomcumlowT} (setting $n=2$), although from \eqref{varscaling}
and \eqref{varT0}
we see it requires $u \gg \log z$,
i.e. an inversion of limits. 

In $d=3$ it has a simple series and integral representation
\begin{equation}\label{eq:var_d=3}
{\cal V}_3(u)=\frac{1}{2\pi^2}\sum_{m=1}^\infty
\left[
\left(1+\frac{m^2}{u^2}\right)
\ln\!\left(1+\frac{u^2}{m^2}\right)
-1
\right] =\frac{1}{\pi^2 u^2}
\int_0^u ds\, s\,
\ln\!\left(\frac{\sinh(\pi s)}{\pi s}\right) =
\begin{cases} \frac{u^2}{24}
-\frac{\pi^2 u^4}{1080}
+O(u^6)  \\
\frac{u}{3\pi} - \frac{1}{2 \pi^2} \log(2 \pi u) + O(1) \end{cases} 
\end{equation}
The integral can be evaluated explicitly in terms of polylogarithm functions. 
Finally, for $d=2$ the simplest representation seems to be 
\bea \label{eq:V2_var}
{\cal V}_2(u)= \frac{2}{\pi} \int_0^{+\infty} dk J_1(k)^2 \int_0^{\frac{\pi}{2}} 
d\theta \frac{\cos \theta }{e^{2 k \cos \theta/u}-1} \;.
\eea 

\begin{figure}
\centering
\includegraphics[width=0.45\linewidth]{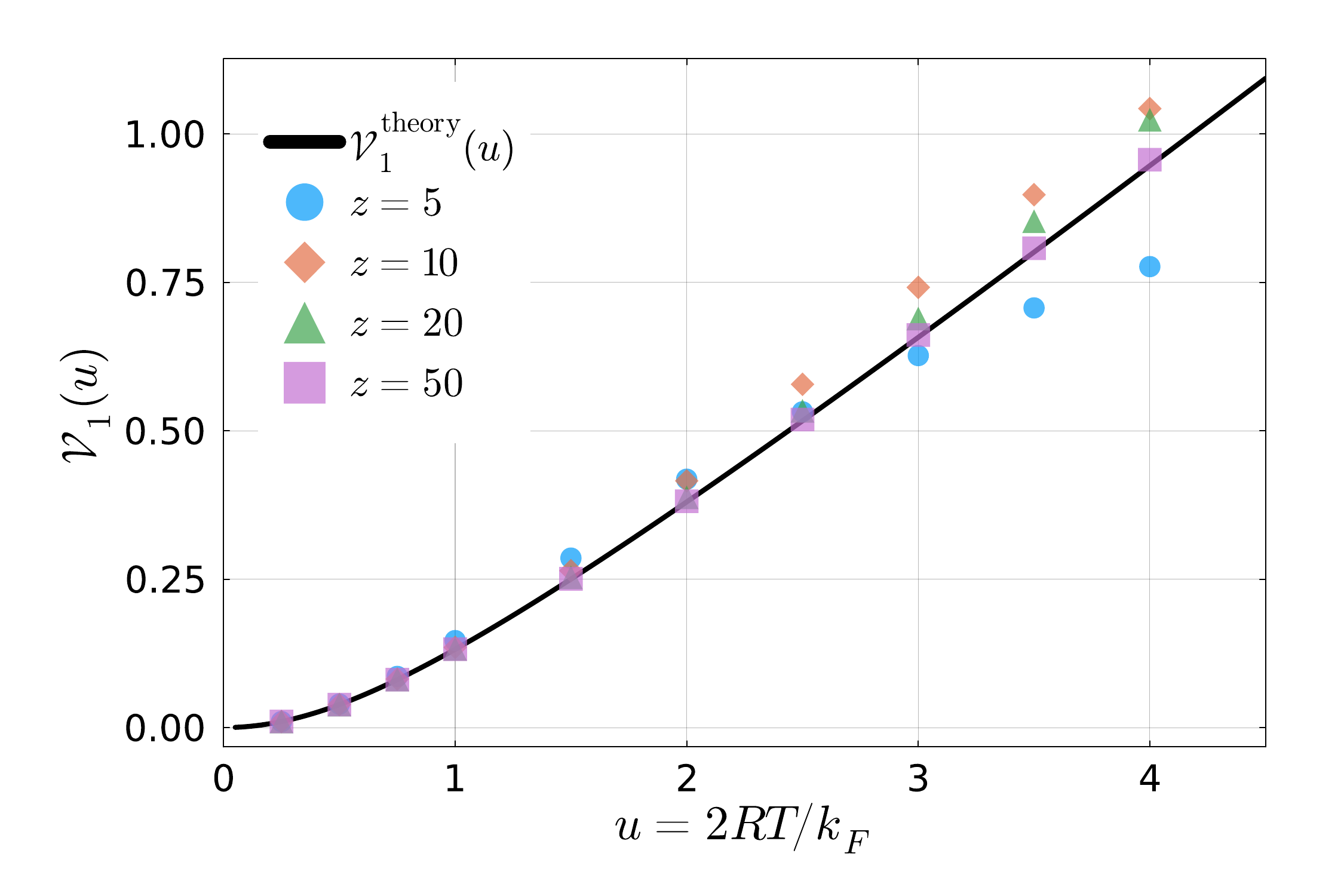}
\includegraphics[width=0.45\linewidth]{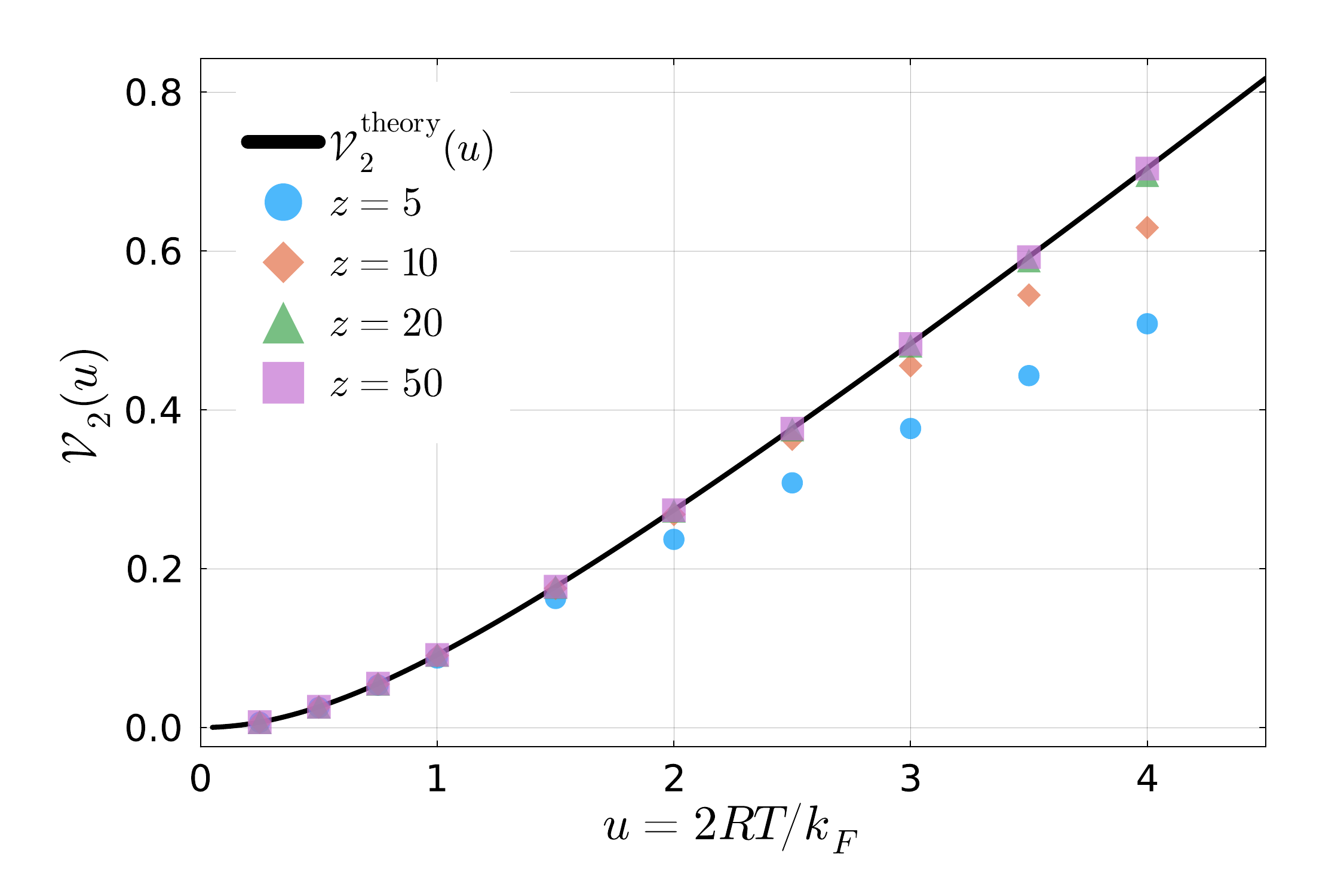}
\includegraphics[width=0.45\linewidth]{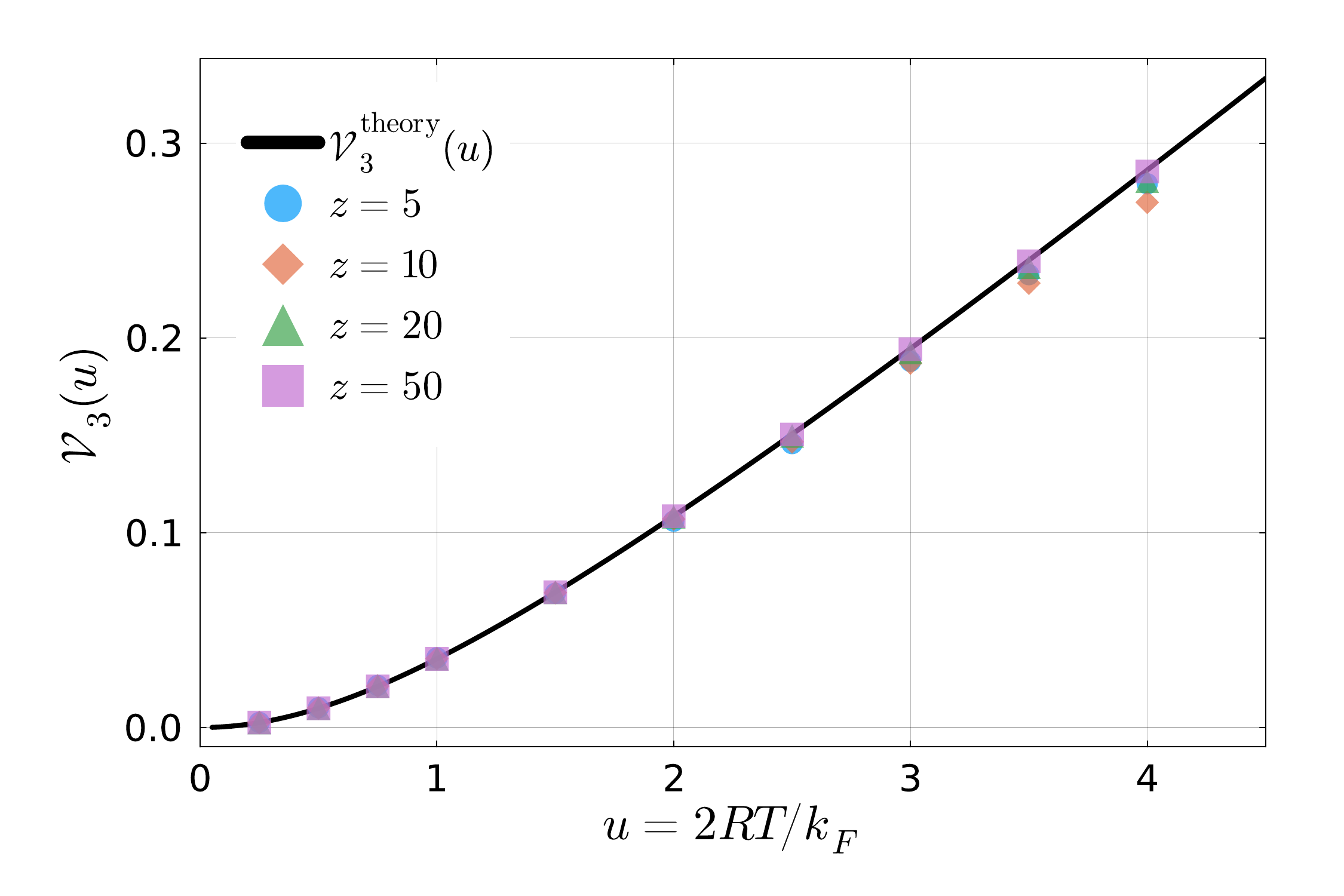}
    \caption{Scaling functions ${\cal{V}}_d(u)$ in Eq.~\eqref{varscaling} describing the variance crossover at low temperature for non-interacting fermions in $d=1,2,3$. Marker points are obtained by integrating numerically Eq.~\eqref{var1} and subtracting the zero temperature variance $\langle {\cal N}_R^2 \rangle^c|_{T=0}$ in Eq.~\eqref{varT0} while continuous lines are the theoretical predictions.}
    \label{fig:variance1}
\end{figure}

{\bf Entanglement entropy}. Let us recall the definition of the bipartite Renyi's entropies $S_q$ for non-interacting fermions, relative to a domain ${\cal D}$. They
are given by $S_q=\sum_j h_q(K_{\cal D}) = \sum_j h_q(\nu_j)$, where 
$h_q(\nu)=\frac{1}{1-q}\ln\!\big[\nu^q+(1-\nu)^q\big]$, for $0<\nu<1$ and $q>0$,
and $\nu_j$ are the eigenvalues of the correlation matrix, i.e.
of the kernel $K_{\cal D}$ restricted to the domain ${\cal D}$. 
For $q \to 1$ it gives the Von Neumann entropy $S_1$. For any non-interacting fermion system prepared in 
a Gaussian state (i.e. forming a DPP), which includes finite temperature in the GC ensemble, there is an exact relation between the Renyi's entropies $S_q$ and the FCS (the cumulants) \cite{KlichLevitov2009,LeHur2011}, see below. At zero temperature, and in fact in the full low temperature regime, the variance dominates, and one has 
\be S_q \simeq \frac{\pi^2}{6} (1 + \frac{1}{q}) 
\langle {\cal N}_R^{2} \rangle^c
\ee 

We can thus check that the result \eqref{vard1} in $d=1$ is consistent with the 
finite temperature entropy formula of Calabrese and Cardy \cite{CalabreseCardy2004},
which is universal and valid for any model described at large scale by the conformal field theory (CLT)
with central charge $c$
\be
S_q = \frac{c}{6} (1 + \frac{1}{q}) \log( \frac{v_F \beta}{\pi a} \sinh( \frac{\pi \ell}{v_F \beta})) + c_q + o(1) 
\ee 
for ${\cal D}$ being an interval of size $\ell$ and $a$ an UV cutoff (we have restored the sound wave velocity, here $v_F$). Substituting $c=1$ for free fermions and $\ell=2 R$
we see that $\ell/(v_F \beta)=\ell/\xi_T=u$ and 
the two formula are consistent with $a \propto 1/k_F$. 

For the Von Neumann entropy $q=1$ the exact relation between the FCS and $S_1$ 
can be written as \cite{Gamayun2020}
\be 
S_1= \sum_{m \geq 1} 2 \zeta(2 m) \langle {\cal N}_R^{2 m} \rangle^c
= \int_{-\infty}^{\infty} ds K_1(s) \chi(s) \quad , \quad K_1(s)= 
\frac{1}{4 \sinh^2(s/2)} 
\ee 
where to make the integral well defined one can replace $ \chi(s) \to \chi(s) + s \langle {\cal N}_R\rangle $, or symmetrize as $
 \chi(s) \to \frac{1}{2} (\chi(s)+\chi(-s))$. The last equality in the r.h.s. is a non-perturbative result. One can check that it reproduces the
series from the identity $\int_{-\infty}^{\infty} ds \frac{s^{2m}}{(2m)!} \frac{ \chi(s)}{4 \sinh^2(s/2)}= 2 \zeta(2 m)$. The non-perturbative proof is based on the following identity 
\be 
\int_{-\infty}^{\infty} ds\; K_1(s) \left( 
\ln(1-\nu+\nu e^{-s})+\nu s \right) = - ((1-\nu) \log(1-\nu) + \nu \log \nu) 
\ee 
We can thus compute the Von Neumann entropy in the high temperature Widom regime
$T=O(T_F)$. Using the identity 
\bea 
\frac{1}{2} \int_{-\infty}^{+\infty} ds \frac{1}{4 \sinh^2(s/2)} \log(\frac{1 + A^2 + 2 A \cosh(s)}{(1+A)^2})
= \log(1+ A) - \frac{A}{A+1} \log A 
\eea 
and the result for $\chi(s)$ in \eqref{chiscum} we obtain, with $V_d=S_d/d= \pi^{d/2}/\Gamma(1+d/2)$
and $p=k^2/2$ 
\bea  \label{entropy1}
&& S_1 \simeq  
 (R \sqrt{T})^d V_d \frac{S_d}{(2 \pi)^d}  2^{\frac{d-2}{2}}
 \int_0^{+\infty} dp p^{\frac{d-2}{2}} 
\bigg[ 
 \log (1 + e^{ \beta \tilde \mu - p} ) 
 - \frac{\beta \tilde \mu- p}
 {1 + e^{ p - \beta \tilde \mu}} 
\bigg] \\
&& \simeq  \frac{(R \sqrt{T})^d}{ 2^{1 + \frac{d}{2}} \Gamma(1 + \frac{d}{2}) }
\left(2 \beta \tilde \mu
   \text{Li}_{\frac{d}{2}}\left(-e^{\beta \tilde \mu}\right)-(d+2)
   \text{Li}_{\frac{d}{2}+1}\left(-e^{\beta \tilde \mu}\right)\right)
\eea 

{\bf Remark}. One can check that, to leading order at large $R$ and in the fixed $T$ regime described by Widom's formula, the bipartite entanglement entropy is given by 
\be 
S_1 = s(T,\tilde \mu) V_d R^d + O(R^{d-1})
\ee 
where $s(T,\tilde \mu)$ is the thermodynamic entropy density for the free Fermi gas 
\bea 
s(T,\mu) = \partial_T|_{\tilde \mu} p(\tilde \mu,T) = \partial_T|_{\tilde \mu} \left( \frac{T}{V} \log Z_{\rm gc}(\tilde \mu,T) \right) 
= \partial_T|_{\tilde \mu}  \left( T \int \frac{d^d k}{(2 \pi)^d} \log(1 + e^{- \beta (\frac{k^2}{2} - \tilde \mu)}) \right) 
\eea 
where $Z_{gc}(\tilde \mu ,T)$ is the grand canonical partition function.
This relation and the subleading terms for $S_1$ have been studied 
in e.g. \cite{entropy_spitzer}. 
\\

\section{Numerical implementation}
\label{app:numerical}

In this appendix we describe the numerical pipeline used to calculate the hole probability and all the figures appearing in the main text and in this Supplementary Material. The algorithm developed might be of independent interest and applicable to other Fredholm determinants with different kernels, notably the Airy kernel describing the edge of the eigenvalues distribution in classical RMT ensembles and non-interacting fermions in a trap.

The basics tools for evaluating such determinants can be found in Ref. \cite{Bornemann_2009}. We have found that a naive application of the methods presented there are not sufficient to capture correctly the scaling function $\Phi_d(u)$ (Eq.~\eqref{phid} in the main text) we are interested in. Therefore, we will explain our algorithm in detail.

\subsubsection{General setup}

There are two stages to compute numerically the scaling function. First, the \emph{Fredholm determinant solver} which computes numerically the Fredholm determinant of a one dimensional kernel. This is explained in Sec.~\ref{app:fredholm_solver}. In dimension $d=1$, this is enough. For $d\ge2$ we need to compute several one dimensional Fredholm determinants and sum over the angular sectors (see eq.~\eqref{Pell} in the main text). This part is the subject of Sec.~\ref{app:gend}. In Sec.~\ref{app:scaling_limit} we give the details of how to take the scaling limit relevant to the numerical calcualation of the scaling function $\Phi_d(u)$. The second stage of the whole algorithm is a post-processing pipeline that fits the scaling function taking into account corrections to the scaling limit. This is explained in Sec.~\ref{app:data_analysis}.

As anticipated above, the basic ingredient is the calculation of the on Fredholm determinant of a one dimensional kernel. To be concrete,  let us consider a kernel $\bar K$ acting on functions defined on $\mathbb{R}$ as an integral operator with matrix elements $\bar K(x,y)$. Then we want to compute the determinant, that we call $\bar P$, of its projection to the interval $[a,b]\subset \mathbb{R}$, namely
\begin{equation}\label{eq:num_fred}
\bar P=\Det(1-\Pi_{[a,b]}\bar K),
\end{equation}
where $\Pi_{[a,b]}$ is the projector on the interval $[a,b]$. The algorithm we are going to describe is valid for any kernel whose matrix elements in the position basis are explicitly given by
\begin{equation}\label{eq:gen_ker}
    \bar K(x,x') = \int_0^{+\infty}\frac{\dd k}{\pi}\sigma(k) g(k;x)h(k;x')
\end{equation}
where $\sigma(k)$ is the Fermi factor
\begin{equation}\label{eq:fermi_factor}
\sigma(k)=\frac{1}{1+e^{\beta(k^2/2-\tilde \mu)}}
\end{equation}
and $g(k;x)$ and $h(k;x')$ are arbitrary sufficiently regular functions. In Eq.~\eqref{eq:fermi_factor} $\tilde \mu$ is the chemical potential and $\beta = 1/T$ the inverse temperature.

For the present paper the relevant cases are 
\begin{subequations}\label{eq:g_func}
\begin{equation}
    g(k;x) = h(k;-x) = e^{ik x}\quad \text{when}\quad d=1
\end{equation}
\begin{equation}
    g(k;x) = h(k;x) = \sqrt{x\,k}J_\nu(k\, x)\quad \text{when}\quad d\ge 2
\end{equation}    
\end{subequations}
where $\nu = \ell + d/2-1$. For $d=1$, Eq.~\eqref{eq:g_func} gives the finite temperature Sine kernel in Eq.~\eqref{def_sineK} while for $d\geq 2$ it gives the finite temperature Bessel kernels of index $\nu$ in Eq.~\eqref{hat_BesselKernel_txt} in the main text.
According to Eq.~\eqref{eq:gen_ker}, these kernels are given as momentum integrals over the Fermi factor. 

There are four steps to compute numerically the Fredholm determinant in Eq.~\eqref{eq:num_fred}: the first step is to approximate the functional determinant with a standard determinant of matrix $\bar K_{ij}$ related to the kernel $\bar K(x,x')$ discretizing the spatial integrals; the second step is to obtain the matrix elements $\bar K_{ij}$ for a given choice of the functions $g(k;x)$ and $h(k;x)$ in Eq.~\eqref{eq:gen_ker} discretizing the momentum integrals; third,  we will show that, for kernels of the form Eq.~\eqref{eq:gen_ker}, the use of the Sylvester identity (in Eq.~\eqref{eq:sylv}) allows to rewrite the determinant giving the hole probability as the determinant of a positive definite and symmetric Gram matrix. This has the effect of reducing the size of the matrices used to approximate the kernel; finally, since we are interested in the logarithm of this determinant (giving the scaling function $\Phi_d(u)$ up to a factor) the symmetry of the Gram matrix allows a clever use of its Cholesky factorization bypassing the calculation of the eigenvalues and the eigenvectors. These steps are the subject of the next Section.

\subsection{One dimensional Fredholm determinant solver}
\label{app:fredholm_solver}

\subsubsection{Spatial quadrature grids}

We begin with the approximation of the matrix elements in Eq.~\eqref{eq:gen_ker} with a finite dimensional matrix (finite rank operator). This part is standard \cite{Bornemann_2009}.
The first observation is that the operator $\Pi_{[a,b]}\bar K$ in Eq.~\eqref{eq:gen_ker} acts on suitable functions as an integral operator. This action can be approximated by a standard $m$-point Gaussian quadrature rule as follows
\begin{equation}\label{eq:ker_gauss}
    (\bar K f)(y) = \int_a^b \bar K(x,x') f(x') \dd x'\approx \sum_{i=1}^m \bar K(x,x_i)f(x_i)\omega_i^x
\end{equation}
where $m$ is the number of nodes chosen and controls the error of the quadrature rule \cite{Bornemann_2009}. The nodes $x_i$ and the weights $\omega_i^x$ depend explicitly on the  spatial interval $[a,b]$, as explained below. This discretization scheme allows to compute approximately the Fredholm determinant in Eq.~\eqref{eq:num_fred} via
\begin{equation}\label{eq:app_det}
    \Det(1 - \bar K) \approx \det_{1\le i,j,\leq m}( 1- \bar K_{ij})
\end{equation}
where  the right hand side is now an ordinary determinant the rank $m$ matrix
\begin{equation}\label{eq:K_disc_naive}
    \bar K_{ij} \equiv \bar K(x_i,x_j) \omega_j^x\,.
\end{equation}
Below, we will suppress the indices $i,j$ from the left hand side and indicate the finite dimensional matrix by the same symbol used for the kernel $\bar K$ understanding that $\det$ is an ordinary determinant while $\Det$ is a Fredholm determinant.
In Eq.~\eqref{eq:K_disc_naive} the $\{x_i,\omega_i^x\}_{i=1}^m$'s are the nodes and the weights of the Gaussian quadrature rule in $[a,b]$ appearing in Eq.~\eqref{eq:ker_gauss}. A first trick that is useful for numerical stability is \cite{Bornemann_2009} to instead compute the determinant of the symmetric matrix
\begin{equation}\label{eq:K_disc_symm}
    \bar K^s_{ij} \equiv \sqrt{\omega_i^x}\bar K(x_i,x_j)\sqrt{\omega_j^x}\,.
\end{equation}
The relation between $\bar K_{ij}$, in Eq.~\eqref{eq:K_disc_naive}, and $\bar K^s_{ij}$ , in Eq.~\eqref{eq:K_disc_symm}, is just a similarity transformation that preserves the determinant in Eq.~\eqref{eq:app_det}.
Standard numerical software produce the nodes $t_i$ and the weights $\omega_i$ relative to the interval $[-1,1]$, i.e.,
\begin{equation}
    \int_{-1}^1 f(t) \dd t \approx \sum_{i=1}^m f(t_i) \omega_i\,.
\end{equation}
The transformation
\begin{equation}
    t = f(x) = \frac{1}{b-a}(2x - a - b)
\end{equation}
maps the interval $[a,b]$ bijectively to the interval $[-1,1]$. This allows to extract the nodes and the weights for a general in interval $[a,b]$, i.e. $\{x_i,\omega_i^x\}_{i=1}^m$, from those of $[-1,1]$, i.e. $\{t_i,\omega_i\}_{i=1}^m$. A change of variables in Eq.~\eqref{eq:ker_gauss} shows that 
\begin{equation}\label{eq:mapping_nodes}
    x_i = \frac{b-a}{2}t_i + \frac{a+b}{2}\quad , \quad \omega_i^x = \frac{b-a}{2}\omega_i \,.
\end{equation}
The chioces that we made in this paper are explained just below.

\paragraph{\bf{$d=1$ finite temperature Sine kernel and single interval $[-R,R]$.}}
In one dimension we are interested in the interval $[a,b]=[-R,R]$ (this is a `ball' in one dimension). In this case, the kernel of interest is the Sine kernel, see Eq.~\eqref{eq:g_func}. Owing to translation invariance, it can be equivalently evaluated on an interval of
the same length, namely $[0,X]$ with
\begin{equation}
X=2R.
\end{equation}
The nodes and weights $\{x_i,\omega_i^x\}_{i=1}^m$ appropriate for this interval $[0,X]$ are obtained from setting $a=0$, $b=X$ in Eq.~\eqref{eq:mapping_nodes}. The result is
\begin{equation}\label{eq:spat_weights1}
x_i=\frac{X}{2}(t_i+1),
\qquad
\omega_i^{x}=\frac{X}{2}\,\omega_i
\end{equation}
where $\{t_i,\omega_i\}_{i=1}^m$ are the nodes and weights for the interval $[-1,1]$.

\paragraph{\bf{$d\geq 2$ finite temperature Bessel kernel and spherical ball of radius $R$.}}
In this case, the Bessel kernel acts on functions defined on $[0,+\infty]$ and the projector restricts the radial variable to $[0,R]$ with
\begin{equation}\label{eq:spat_weightsd}
r_i=\frac{R}{2}(t_i+1),
\qquad
\omega_i^{r}=\frac{R}{2}\,\omega_i
\end{equation}
where $r_i$ and $\omega_i^r$ are the radial nodes and weights repectively, obtained from those for the interval $[-1,1]$ again.

\subsubsection{Momentum quadrature grids}
We recall, that at finite temperature, the kernels governing the hole probability are given by integrals over the momentum variable $k$ ranging in $[0,+\infty]$ (after exploiting the symmetry of the Fermi factor, see below). These integrals are approximated by a second Gaussian quadrature rule. We use an $m'$-point rule and cut-off the infinite integration domain with
$[0,k_{\max}]$.  The cutoff $k_{\max}$ is chosen from the condition that
the Fermi factor in eq.~\eqref{eq:fermi_factor} is below a prescribed tolerance $\varepsilon_k$, i.e., solving $\sigma(k_{\max}) = \varepsilon_k$. The result is
\begin{equation}
k_{\max}
=
\sqrt{2\left(\tilde \mu+\frac{1}{\beta}\log\frac{1-\varepsilon_k}{\varepsilon_k}\right)},
\end{equation}
up to a fixed safety factor.  The mapped momentum nodes from $[-1,1]$ to $[0,k_{\max}]$ and the corresponding weights are
denoted by
\begin{equation}\label{eq:mom_weights}
\{k_a, w_a\}_{a=1}^{m'}    
\end{equation}
and are calculated from those of the interval $[-1,1]$ via the mapping in Eq.~\eqref{eq:mapping_nodes} replacing $x_i$ with $k_a$ and $\omega_i^x$ with $\omega_a$.
Thus, up to now the determinant computation uses two discretizations:
one in the spatial variable with $m$ nodes, and one in momentum $k$ with $m'$ nodes. Next we specialize to the finite temperature Sine kernels in $d=1$ and show how to compute the finite dimensional determinant in Eq.~\eqref{eq:app_det} efficiently using the Sylvester identity. In Sec.~\ref{app:gend} we generalize to higher dimensions as in that case we just need to sum over angular sectors.

\subsubsection{Finite temperature sine-kernel and rank reduction via the Sylvester identity}

We now apply the quadrature rules explained above to the $d=1$ Sine kernel corresponding to the first line of Eq.~\eqref{eq:g_func}. After that, we show how the form of the kernel in Eq.~\eqref{eq:gen_ker} allows to use the Sylvester identity and reduce the rank of the finite dimensional matrix used to approximate the Fredholm determinant in Eq.~\eqref{eq:num_fred}.

Due to Eq.~\eqref{eq:g_func}, in $d=1$ the kernel $\bar K$ in Eq.~\eqref{eq:gen_ker} is just the Sine kernel. This was defined in Eq.~\eqref{def_sineK} so we set $\bar K = \hat K$. Using the symmetry of the Fermi factor $\sigma(k) = \sigma(-k)$ in Eq.~\eqref{eq:fermi_factor} and using standard trigonometric identities, Eq.~\eqref{eq:gen_ker} can be written as
\begin{equation}
\hat K(x,y)
=
\int_0^\infty \frac{dk}{\pi}\,\sigma(k)
\Bigl[\cos(kx)\cos(ky)+\sin(kx)\sin(ky)\Bigr]\,.
\end{equation}
Applying the spatial and momentum quadrature rules to this kernel leads to two rectangular
matrices $U^c$ and $U^s$ of size $m\times m'$:
\begin{equation}
U^c_{ia}
=
\sqrt{\omega_i^x}\,
\sqrt{\frac{w_a\,\sigma(k_a)}{\pi}}\,
\cos(k_a x_i),
\qquad
U^s_{ia}
=
\sqrt{\omega_i^x}\,
\sqrt{\frac{w_a\,\sigma(k_a)}{\pi}}\,
\sin(k_a x_i)
\end{equation}
where we recall that $\{x_i,\omega_i^x\}_{i=1}^m$ and $\{k_a,\omega_a\}_{a=1}^{m'}$ are the nodes and the weights corresponding to the spatial and the momentum Gaussian quadrature rules respectively (see Eq.~\eqref{eq:spat_weights1} and Eq.~\eqref{eq:mom_weights}).
Defining the block matrix
\begin{equation}
U=[U^c\,|\,U^s]
\end{equation}
obtained stacking horizontally the matrices $U^c$ and $U^s$,
the discretized symmetric kernel matrix in Eq.~\eqref{eq:K_disc_symm} takes the form
\begin{equation}
\bar K^s=UU^T
\end{equation}
where $U^T$ denotes the matrix transpose of $U$.
The standard Bornemann's approach \cite{Bornemann_2009} would construct the full $m\times m$ matrix
$\bar K$ and compute
\begin{equation}
\log\bar P=\log\det(1-\bar K^s)\,.    
\end{equation}
This costs $O(m^2)$ memory and $O(m^3)$ operations at the
determinant stage by a naive calculation using a standard LU method. Instead, using Sylvester's identity (see \cite{matrix_analysis} p. 73 Exercise 1.3.P28)
\begin{equation}\label{eq:sylv}
\det(1-UU^T)=\det(1-U^TU),
\end{equation}
we reduce the problem to the much smaller Gram matrix
\begin{equation}
G=U^TU
\end{equation}
whose size is controlled by th momentum quadrature. Indeed, 
in $d=1$ this is a $2m'\times 2m'$ matrix with block structure
\begin{equation}
G=
\begin{pmatrix}
(U^c)^TU^c & (U^c)^TU^s\\
(U^s)^TU^c & (U^s)^TU^s
\end{pmatrix}.
\end{equation}
The log-determinant is then computed as
\begin{equation}\label{eq:log_det_G}
\log \bar P=\log\det(1-G).
\end{equation}
Even when $2m'$ (two times the number of points used for the momentum quadrature) and $m$ (the number of points used for the spatial discretization)
are of comparable size, the Gram reduction is still advantageous
because the determinant is taken on a matrix whose size is controlled by
the momentum quadrature (whose error is controlled by the smoothness in $k$) rather than by the full spatial discretization.

\subsubsection{Cholesky evaluation of the log-determinant}

This is the last technical step of the algorithm to calculate the hole probability and the scaling function. We need to evaluate the log-determinant in Eq.~\eqref{eq:log_det_G}. A good algorithm is one that exploits the symmetry of the problem. In our case, the matrix $1-G$ is symmetric and, at the exact level, positive
definite.  The code therefore computes a Cholesky factorization (see \cite{num_recipes} p. 96)
\begin{equation}
1-G=LL^T,
\end{equation}
where $L$ is the Cholesky factor, i.e., a lower triangular matrix with (always) real and positive diagonal entries. Then the sum over the diagonals
\begin{equation}\label{eq:cholesky}
\log \bar P
=
2\sum_i \log L_{ii}
\end{equation}
is evaluated, bypassing the computation of the LU decomposition which computes both eigenvalues and eigenvectors and is typically unstable \cite{foot_unstable}
The code also records the largest eigenvalue of $G$ as a stability
diagnostic.

\subsection{Generalization to $d\ge 2$}
\label{app:gend}
For $d\ge2$ the philosophy is the same but we need to sum over the angular sectors. Indeed, rotational invariance allows us to decompose the determinant
into angular-momentum channels. In the notations of the main text (Eq.~\eqref{Pell}) we need to compute
\begin{equation}\label{eq:chann_sum}
\log P(R,T)=\sum_{\ell\ge0} g_d(\ell)\,\log P^{(\ell)}(R,T),
\end{equation}
where $g_d(\ell)$ is the degeneracy factor,
\begin{equation}
g_d(0)=1,
\qquad
g_d(\ell)=\frac{(2\ell+d-2)\Gamma(\ell+d-2)}
{\Gamma(\ell+1)\Gamma(d-1)}
\qquad (\ell\ge1)
\end{equation}
and 
\begin{equation}
    P^{(\ell)}(R,T) = \Det(1 - \Pi_{R} \hat K_\nu)\quad \text{where}\quad \nu = \ell + \frac{d}{2} - 1
\end{equation}
where $\Pi_R$ is the projector on $[0,R]$ and
the kernel, reported here for convenience, is (see Eq.~\eqref{hat_BesselKernel_txt} of the main text)
\begin{equation}
\hat K_\nu(r,r')
=
\sqrt{rr'}
\int_0^\infty dk\,k\,\sigma(k)\,J_\nu(kr)J_\nu(kr')\,.
\end{equation}
After Gaussian quadrature in $r$ and $k$ as explained in Sec.~\ref{app:fredholm_solver}, the channel is represented by a matrix
$U^{(\ell)}$ of size $m\times m'$:
\begin{equation}
U^{(\ell)}_{ia}
=
\sqrt{\omega_i^r}\,\sqrt{r_i}\,
\sqrt{w_a\,k_a\,\sigma(k_a)}\,J_\nu(k_a r_i)\,.
\end{equation}
Then
\begin{equation}
G^{(\ell)}=(U^{(\ell)})^T U^{(\ell)},
\qquad
\log P^{(\ell)}=\log\det(1-G^{(\ell)})\,.
\end{equation}
Again the determinant is evaluated by Cholesky factorization as in Eq.~\eqref{eq:cholesky}.

At the level of one channel, the complexity is analogous to the
one-dimensional case, except that now the reduced matrix has size
$m'\times m'$ instead of $2m'\times 2m'$.  The total cost is obtained
by summing over the channels retained in the angular-momentum
decomposition.
\subsection{Scaling limit in $d$-dimension}\label{app:scaling_limit}

To extract the scaling function numerically we need to take a particular scaling limit as explained in the main text. Indeed, as shown also in Ref. \cite{Gabriel_gap}, in Eq.~\eqref{eq:chann_sum} we must sum over $\ell \sim z$ channels to access the large deviation regime. The parameter $z$ is defined just below in Eq.~\eqref{eq:sim_params}. Thus, in Eq.~\eqref{eq:chann_sum} we adaptively trunacate the summation choosing a safe lower bound for $\ell_{\rm min}$ for $\ell$ as follows
\begin{equation}
\ell_{\min}
=
\Bigl\lceil z+3\sqrt{\max(z,1)}+10\Bigr\rceil
\end{equation}
ensuring $\ell =O(z)$. Above $\lceil \cdot\rceil$ is the ceiling function.
After that, channels are added until a prescribed number of successive
contributions become smaller than a fixed tolerance.  The stopping index
is recorded as a diagnostic, denoted by $\ell_{\mathrm{stop}}$ and used for future reference.

The final raw output of the solver 
\begin{equation}\label{eq:phi_numd}
\Phi_{d}(u;z)\equiv-\frac{\log P(R,t)}{z^{d+1}}\quad , \quad 
 d=1,2,3.
\end{equation}
where $z=k_F R$ and $t=T/T_F$ with $R$ the radius of the spherical region and $T$ the temperature. The (temperature dependent) Fermi momentum $k_F$ and Fermi temperature are defined as $k_F = \sqrt{2\tilde \mu}$ and $T_F = k_F^2 / 2$ where $\tilde \mu$ is the chemical potential appearing in the Fermi factor in Eq~\eqref{eq:fermi_factor}.
As $z\to\infty$, $t\to 0$ with $u = zt = O(1)$ fixed, the right hand size of Eq.~\eqref{eq:phi_numd}, should tend to the scaling function calculated theoretically $\Phi_d(u)$, i.e., $\Phi_d(u;\infty) = \Phi_d(u)$.  Throughout the numerical routine the chemical potential is fixed for convenience to
\begin{equation}
\tilde \mu=\frac12\,.
\end{equation}
Hence
\begin{equation}\label{eq:sim_params}
k_F=\sqrt{2\tilde \mu}=1,\qquad z=k_F R=R,
\qquad
t=\frac{T}{T_F},
\qquad
T_F=\frac12.
\end{equation}
For each pair $(u,z)$, the code reconstructs the inverse temperature as
\begin{equation}
\frac{1}{T}=\beta=\frac{2z}{u},
\end{equation}
which is equivalent to $u=zt$ since $k_F=1$. Notice that $u = 2 R T / k_F$. This choice is convenient to numerically calculate the scaling function $\Phi_d(u)$.
\subsection{Data analysis and extraction of the scaling function}
\label{app:data_analysis}
Here we explain the second stage of the numerical procedure. The output of the Fredholm determinant solver is the function $\Phi_d(u;z)$ defined in Eq.~\eqref{eq:phi_numd}. In the scaling limit described in Sec.~\ref{app:scaling_limit} we expect $\Phi_d(u;\infty)= \Phi_d(u)$ where $\Phi_d(u)$ is the theoretical prediction in Eq.~\eqref{phid} in the main text.
Since we cannot take $z=\infty$ numerically we have computed $\Phi_d(u;z)$ in Eq.~\eqref{eq:phi_numd} on a two dimensional grid $(u_i,z_j)$
\begin{equation}\label{eq:grid}
\Phi_{d,\mathrm{num}}(u_i;z_j)
\quad \text{for}\quad 
z_j\in\{10,15,20\}\quad \text{and}\quad u_i \in \{i \cdot 0.1, \, \, i=1,\dots u_{\max}/0.1\}
\end{equation}
where $u_{\max}$ can be chosen freely. The parameters of the kernels are described in Sec.~\ref{app:scaling_limit} and together with the grid in Eq.~\eqref{eq:grid} they correctly capture the scaling regime to get $\Phi_d(u)$. Nevertheless, we have found that finite $z$ effects are important, especially for large $u$. This can be seen in Fig.~\ref{fig:errors} where the relative errors for $z=20$ can go up to $~50\%$ for $u_{\max}\approx6$. Equivalently, in Fig.~\ref{Fig_conv} and Fig.~\ref{Fig_conv2} one can see that the markers corresponding to finite $z$ numerical evaluation of the scaling function significantly depart from the theoretical prediction. 

To account for this, one has to take into account the corrections in the large $z$ limit. We found numerically that the first non-trivial correction seems to be of order $O(z^{-2})$. This is corroborated by a
calculation in the large $u$ regime, see Remark below. 
Thus, for each fixed $u$, the code assumes the expansion
\begin{equation}\label{eq:extrapolation}
\Phi_{d,\mathrm{num}}(u;z)
=
\Phi_{d}^{\mathrm{fit}}(u;\infty)
+\frac{\Psi_d^{\mathrm{fit}}(u)}{z^2}.
\end{equation}
Since only the three values $z=10,15,20$ are used, we fit using a
linear regression in the variable $x=z^{-2}$.
In particular, for each fixed $u_i$ in the grid in Eq.~\eqref{eq:grid} we fit
\begin{equation}
\Phi_{ij}=a_i+b_i x_j,
\qquad
\Phi_{ij}=\Phi_{d,\mathrm{num}}(u_i;z_j) \;,
\qquad
x_j=z^{-2}_j \;,
\end{equation}
with $z_j\in\{10,15,20\}$.
The intercept and slope are then identified as
\begin{equation}
\Phi_{d}^{\mathrm{fit}}(u_i;\infty)=a_i,
\qquad
\Psi_d^{\mathrm{fit}}(u_i)=b_i\,.
\end{equation}
Then we compare the extrapolated value $\Phi_{d}^{\mathrm{fit}}(u_i)$ to the theoretical prediction $\Phi_d(u)$ (Eqs.~\eqref{phid}-\eqref{phi_dp}-\eqref{phi_dm}-\eqref{phi_dm2} in the main text) for different dimensions. This results in the markers data in Fig.~\ref{Fig} of the main text. Fig.~\ref{Fig_conv} and Fig.~\ref{Fig_conv2} show the comparison between the raw finite $z$ data, the extrapolated ones and the theoretical predictions in $d=1,2,3$.
As discussed above, we have also analyzed the (percentage) relative errors. These are defined as
\begin{equation}
\mathrm{Rel}_{\mathrm{num}}(u_i;z_j)
=
100\,
\frac{\Phi_{d,\mathrm{num}}(u_i;z_j)-\Phi_d(u_i)}{\Phi_d(u_i)}\,.
\end{equation}
The lowest error $\mathrm{Rel}_{\mathrm{num}}(u_i;\infty)$ corresponding to the extrapolated scaling function $\Phi_d(u;\infty)$ as in Eq.~\eqref{eq:extrapolation} is indicated by the black dotted line in Fig.~\ref{fig:errors}.
In the same figure, one can clearly see that after extrapolation using Eq.~\eqref{eq:extrapolation} these relative errors drop to $~1\%$ (the dotted black line).

\begin{figure}[h]
\includegraphics[width = 0.45\linewidth]{01_d2_scaling_with_extrapolation.pdf}
\includegraphics[width = 0.45\linewidth]{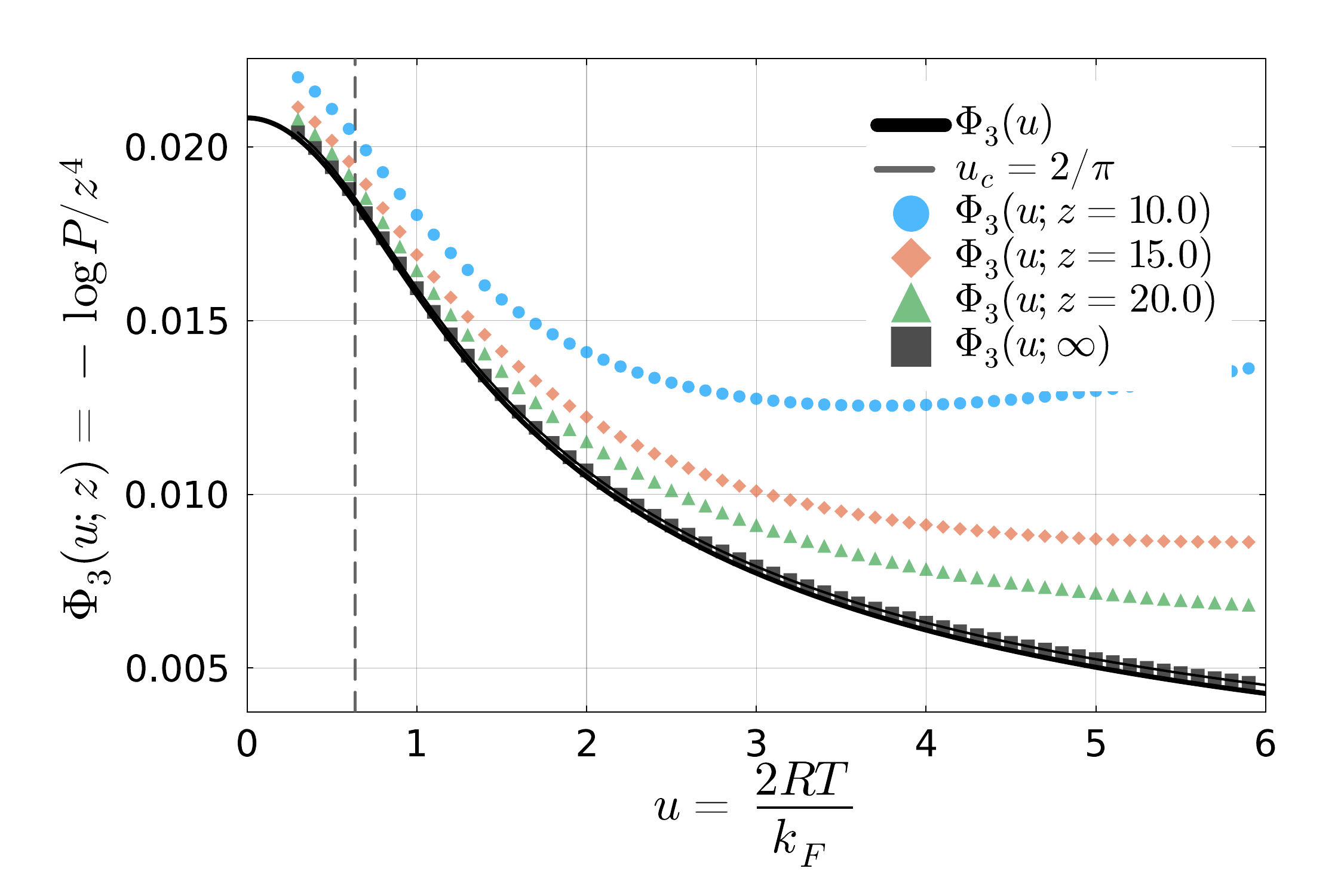}
\caption{Plots of the $d=2$ (left) and $d=3$ (right) scaling functions compared to the numerical simulations via Fredholm determinant \cite{SM}. Notice how the convergence worsen as $u$ grows. The extrapolated value $\Phi_{d}(u;\infty)$ is obtained taking into account the finite $z$ corrections as explained in \cite{SM}.}\label{Fig_conv2}
\end{figure}

\noindent {\bf Remark}. In the regime $u \gg 1$ one can, using matching on the high $T$ Widom regime, obtain the form of the finite $z$ correction. A detailed calculation based on the formula given in Section~\ref{sec:widom} leads to 
\bea \label{limits2}
&& A_d(t) (R \sqrt{T})^{d} \simeq z^{d+1}  \Phi_d(u)|_{1/u} ( 1 + \frac{\pi^2 (d+2)}{12} \frac{u^2}{z^2} ) \\
&& B_d(t) (R \sqrt{T})^{d-1} \simeq z^{d+1}  \Phi_d(u)|_{1/u^2} + \frac{z^{d-1}}{6 \Gamma(d)} \log(u/z)  \\
&& = z^{d+1}  \Phi_d(u)|_{1/u^2} (1 + \frac{(d^2 + 4 d + 3) u^2}{48 z^2} \log(u/z) ) \;. 
\eea 
Hence, up to the logarithmic corrections, it is consistent with our ansatz \eqref{eq:extrapolation}, at least for $u>u_c$. 

\begin{figure}[t!]
    \centering
\includegraphics[width=0.45\linewidth]{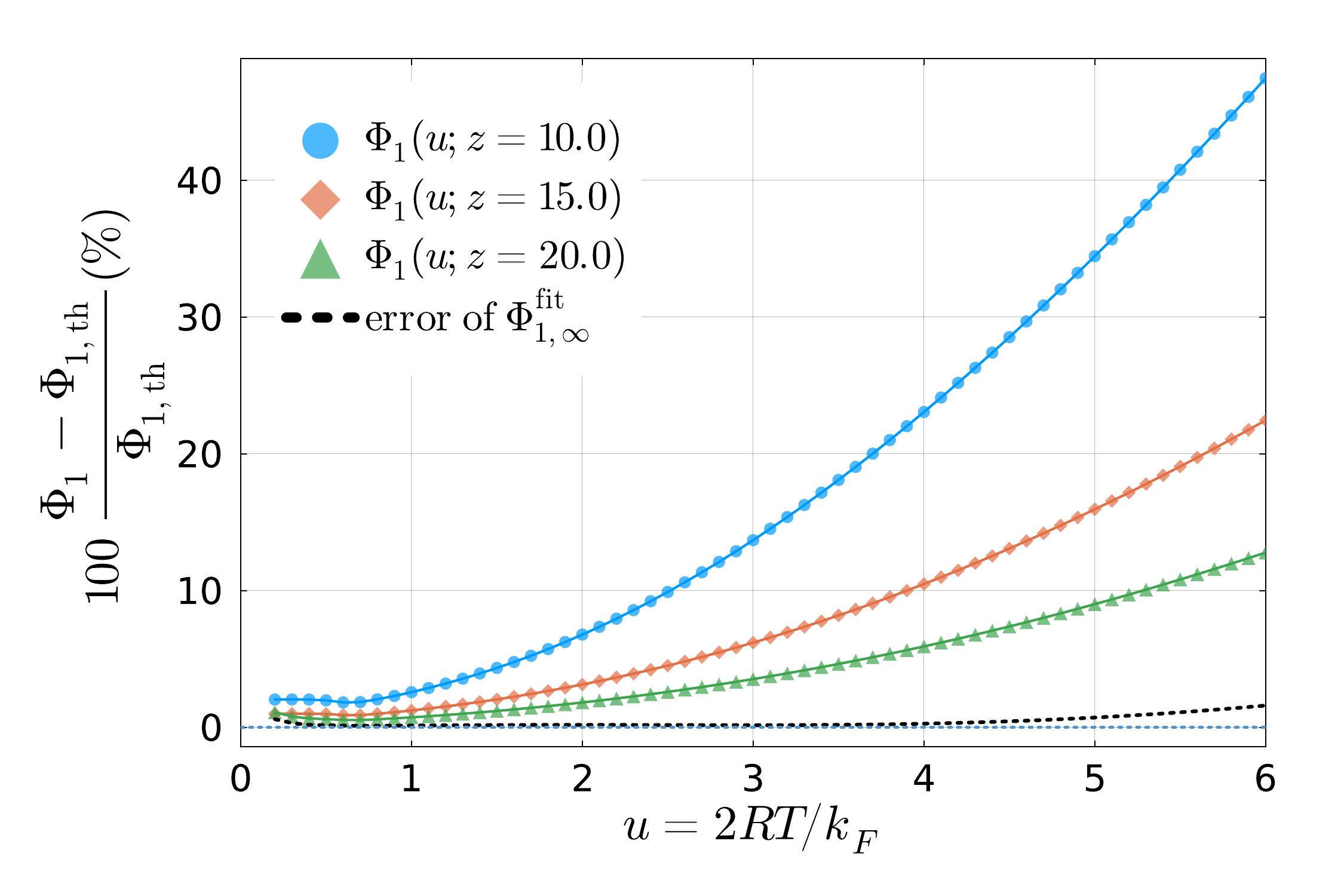}
\includegraphics[width=0.45\linewidth]{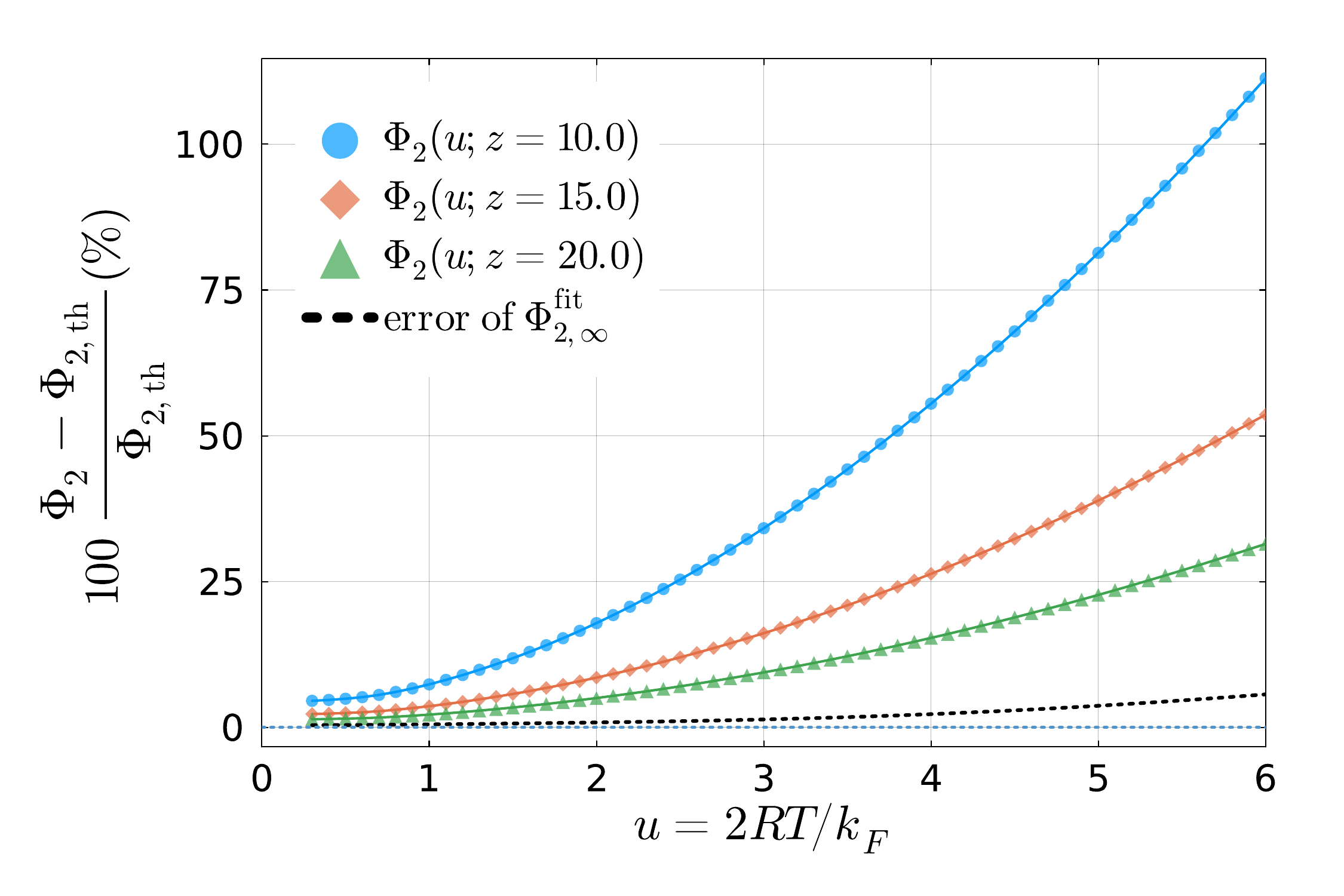}
\includegraphics[width=0.45\linewidth]{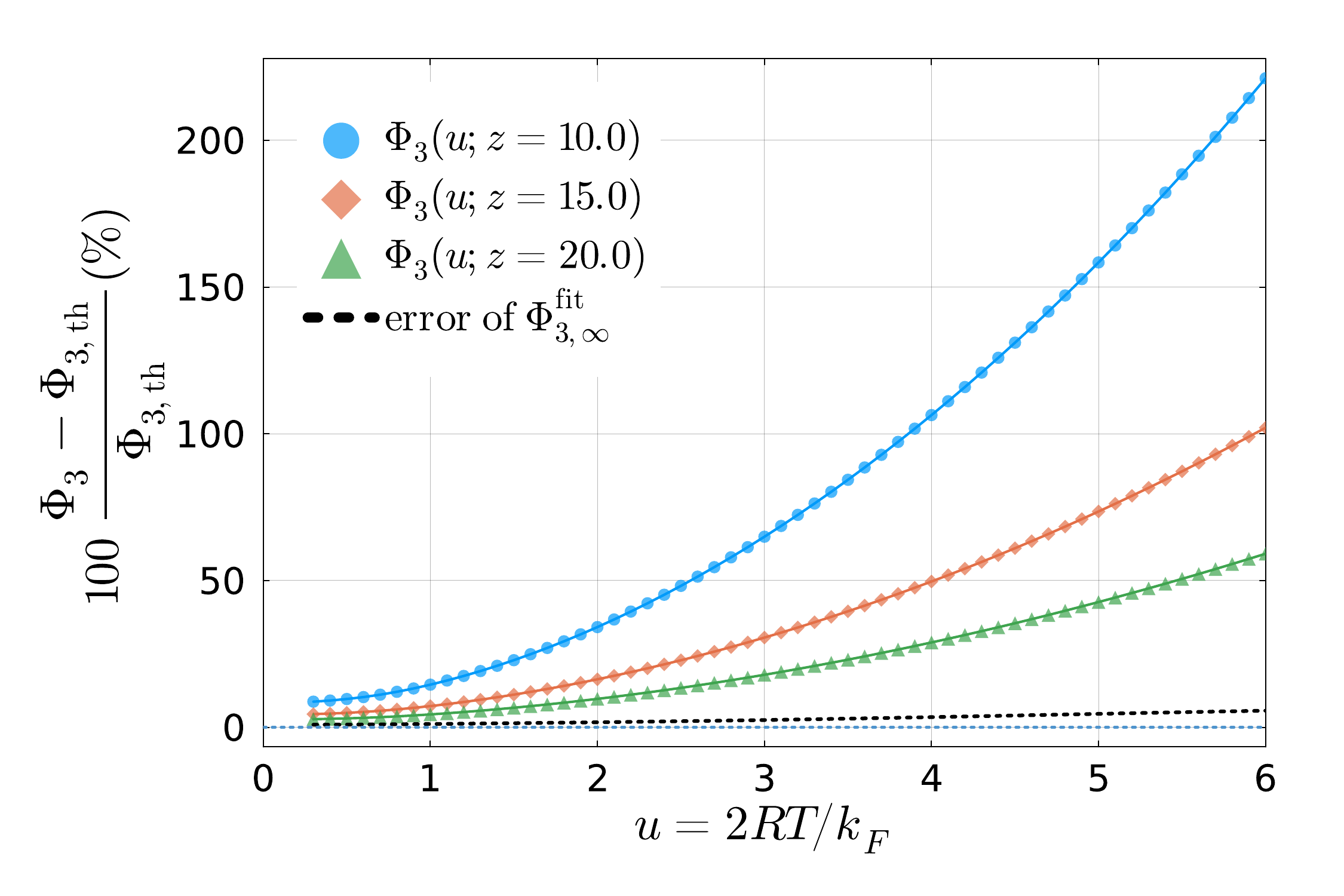}
    \caption{Percentage errors between the theoretical prediction for $\Phi_d(u)$ (Eq.~\eqref{phid} in the main text) and the numerically computed values using our algorithm $\Phi_d^{\rm num}(u)$ before (markers) and after the extrapolation to $z=+\infty$. The dashed black line shows the drastic improvement of the relative error after extrapolation from $\sim50\%$ at $u_{\max}\approx 6$ for $z=20$ up to $\sim 1\%$ at the same $u_{\max}$.}
    \label{fig:errors}
\end{figure}

\subsection{Numerical details on the number variance}
In this short Section we present some additional figures on the scaling function ${\cal V}_d(u)$ describing the number variance crossover from low to high temperatures. The theoretical predictions are presented in Eq.~\eqref{eq:varsd1} for $d=1,2,3$ respectively. These are compared to the direct numerical quadrature of Eq.~\eqref{var1} after subtracting the zero temperature variance in eq.~\eqref{varT0}. In practice, in Fig.~\ref{fig:variance_collapse} we plot
\begin{equation}
    {\cal V}_d(u) = z^{1-d}\left(\langle {\cal N}_R^2 \rangle^c-\langle {\cal N}_R^2 \rangle^c|_{T=0}\right)\,.
\end{equation}
See Eq.~\eqref{varscaling} and Section~\ref{app:var_scaling} for details.
From Fig.~\ref{fig:variance_collapse}, we can see that as $z$ grows the data (markers) collapse on the theoretical predictions (full line) in all dimensions. 
\begin{figure}[b!]
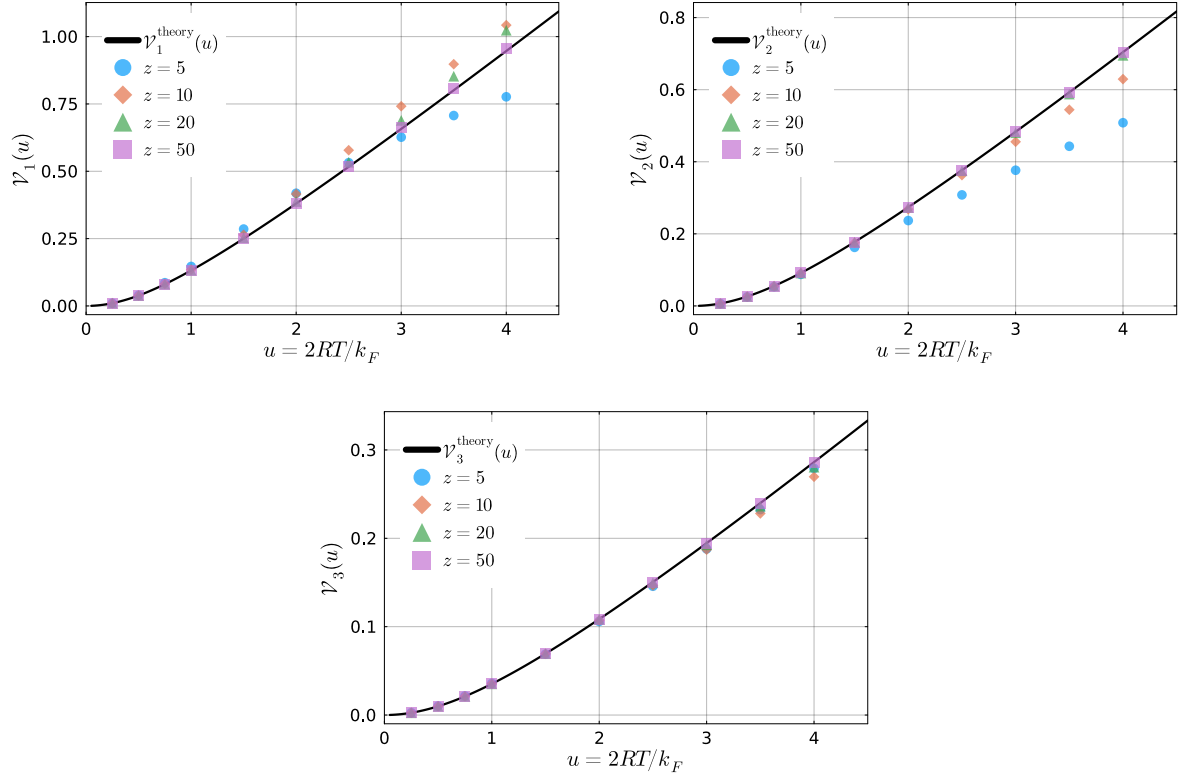

    \centering
    \includegraphics[width=0.45\linewidth]{collapse_d1.pdf}
    \includegraphics[width=0.45\linewidth]{collapse_d2.pdf}
    \includegraphics[width=0.45\linewidth]{collapse_d3.pdf}
    \caption{Scaling collapse of the functions ${\cal V}_d(u)$ describing the crossover of the variance from the extensive high temperature result to the sub-extensive low temperature result. Theoretical predictions correspond to the black full line while the numerical data for different values of $z = k_F R$ corresponds to the markers. As $z$ grows, we enter the scaling regime and the data collapse on the theoretical curves as predicted by Eq.~\eqref{eq:varsd1} valid for dimensions $d=1,2,3$ respectively.}
    \label{fig:variance_collapse}
\end{figure}


\end{widetext}

\begin{thebibliography}{99}
\bibitem{thermo} 
{The volume term can be obtained from simple thermodynamics, i.e. $T^{d/2} A_d(t)= \beta V_d (p(\tilde \mu=-\infty,T)-p(\tilde \mu,T))$ where $p(\tilde \mu,T)$ is the pressure of the Fermi gas, see Eq.~\eqref{thermowidom}
in \cite{SM}.}
\bibitem{lambda}
The upper limit $\lambda<1$ arises since the centrifugal energy $\frac{\nu^2-1/4}{2 R^2}$
is bounded by $\mu$.
\end{thebibliography}
\end{document}